\documentclass[10pt]{article}
\usepackage[left=.75in,right=.75in,top=.75in,bottom=.75in]{geometry}
\usepackage[utf8]{inputenc}
\usepackage{amssymb,amsmath,latexsym,amsthm,amsfonts,mathtools}

\usepackage{mathrsfs}
\usepackage{multirow}
\usepackage{graphicx}
\usepackage{textcomp,siunitx}
\usepackage{float}
\usepackage{subcaption}
\usepackage{booktabs}
\usepackage{enumitem}

\usepackage[hyperref]{xcolor}
\definecolor{ao(english)}{rgb}{0.0, 0.5, 0.0}
\usepackage{hyperref}
\hypersetup{colorlinks, breaklinks, citecolor=blue, linkcolor=ao(english), urlcolor=blue}
\usepackage[nameinlink,noabbrev,sort&compress]{cleveref}
\Crefname{equation}{Eq.}{Eqs.}
\Crefname{figure}{Fig.}{Figs.}
\Crefname{assumption}{Assumption}{Assumptions}
\Crefname{algocfline}{Algorithm}{Algorithms}
\usepackage{setspace}
\usepackage{cite}
\allowdisplaybreaks

\theoremstyle{plain}
\newtheorem{theorem}{Theorem}
\newtheorem{lemma}{Lemma}

\theoremstyle{remark}
\newtheorem{remark}{Remark}

\theoremstyle{definition}
\newtheorem{definition}{Definition}

\newcommand{\sign}{\mathrm{sign}}
\newcommand{\diag}{\mathrm{diag}}
\newcommand{\U}{\mathcal{U}}

\usepackage[linesnumbered,ruled]{algorithm2e}
\SetKwInput{KwInput}{Input}          
\SetKwInput{KwOutput}{Output}

\usepackage{newtxmath}

\usepackage{anyfontsize}
\usepackage{accents}

\usepackage{scalerel}
\usepackage{tikz}
\usetikzlibrary{automata, shapes, arrows, calc, arrows.meta, fit, positioning}
\usetikzlibrary{svg.path}
\definecolor{orcidlogocol}{HTML}{A6CE39}
\tikzset{
	orcidlogo/.pic={
		\fill[orcidlogocol] svg{M256,128c0,70.7-57.3,128-128,128C57.3,256,0,198.7,0,128C0,57.3,57.3,0,128,0C198.7,0,256,57.3,256,128z};
		\fill[white] svg{M86.3,186.2H70.9V79.1h15.4v48.4V186.2z}
		svg{M108.9,79.1h41.6c39.6,0,57,28.3,57,53.6c0,27.5-21.5,53.6-56.8,53.6h-41.8V79.1z M124.3,172.4h24.5c34.9,0,42.9-26.5,42.9-39.7c0-21.5-13.7-39.7-43.7-39.7h-23.7V172.4z}
		svg{M88.7,56.8c0,5.5-4.5,10.1-10.1,10.1c-5.6,0-10.1-4.6-10.1-10.1c0-5.6,4.5-10.1,10.1-10.1C84.2,46.7,88.7,51.3,88.7,56.8z};
	}
}

\newcommand\orcidicon[1]{\href{https://orcid.org/#1}{\mbox{\scalerel*{
				\begin{tikzpicture}[yscale=-1,transform shape]
					\pic{orcidlogo};
				\end{tikzpicture}
			}{|}}}}

\title{Three-dimensional Nonlinear Path-following Guidance with Bounded Input Constraints}

\author{Saurabh~Kumar\textsuperscript{\orcidicon{0000-0002-8344-5966}}
	\thanks{Corresponding author.\newline Saurabh Kumar and Shashi Ranjan Kumar are with the Intelligent Systems \& Control Lab, Department of Aerospace Engineering, Indian Institute of Technology Bombay, Powai-- 400076, Mumbai, India. Abhinav Sinha is with the Guidance, Autonomy, Learning and Control for Intelligent Systems (GALACxIS) Lab, Department of Aerospace Engineering and Engineering Mechanics, University of Cincinnati, Cincinnati, OH, 45221,  USA
 e-mails: saurabh.k@aero.iitb.ac.in, srk@aero.iitb.ac.in, abhinav.sinha@uc.edu.}	
    \and Shashi~Ranjan~Kumar\textsuperscript{\orcidicon{0000-0001-6446-7281}}
	\and Abhinav~Sinha\textsuperscript{\orcidicon{0000-0001-6419-2353}}}	
 
\date{}
\begin{document}
\maketitle
\doublespacing

\begin{abstract}
In this paper, we consider the tracking of arbitrary curvilinear geometric paths in three-dimensional output spaces of unmanned aerial vehicles (UAVs) without pre-specified timing requirements, commonly referred to as path-following problems, subjected to bounded inputs. Specifically, we propose a novel nonlinear path-following guidance law for a UAV that enables it to follow any smooth curvilinear path in three dimensions while accounting for the bounded control authority in the design. The proposed solution offers a general treatment of the path-following problem by removing the dependency on the path's geometry, which makes it applicable to paths with varying levels of complexity and smooth curvatures. Additionally, the proposed strategy draws inspiration from the pursuit guidance approach, which is known for its simplicity and ease of implementation. Theoretical analysis guarantees that the UAV converges to its desired path within a fixed time and remains on it irrespective of its initial configuration with respect to the path. Finally, the simulations demonstrate the merits and effectiveness of the proposed guidance strategy through a wide range of engagement scenarios, showcasing the UAV's ability to follow diverse curvilinear paths accurately.
	\medskip
 
	\noindent \emph{\textbf{Keywords}}--- UAV, nonlinear guidance, path-following, bounded input, fixed-time convergence, input saturation.
\end{abstract}

\section{Introduction}\label{introduction}
With the advancement in guidance, navigation, and control technology, the applicability of unmanned aerial vehicles (UAVs) has now been expanded to many civilian and disaster management tasks in addition to its earlier military applications. This is due to the fact that UAVs offer more flexibility for repetitive and dangerous missions. Some of the applications include search and rescue, aerial mapping, pipeline inspection, boundary line surveillance, forest fire monitoring, and so on (see, for example, \cite{doi:10.2514/1.I010843,10003994,doi:10.1177/09544100221115238}). From a guidance and control perspective, the primary focus of deploying UAVs in any mission is to first generate a feasible path and then control the UAV's motion to follow this path. The task of controlling the UAV's motion is known as the motion control problem, and it has garnered significant attention due to its broad range of applications.

There have been three methods utilized to deal with the motion control problem, namely, set-point tracking, trajectory-tracking, and path-following. The set-point tracking problem mainly deals with scenarios in which the vehicle has to reach a particular location and maintain itself there. In the trajectory-tracking control, the temporal evolution of each space coordinate is provided as feedback to the vehicle to generate the necessary control inputs to track these coordinates. On the other hand, path-following relaxes the temporal requirements, and as a consequence, the speed of the vehicle becomes path-independent. This, in turn, provides an extra degree of freedom to the designer, which can be adjusted depending on the mission requirements. Additionally, as shown in \cite{doi:10.1016/j.automatica.2003.10.010,1393141,doi:10.1016/j.automatica.2007.06.030}, the path-following control offers significant advantages over the trajectory-tracking control, such as better transient performance, smoother convergence, and better steady-state behavior in the presence of unstable zero dynamics. A survey on path-following guidance laws can be found in \cite{6712082}. However, the path-following control problem still remains an open and challenging problem, mainly due to three reasons: (1) most of the existing solutions are in planar settings, (2) most solutions are developed for special path constructs, for example, straight-line and circular paths, and (3) no consideration of bounded control authority in the design.

Various linear and nonlinear strategies have been used to achieve a precise path-following of the UAVs. For example, the work in \cite{yoshitani2010} discussed a proportional-integral-derivative (PID) controller for autonomous path-following of a fixed-wing aircraft wherein the desired path was given as a sequence of segments of straight-line and circular arc. A tight path-following algorithm based on a PID controller with feedforward compensation was devised in \cite{5602615}. First, several waypoints on the desired path were taken, and then the path between these waypoints was approximated by a cubic spline. It is important to note that these methods relied on linearizing the dynamics around a specific operating point, offering a satisfactory performance as long as deviations from the operating point were minimal. However, they may become ineffective under varying operating conditions and larger deviations. Note that the nonlinear path-following control techniques can be broadly characterized as error-regulation methods, vector field methods, and virtual target point-based methods. These approaches could either be control theoretic or geometric.

The key idea of error-regulation-based approaches, for example, \cite{doi:10.2514/6.2003-6626,doi:10.2514/1.46679,doi:10.2514/6.2023-1055,doi:10.1016/j.ifacol.2023.03.036,doi:10.2514/1.42056,9721593}, is to first define a suitable error variable such as cross-track error, heading angle error, course angle error, etc., and then subsequently design a nonlinear controller to nullify these error variables to zero. For example, a \emph{good helmsman} problem-based strategy was presented in \cite{doi:10.2514/6.2003-6626}, where the helmsman behavior was modeled using the desired relative course angle and the cross-track distance. Thereafter, the commanded course angle was designed to follow straight-line and circular paths based on these variables. Another approach considering a moving {ghost vehicle} onto the desired path was discussed in \cite{doi:10.2514/1.46679}, and exponential convergence of the error variables was shown except when the velocity vector becomes perpendicular to the desired path.  A lateral cross-track error-based approach was presented in \cite{doi:10.2514/6.2023-1055,doi:10.1016/j.ifacol.2023.03.036}. To achieve the path-following objective, a design consisting of guidance and control subsystems was presented in \cite{doi:10.2514/1.42056}.  The work in \cite{9721593} used cross-track error in the horizontal plane and linearized dynamics to meet the path-following objective. 

In vector field-based methods such as \cite{1657648,doi:10.2514/6.2006-6467,doi:10.2514/1.34896,doi:10.2514/1.G004053,6494384}, a vector field that surrounds the desired path by matching its desired course angle, velocity, etc., was first constructed, and then the guidance laws were designed in such a way that the vehicle converges to the desired path along it. For instance, the work in \cite{1657648} constructed a vector field around the desired path (straight-line and circular), and then the commanded heading angles were obtained using the inverse tangent function and position error. Later, the design was extended for curve paths in \cite{doi:10.2514/6.2006-6467}. A method for the construction of a globally attractive vector field, which inherently has the Lyapunov stability properties, was discussed in \cite{doi:10.2514/1.34896}. The authors in \cite{6494384} augmented the Lyapunov vector field with a tangent vector field and presented a two-stage vector field algorithm as a tangent-plus-Lyapunov vector field. The work in \cite{8385190} discussed a guidance law by integrating vector field with the input-to-state stability notation. A gradient vector field-based path following strategy was discussed in \cite{doi:10.2514/1.G004053}. It is worth mentioning that these methods offered global convergence of the UAV to its desired path. However, these strategies were primarily focused on special types of path constructs (mostly straight-line and circular), and in the case of multiple path segments, they require a separate switching logic, making them complicated. As a matter of fact, they cannot be directly extended for a generic curvilinear path-following problem. Additionally, only asymptotic or exponential stability was achieved with these methods. 

In the virtual target point-based approaches, such as those in \cite{doi:10.2514/1.28957,doi:10.2514/1.G001060,doi:10.2514/6.2024-1593,doi:10.1016/j.ast.2024.109225}, a virtual target point was chosen on the desired path first, and then the guidance laws were designed to steer the vehicle toward this target, resulting in effective path-following behavior. The work in \cite{doi:10.2514/1.28957} introduced a nonlinear path-following guidance law by projecting the UAV onto the desired path and choosing a pseudo-target at $L$ distance ahead of the projection. While the method had similarity with the proportional-navigation guidance, the choice of this $L$ was restrictive. The method ensured the asymptotic convergence of the UAV to its desired path. Using the differential geometry of the space curves, the authors in \cite{doi:10.2514/1.G001060} presented a three-dimensional path-following strategy. Another nonlinear path-following guidance strategy was presented in \cite{doi:10.2514/6.2024-1593} and later extended in the presence of wind in \cite{doi:10.1016/j.ast.2024.109225}. A path generation and tracking strategy for the UAV in three dimensions was presented in \cite{4908914}. The authors in \cite{doi:10.2514/1.58259} introduced a three-dimensional path-following technique by projecting the UAV dynamics into three separate channels: roll, pitch, and yaw. The work in \cite{9610125} discussed a three-dimensional guidance strategy for combat UAVs based on the concept of a look-ahead vector.

While there is a wide body of literature focused on the challenge of path-following, most of these methods are limited to two-dimensional settings, and their extension to three-dimensional settings is non-trivial. Some works, such as \cite{4908914,doi:10.2514/1.58259,9610125}, have addressed the three-dimensional path-following problem, but they typically do so by decoupling the UAV's motion into multiple planes, effectively reducing it to a two-dimensional path-following problem. Furthermore, almost all of the above-mentioned works do not consider the bounds on the available control inputs in the design. The UAVs always have limited control authority due to the physical constraints of the actuators. Consideration of bounded control input during design will bring it closer to real-world applications. Some of the works, for example, \cite{6740857,7172278}, considered the input saturation in the design within a two-dimensional setting. Motivated by these, in this work, we propose a solution to the three-dimensional path-following problem of the UAV with the limited available control authority. We summarize our contributions below:

We develop a novel path-following guidance law for a UAV that steers it onto its desired path in a three-dimensional setting while accounting for limited control authority in the design. This makes the proposed solution more practical and suitable for real-world deployment. To the best of the authors' knowledge, this work is the first to propose a path-following guidance law while accounting for the actuator saturation in the design for the three-dimensional generic path.

The proposed guidance law steers the UAV onto its desired path using limited information about the path,  which is another alluring feature of the proposed strategy. The proposed three-dimensional path-following solution is expected to offer greater flexibility and adaptability in UAV motion compared to two-dimensional alternatives, enabling the UAV to perform more complex tasks in various scenarios, such as urban air mobility and space exploration.

Unlike all of the existing methods, the proposed guidance law offers a more generalized approach to the path-following problem by eliminating the reliance on the path's geometry and vehicle's class. The majority of the previous approaches often assume the path to be a straight line, a circular arc, or a combination of these two primitives. Additionally, when the desired path consists of multiple segments, a separate and often complex mid-way guidance strategy is required.  In contrast, the proposed guidance strategy is independent of the path type, allowing smoother convergence to the paths without the need for switching, making it inherently more versatile. The proposed guidance strategy remains applicable to vehicles capable of either maintaining zero linear speed (for example, multi-rotors) or requiring a non-zero minimum speed (for example, fixed-wing aircraft). Consequently, the proposed guidance laws are \emph{adaptable} to various classes of vehicles.
     
The proposed method considers a persistent influence of UAV's longitudinal motion on lateral motion and vice versa during the design of guidance strategy, unlike the previous methods, such as in \cite{4908914,doi:10.2514/1.58259,9610125} where the system dynamics were decoupled, and the three-dimensional problem was treated as separate two-dimensional problems. Therefore, the proposed approach preserves the strong cross-coupling between the yaw and pitch channels. To the best of the authors' knowledge, these factors have not yet been fully integrated into the general path-following problem. However, their inclusion is crucial for effectively addressing a wide range of motion planning challenges. 
      
The proposed path-following guidance law guarantees global convergence to the path by achieving fixed-time convergence of the concerned error variables. This essentially means that the UAV will converge to its desired path within a fixed time and remain on it thereafter, regardless of its initial configuration with respect to the path. Additionally, the settling time for different variables could be customized according to specific requirements, using only the design parameters. This flexibility provides an additional degree of freedom to the designer.

 Unlike the method in \cite{doi:10.2514/1.G007964}, the proposed approach regulates two error variables, the UAV's lead angles in the azimuth and elevation directions, to zero, rather than just one error variable, the UAV's effective heading angle. This approach eliminates the need for a potentially complex control allocation scheme. As a result, it is expected to be less computationally expensive than the method discussed in \cite{doi:10.2514/1.G007964}. Additionally, setting the individual lead angles in the pitch and yaw directions to zero is more intuitive and provides a direct approach to regulating these variables using the corresponding angular velocities in each channel. Furthermore, the design is \cite{doi:10.2514/1.G007964} does not consider bounds on the control inputs during design, which is unlike the present work.

The rest of this paper is structured as follows: After an overview of the existing works in \Cref{introduction}, \Cref{sec:problem} introduces the UAV's kinematic model and outlines the problem statement. The bounded-input guidance law is developed in \Cref{sec:mainresults}, followed by performance evaluations of the developed guidance laws in \Cref{sec:simulation}. Finally, \Cref{sec:conclusion} offers concluding remarks and suggests some directions for future research.

\section{Dynamical Model}\label{sec:problem}
\begin{figure}[!ht]
\centering
\includegraphics[width=0.75\linewidth]{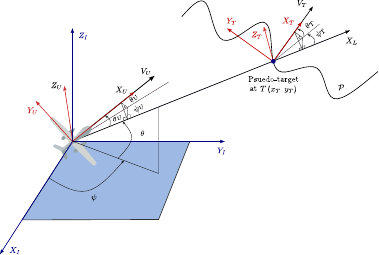}
  \caption{UAV-target path following scenario.}
  \label{fig:engagement}
\end{figure}

In this section, we first present the three-dimensional path-following scenario and then formulate the problem addressed in this paper. Consider a scenario between a UAV and a predefined path ($\mathcal{P}$) in the three-dimensional space as illustrated in \Cref{fig:engagement}. We assume the path to be a smooth three-dimensional curve whose curvature is unknown but bounded. The UAV is also assumed to be a point-mass non-holonomic vehicle. To characterize the UAV's equations of motion, four reference frames-- inertial, line-of-sight (LOS), body, and target frame of references are considered. The target, body, and inertial frame of references are denoted by the mutually orthogonal axes, $(X_T, Y_T, Z_T)$, $(X_U, Y_U, Z_U)$, and $(X_I, Y_I, Z_I)$, respectively. The LOS frame of reference is denoted using the subscript L, and it is defined with respect to the inertial frame using elevation and azimuth angles, $\theta$ and $\psi$, respectively. The linear speed of the UAV is represented by $V_U$, and initially, it is assumed to be aligned with the body frame's $X$-axis. The orientation of $V_{U}$ in the azimuth and the elevation direction is defined from the LOS frame using the lead angles $\psi_{U}$ and $\theta_{U}$, respectively. 

Note that the desired path $\mathcal{P}$ can be thought of as a continuous progression of a moving target's position. In other words, it can be regarded as the trajectory of an arbitrarily moving target, which we will refer to as a virtual target or pseudo-target from here on. Let $T$ denote the said pseudo-target on the desired path, moving with speed $V_T$ (either constant or time-varying) and having lead angles in azimuth and elevation directions as $\psi_T$ and $\theta_T$. It is important to note that the rate of change of these lead angles, $\dot{\psi}_T$ and $\dot{\theta}_T$, is unknown to the UAV's guidance systems. As a result, the path curvature information remains unavailable to the UAV. Therefore, the predefined path $\mathcal{P}$ represents a generic path with an unknown curvature that the vehicle needs to follow. Since we are dealing with smooth paths, it is reasonable at this point to assume that the path's curvature is bounded. The dynamics that describe the relative motion between the UAV and the pseudo-target along the desired path in such scenarios can be expressed as follows:
\begin{subequations}\label{eq:engkinematics}
		\begin{align}
			\dot {r}&=V_T\cos{\theta_T}\cos{\psi_T}-V_U\cos{\theta_U}\cos{\psi_U},\label{eq:rdot}\\
			\dot{\theta}&=\dfrac{V_T\sin{\theta_T}-V_U\sin{\theta_U}}{r},\label{eq:thetadot}\\
            \dot{\psi}&=\dfrac{V_T\cos\theta_T\sin\psi_T-V_U\cos\theta_U\sin\psi_U}{r\cos\theta},\label{eq:psidot}\\ 
			\dot{\theta}_U&=\omega_U^z-\dot{\psi}\sin\theta\sin\psi_U-\dot{\theta}\cos\psi_U,\label{eq:thetaUdot}\\
		\dot{\psi}_U&=\dfrac{\omega_U^y}{\cos\theta_U}+\dot{\psi}\tan\theta_U\cos\psi_U\sin\theta-\dot{\psi}\cos\theta-\dot{\theta}\tan\theta_U\sin\psi_U,\label{eq:psiUdot}
		\end{align}
	\end{subequations}
where $r$ and $\theta$ denote the relative range and the LOS angle of the UAV with respect to the pseudo-target. We refer to $\dot{r}$ and $\dot{\theta}$ as the range rate and the LOS rate, respectively, from hereafter. The terms $\omega_U^y$ and $\omega_U^z$ represent the angular velocities in mutually orthogonal planes, which, together with the linear speed $V_{U}$, serve as control inputs to the UAV. One may observe from \Cref{eq:engkinematics} that the relative range and the UAV's lead angles ($\theta_U$, $\psi_U$) dynamics have a relative degree of one with respect to their corresponding control inputs.

Note that the first three expressions (\Cref{eq:rdot,eq:thetadot,eq:psidot}) represent the relative velocity dynamics of the UAV in the LOS frame of reference, while the last two (\Cref{eq:thetaUdot,eq:psiUdot}) denote the UAV's turn rates in the pitch and yaw planes. We further define the UAV's effective heading angle as the angle formed between the $X$-- axis of the body frame and the LOS frame. One can calculate the UAV's effective heading angle in the LOS frame using lead angles $\psi_U$ and $\theta_U$ based on the complementary relationship  $\cos\sigma_U = \cos\psi_U\cos\theta_U$.

Needless to say, the control inputs to the UAV, namely linear speed $V_U$ and angular velocities $\omega_{U}^y$ and $\omega_{U}^z$ are always bounded by the physical constraints of the actuators. The majority of the existing approaches consider the bounds of the inputs in the simulation/implementation in an ad hoc manner, while the control remains unbounded during the design. This can lead to performance degradation during real-time validation. However, accounting for bounded control authority in the design itself not only provides rigorous theoretical guarantees relating to safety but also brings the design closer to real-world deployment.

For the geometry shown in \Cref{fig:engagement}, the primary objective is to devise a simple and effective three-dimensional path-following guidance law that will steer the UAV onto its desired path, regardless of its initial configuration relative to the path, using limited information about the path. In other words, for any arbitrary initial position say $\mathbf{P}(0)\in \mathbb{R}^3$, initial heading angles $\theta_U(0)$ and $\psi_U(0) \in \mathbb{R}$, and a smooth path $\mathcal{P}$, we aim to design a guidance strategy $\mathcal{G}$ (comprising $V_U$, $\omega_U^y$, and $\omega_U^z$) such that $r$, $\psi_U$, and $\theta_U \to 0$ within a fixed time $t_f$. The second objective is to keep the control inputs (linear speed and angular velocities) within a bounded safe set. Note that our design does not decouple the engagement dynamics (as can be observed from \Cref{eq:engkinematics}) into longitudinal and lateral planes. Thus, we consider the persistent influence of longitudinal and lateral dynamics on each other, which, albeit challenging, allows for greater flexibility in a wide range of motion planning scenarios. Additionally, the guidance law will be derived within a nonlinear framework to ensure its effectiveness across a broad range of operating conditions.

We now present certain preliminary results that aid the design in this paper. Consider the system:
\begin{equation}\label{eqn:dyn_system}
    \dot{\varpi}=\mathcal{F}(t,\varpi),
\end{equation}
where $\varpi \in \mathbb{R}^n$ is the system states and $\mathcal{F}:\mathbb{R}_{+} \times \mathbb{R}^n\rightarrow \mathbb{R}^n$ is a nonlinear function. The initial condition for the given system is denoted by $\varpi(0)=\varpi_0$, and the solutions of \Cref{eqn:dyn_system} are understood in the sense of Fillipov. Without loss of generality, we assume the origin to be an equilibrium point of the system given by \Cref{eqn:dyn_system}.
\begin{definition}(\cite{doi:10.1137/S0363012997321358})
    The origin of the system \Cref{eqn:dyn_system} is said to be globally finite-time stable if it is globally asymptotically stable and any solution $\varpi(t,\varpi_0)$ reaches the origin in finite time from an arbitrary initial condition $\varpi(0)$. In other words, $\varpi(t,\varpi_0) = 0$ $\forall$ $t \geq \mathcal{T}(\varpi_0)$, where $\mathcal{T}: \mathbb{R}^n \to \mathbb{R}_+ \cup {0}$ is known as the settling time function
\end{definition}
\begin{definition}(\cite{6104367})
  The origin of the system \Cref{eqn:dyn_system} is said to be fixed-time stable if it is globally finite-time stable and the settling time function remains bounded on $\mathbb R^n$ by a known positive constant, that is, there exists  $\mathcal{T}_{\mathrm max}\in \mathbb R_+$ such that $\forall$ $\varpi \in \mathbb R^n$, $\mathcal{T}(\gamma_0)< \mathcal{T}_{\max}$. In other words, every solution, $\varpi(t,\varpi_0)$ of the system will converge to the origin within $ \mathcal{T}_{\max}$.
\end{definition}
\begin{definition} (\cite{6104367})
    The set $\mathfrak{R}$ is said to be globally finite-time attractive for the system \Cref{eqn:dyn_system} if its any solution $\varpi(t,\varpi_0)$ reaches the set $\mathfrak{R}$ in some finite time $\mathcal{T}(\varpi_0)$ and remains confined within the same set for all time $t\geq \mathcal{T}(\varpi_0)$.
\end{definition}
\begin{definition}(\cite{6104367})
  The set $\mathfrak{R}$ is said to be fixed-time attractive for system \Cref{eqn:dyn_system} if it is globally finite-time attractive and the settling time function remains globally bounded by a known positive constant.   
\end{definition}
\begin{lemma} (\cite{6104367})\label{lem:fixedtime}
    If there exists a continuous radially unbounded function, $\mathcal{V}(\varpi):\mathbb R^n\to \mathbb{R}_+ \cup \{0\}$ such that (1) $\mathcal{V}(\varpi)=0$ $\implies$ $\varpi \in \mathfrak{R}$; (2) any solution  $\varpi(t,\gamma_0)$ of  the system \Cref{eqn:dyn_system} satisfies 
    \begin{equation*}
        \dot{\mathcal{V}}(\varpi)\leq-\left[\kappa_1\mathcal{V}^\alpha(\varpi)+\kappa_2\mathcal{V}^\beta(\varpi)\right],
    \end{equation*}
    for some $\kappa_1,\kappa_2,\alpha,\beta>0$ such that $\alpha>1$ and $0<\beta<1,$ then the origin is fixed-time stable and the settling time, $\mathcal{T}(\varpi_0),$ is estimated by
    \begin{equation}
        \mathcal{T}(\varpi_0)\leq \mathcal{T}_{\max} \triangleq \dfrac{1}{\kappa_1\left(\alpha-1\right)}+\dfrac{1}{\kappa_2\left(1-\beta\right)}.
    \end{equation}
\end{lemma}
 \begin{lemma}(\cite{doi:10.1080/00207179.2013.834484}) \label{lem:inequality_zuo}
For nonnegative numbers $a_1$, $a_{2}$, $\cdots$, $a_n$, and the constants $p>1$ and $0<q\leq 1$ the following inequality holds
\begin{equation}
    \sum_{i=1}^{n} a_{i}^q \geq  \left(  \sum_{i=1}^{n} a_i\right)^q, \;
    \sum_{i=1}^{n} a_{i}^p \geq  n^{(1-p)}\left(  \sum_{i=1}^{n} a_i\right)^p
\end{equation}
 \end{lemma}
\section{Derivation of the Guidance Strategy} \label{sec:mainresults}
In this section, we design the necessary guidance commands for the UAV to steer it onto its desired path in the three-dimensional setting without path curvature information and bounded control authority. The proposed guidance law will eliminate the dependency on the path geometry and the assumption of the infinite available control inputs in the design. The former ensures that the proposed strategy remains valid for any arbitrary smooth path, while the latter brings the design closer to real-world applications, as the maximum available control for any practical system will always be bounded. 

As stated earlier, the desired path can be thought of as a trajectory of an arbitrarily moving pseudo-target, $T$. Therefore, one may think of a portion of the three-dimensional path-following problem as regulating the instantaneous relative separation between the UAV and the pseudo-target to zero. To that end, we first design the UAV's linear speed control input, $V_{U}$, to ensure that the UAV rendezvouses with the pseudo-target. In other words, we aim to design $V_{U}$ such that a target set $\mathscr{T}=\{ \mathscr{S} \rvert$ $r=0$  $\forall$ $t > t_{f} \}$, where $\mathscr{S}$ is the set of states $\theta_{U}$, $\psi_{U}$, $\theta_{T}$, and $\psi_{T}$ and $t_{f}$ is some fixed time, is achieved. 

In practice, the actuators have limited capability, and therefore, they can only provide a certain linear speed to the UAV. Hence to account for that, we consider the available linear speed control to be bounded by known constants, that is, $V\textsubscript{0} \leq V_{U} < V_{U}^{\max}$, where $V\textsubscript{0}$ and $V_{U}^{\max}$ are the minimum and maximum values of UAV's linear speed, respectively. Now inspired by the work in \cite{8401917}, we propose an input saturation model as 
\begin{equation} \label{eq:vu_saturation_model}
    \dot{\U} =  \mathcal{K}\textsubscript{1} \left[ 1 - \left(\dfrac{\U}{\U^{\max}}\right)^{\gamma} \right]   \U^{c} - \mathcal{K}\textsubscript{1}\mathcal{K}\textsubscript{2} \U, 
\end{equation}
where $\U = V_{U} -\dfrac{\left(V\textsubscript{0}+V_{U}^{\max}\right)}{2}$ with $\U(0) = 0$, 
$\U^{\max} := \dfrac{ V_{U}^{\max}-V\textsubscript{0}}{2}$, $\mathcal{K}\textsubscript{1}$ and $\mathcal{K}\textsubscript{2}$ are positive constants, the term $\gamma \geq 2n$ such that $n \in \mathbb{N}$ is a constant, and $\U^c$ denotes the commanded linear speed.

Note that the saturation model discussed in \cite{8401917} can only account for the symmetric input constraints. Therefore, with the method given in \cite{8401917}, the linear speed will remains bounded within $-V_{U}^{\max}< V_{U} < V_{U}^{\max}$. However, the linear speed of the UAV cannot be negative, and the said method cannot be applied directly. Consequently, in order to account for the asymmetric linear speed control, the coordinate transformation for $V_{U}$ to $\U$ as $\U = V_{U} - \dfrac{\left(V\textsubscript{0}+V_{U}^{\max}\right)}{2}$ is necessary.
\begin{theorem}\label{thm:linear_speed_sat_model}
Consider the input saturation model as in \Cref{eq:vu_saturation_model}. If the commanded input, $\U^{c}$, remains bounded for all $t \geq 0$, then the model output, $\U$, remains confined to the set $\mathcal{S}_{u} \in \left\{ \U : \lvert \U\rvert < \U^{\max} \right\}$ for all time $t \geq 0$.
\end{theorem}
\begin{proof}
Please refer to  Appendix I for proof.
\end{proof}

It is worth mentioning that the saturation model considered in \Cref{eq:vu_saturation_model} is affine in the control input $\U^c$, and thus a direct derivation of control input $\U^c$ becomes feasible. Additionally, we can consider an auxiliary system augmented with the relative range dynamics given in \Cref{eq:rdot} to guarantee that the linear speed control remains within a predefined safe set. Now, with the linear speed saturation model given in \Cref{eq:vu_saturation_model}, and using the change of coordinate as $\U = V_{U} - \dfrac{\left(V\textsubscript{0}+V_{U}^{\max}\right)}{2}$ and $\dfrac{ V_{U}^{\max}-V\textsubscript{0}}{2} = \U^{\max}$ , the augmented relative range dynamics becomes
\begin{subequations}\label{eq:rdot_vudot_agumented}
\begin{align}
\dot {r}=&~V_T\cos{\theta_T}\cos{\psi_T}-\left( \U + \U^{\max} +V\textsubscript{0}\right)\cos{\theta_U}\cos{\psi_U},\label{eq:rdot_agumented}\\
\dot{\U} = &~\mathcal{K}\textsubscript{1} \left[ 1 - \left(\dfrac{\U}{\U^{\max}}\right)^{\gamma} \right]   \U^{c} - \mathcal{K}\textsubscript{1}\mathcal{K}\textsubscript{2} \U . \label{eq:vu_agumented}
\end{align}
\end{subequations}
It can be inferred from \Cref{eq:rdot_vudot_agumented} that the relative separation between the UAV and the pseudo-target has a relative degree of two with respect to the commanded linear speed $\U^{c}$.

Now, consider $x = \U - \chi$ as another state variable, where $\chi$ is an auxiliary input that facilitates the design of the UAV's commanded linear speed $\U^{c}$. The following theorem summarises the design of   $\U^{c}$.
\begin{theorem}\label{thm:vu}
Consider the equations of relative motion governing the interaction between the UAV and the pseudo-target and the saturation model given in \Cref{eq:rdot_vudot_agumented}. If the UAV's commanded input is designed as
    \begin{equation}\label{eq:vu} 
    \U^{c} =\dfrac{ \mathcal{K}\textsubscript{1}\mathcal{K}\textsubscript{2} \U + \dot{\chi} + \lvert x \rvert\cos{\theta_U}\cos{\psi_U} - \left(\mathcal{M}\textsubscript{1}  x ^{\alpha\textsubscript{1}} + \mathcal{N}\textsubscript{1}  x  ^{\beta\textsubscript{1}}\right) }{\mathcal{K}\textsubscript{1} \left[ 1 - \left(\dfrac{\U}{\U^{\max}}\right)^{\gamma} \right] },
\end{equation}
with the auxiliary control $\chi$ as
\begin{equation}\label{eq:chi}
    \chi = \dfrac{V_{T}\cos{\theta}_{T}\cos{\psi}_{T} - \left(\U^{\max}+V\textsubscript{0}\right)\cos{\theta_U}\cos{\psi_U} + \left(\mathcal{M}\textsubscript{1}  r ^{\alpha\textsubscript{1}} + \mathcal{N}\textsubscript{1}  r  ^{\beta\textsubscript{1}}\right)}{\cos{\theta}_{U}\cos{\psi}_{U}},
\end{equation}
where $\mathcal{M}\textsubscript{1}$, $\mathcal{N}\textsubscript{1}$, $\alpha\textsubscript{1}$, $\beta\textsubscript{1}>0$ with $\alpha\textsubscript{1}>1$ and $0<\beta\textsubscript{1}<1$, then the UAV rendezvouses with the pseudo-target $T$, within a fixed time $T\textsubscript{1}$, given by
\begin{equation}\label{eq:t1}
    T\textsubscript{1} \leq\dfrac{1}{\bar{\mathcal{M}\textsubscript{1}}(\alpha\textsubscript{1}-1)}+\dfrac{1}{\mathcal{N}\textsubscript{1}(1-\beta\textsubscript{1})},
\end{equation}
with $\bar{\mathcal{M}\textsubscript{1}} = 2^{(1-\alpha\textsubscript{1})}\mathcal{M}\textsubscript{1}$, regardless of the initial UAV-target relative separation.
\end{theorem}
\begin{proof}
Please refer to  Appendix II for proof.
\end{proof}

\Cref{thm:vu} essentially infers that the commanded linear speed $\U^c$ guarantees that the UAV converegs to the desired path $\mathcal{P}$ within a fixed-time $T\textsubscript{1}$ and remains on it thereafter for for all times $t \geq T\textsubscript{1}$. Using the results from \Cref{thm:linear_speed_sat_model} together with \Cref{thm:vu}, implies that $\U$ remains in the predefined set $\mathcal{S}_u$, that is, $-\U^{\max} \leq \U \leq \U^{\max}$.

We know that $\U = V_{U} - \dfrac{\left(V\textsubscript{0}+V_{U}^{\max}\right)}{2}$ and $\U^{\max}=\dfrac{ V_{U}^{\max}-V\textsubscript{0}}{2}$, and thus one may write
\begin{equation*}
   -\dfrac{ V_{U}^{\max}-V\textsubscript{0}}{2} < V_{U} - \dfrac{\left(V\textsubscript{0}+V_{U}^{\max}\right)}{2} < \dfrac{ V_{U}^{\max}-V\textsubscript{0}}{2},
\end{equation*}
which, on solving for $V_U$, yields 
\begin{align*}
   -\dfrac{ V_{U}^{\max}-V\textsubscript{0}}{2} +\dfrac{\left(V\textsubscript{0}+V_{U}^{\max}\right)}{2} <& V_{U}  < \dfrac{ V_{U}^{\max}-V\textsubscript{0}}{2}+\dfrac{\left(V\textsubscript{0}+V_{U}^{\max}\right)}{2},\\
   &~\implies V\textsubscript{0} < V_{U}  < V_{U}^{\max}.
\end{align*}
Therefore, the UAV's linear speed control always remains in the safe set $\mathcal{S}_{v} \in \left\{ V_U: V\textsubscript{0} <V_U < V_U^{\max} \right\}$ for all time $t \geq 0$.
\begin{remark}
    The choice of the minimum available linear speed control $V\textsubscript{0}$ enables the proposed guidance strategy to be applicable to both multi-rotor and fixed-wing aerial vehicles.  For example, setting $V\textsubscript{0}=0$ makes the design suitable for multi-rotor vehicles, which can hover with a zero linear speed. In contrast, fixed-wing vehicles require a nonzero minimum linear speed ($V\textsubscript{0}\neq0$) to generate sufficient lift force using aerodynamic interactions with the environment to counterbalance their weight. Thus, the proposed guidance strategy offers a scalable solution to the path-following problem by also eliminating its dependency to a certain extent on the vehicle type, which is another alluring feature of the proposed design.
\end{remark}

Note that for various values of UAV/pseudo-target speeds, $V_{U}/V_{T}$, the UAV may choose a pursuit behavior to rendezvous with the pseudo-target on the desired path by always looking in the direction of the UAV-target LOS. In other words, if the UAV maintains its velocity lead angles, $\psi_{U}$ and $\theta_{U}$, to zero, then it can be ensured that the UAV always follows a pursuit course toward the pseudo-target to rendezvous with the latter. In due course, the UAV mimics the trajectory of the pseudo-target, which leads to a path-following behavior. In short, the UAV first aligns itself with the UAV-target LOS ($\psi_{U}$, $\theta_{U} \to 0$ ) and then rendezvous with the pseudo-target ($r\to 0$) by \emph{pursuing} or carrying out a \emph{tail-chase} motion toward it. Hence, it is meaningful to regulate the UAV's lead angles in azimuth and elevation directions to zero to bring the UAV onto the desired pursuit course. 

\begin{remark}
Note that the proposed guidance strategy shares philosophical ties with pursuit guidance. However, it offers significant advantages over classical pure pursuit guidance, particularly in its ability to control the UAV's linear speed, a feature typically absent in pure pursuit guidance.
\end{remark}

\begin{remark}
 The method in \cite{doi:10.2514/1.G007964} regulated the UAV's effective heading angle $(\sigma_U)$ to zero to guide the vehicle onto the desired pursuit course. However, this approach required a separate control allocation scheme to distribute control effort between the pitch and yaw channels. While feasible and effective, the control allocation scheme adds complexity, particularly given the limited computational power of the UAV's guidance system. In contrast, the proposed strategy offers a more elegant solution by allowing different convergence times for the individual lead angles $(\psi_U, \theta_U)$ based on the available angular velocities in each plane, while also eliminating the need for a separate control allocation scheme.
\end{remark}
As stated earlier, the physical capabilities of the actuators restrict the control input that the UAV can generate to perform any maneuver. Therefore, we consider the UAV's angular velocities in the pitch and yaw channels to be limited by a known constant $\omega_{U}^{\max}$, that is, $\lvert\omega_{U}^y \rvert < \omega_{U}^{\max}$ and $\lvert\omega_{U}^z \rvert < \omega_{U}^{\max}$. Note that we have assumed the maximum angular speed in both channels to be the same; however, one may consider them to be different without affecting the design. To that end, we propose a smooth angular velocity saturation model as
\begin{equation}\label{eq:omega_saturation_model}
\dot{\mathbf{\Gamma}} =  \left( \mathcal{I}\textsubscript{2} - \mathcal{F} \right) \mathcal{K}\textsubscript{3} \mathbf{\Gamma}^{c} - \mathcal{K}\textsubscript{3}\mathcal{K}\textsubscript{4} \mathbf{\Gamma}; \quad \mathbf{\Gamma}(0)=0,\;\quad \mathbf{\Omega} = \mathbf{\Gamma},
\end{equation}
where $\mathbf{\Gamma}=\left[\omega_{U}^y \quad \omega_{U}^z\right]^\top $, $\mathbf{\Gamma}^{c}=\left[\omega_{U}^{c,y} \quad \omega_{U}^{c,z}\right]^\top $, and $\mathbf{\Omega} \in \mathbb{R}\textsuperscript{2}$ are the states, inputs, and outputs of the saturation model, respectively. The terms $\mathcal{K}\textsubscript{3},\mathcal{K}\textsubscript{4}>0$ are some non-negative constants, $\mathcal{I}\textsubscript{2}$ denotes the identity matrix of dimension two, $\mathcal{F}=\diag(f_{\ell})$ with $f_{\ell}= \left(\dfrac{\omega_{U}^{\ell}}{\omega_{U}^{\max}}\right)^{\gamma}$ for $\ell= y$, $z$ and $\gamma \geq 2n$, $n \in \mathbb{N}$, is a diagonal matrix with entries $f_\ell$.
\begin{theorem}\label{thm:omega_saturation}
Consider the saturation model given in \Cref{eq:omega_saturation_model}. If the commanded angular velocities $\omega_{U}^{c,y}$ and $\omega_{U}^{c,z}$ remain bounded for all $t \geq 0$, then the UAV's angular velocities, $\omega_{U}^y$ and $\omega_{U}^z$, remain confined to the set $\mathcal{S}_{\omega} \in \left\{\omega_{U}^{\ell}: \lvert \omega_{U}^{\ell} \rvert < \omega_{U}^{\max} \right\}$ for $\ell = y$, $z$ $\forall$ $t \geq 0$.  
\end{theorem}
\begin{proof}
Proof of \Cref{thm:omega_saturation} is similar to the proof of \Cref{thm:linear_speed_sat_model}, and thus omitted here.
\end{proof}
With the saturation model given in \Cref{eq:omega_saturation_model}, the augmented system dynamics (in the pitch and yaw planes) becomes
\begin{subequations}\label{eq:omega_u_z_agumneted}
  \begin{align}
\dot{\theta}_U&=\omega_U^z-\dot{\psi}\sin\theta\sin\psi_U-\dot{\theta}\cos\psi_U,\label{eq:thetaUdot_agumented}\\
\dot{\omega}_{U}^{z} & =\left( 1- f_{z} \right) \mathcal{K}\textsubscript{3} \omega_{U}^{c,z} - \mathcal{K}\textsubscript{3}\mathcal{K}\textsubscript{4} \omega_{U}^{z}, \label{eq:dot_omega_u_z}
  \end{align}
\end{subequations}
and
\begin{subequations}\label{eq:omega_u_y_augmented}
\begin{align}
\dot{\psi}_U&=\dfrac{\omega_U^y}{\cos\theta_U}+\dot{\psi}\tan\theta_U\cos\psi_U\sin\theta-\dot{\psi}\cos\theta-\dot{\theta}\tan\theta_U\sin\psi_U,\label{eq:psiUdot_agumented}\\
\dot{\omega}_{U}^{y} & =  \left( 1- f_{y} \right)\mathcal{K}\textsubscript{3} \omega_{U}^{c,y} - \mathcal{K}\textsubscript{3}\mathcal{K}\textsubscript{4} \omega_{U}^{y}. \label{eq:dot_omega_u_y}
\end{align}
\end{subequations}
One may observe from \Cref{eq:omega_u_z_agumneted,eq:omega_u_y_augmented} that the turn rates in pitch and yaw planes have a relative degree of two with respect to their commanded angular velocities, $\omega_{U}^{c,z}$ and $\omega_{U}^{c,y}$.

Next, we endeavor to design the said commanded angular velocities to align the UAV onto the LOS joining it and the pseudo-target. To that, we first consider a state as $z=\omega_{U}^{z} - \eta $, where $\eta$ is the auxiliary input which will be designed in the following theorem.
\begin{theorem}\label{thm:omega_u_z_c}
Consider the UAV's pitch rate dynamics as in \Cref{eq:omega_u_z_agumneted}. If the commanded angular velocity is designed as
\begin{equation}\label{eq:omega_z_c}
    \omega_{U}^{z,c} = \dfrac{\mathcal{K}\textsubscript{3}\mathcal{K}\textsubscript{4} \omega_{U}^{z} + \dot{\eta} -\lvert z \rvert \sign(\theta_{U}) - \left(\mathcal{M}\textsubscript{2}  z  ^{\alpha\textsubscript{2}} + \mathcal{N}\textsubscript{2}   z   ^{\beta\textsubscript{2}}\right) } {\mathcal{K}\textsubscript{3} \left( 1- f_{z} \right)}
\end{equation}
with the auxiliary control $\eta$ as
\begin{equation}\label{eq:eta}
    \eta = \dot{\psi} \sin{\theta} \sin{\psi}_{U} + \dot{\theta} \cos{\psi}_{U} - \left(\mathcal{M}\textsubscript{2}   \theta_{U}  ^{\alpha\textsubscript{2}} + \mathcal{N}\textsubscript{2}   \theta_{U}  ^{\beta\textsubscript{2}}\right) ,
\end{equation}
where $\mathcal{M}\textsubscript{2}$, $\mathcal{N}\textsubscript{2}$, $\alpha\textsubscript{2}$, $\beta\textsubscript{2}>0$ with $\alpha\textsubscript{2}>1$ and $0<\beta\textsubscript{2}<1$, then the lead angle $\theta_{U}$ nullifies to zero within a fixed-time given by
\begin{equation}\label{eq:t2}
    T\textsubscript{2} \leq\dfrac{1}{\bar{\mathcal{M}\textsubscript{2}}(\alpha\textsubscript{2}-1)}+\dfrac{1}{\mathcal{N}\textsubscript{2}(1-\beta\textsubscript{2})},
\end{equation}
with $\bar{\mathcal{M}\textsubscript{2}} = 2^{(1-\alpha\textsubscript{2})}\mathcal{M}\textsubscript{2}$, regardless of its initial value.
\end{theorem}
\begin{proof}
Please refer to  Appendix III for proof.
\end{proof}

Now we aim to design the commanded angular velocity $\omega_{U}^{y,c}$, which will regulate the UAV's lead angle $\psi_{U}$ to zero within a fixed time and maintain it there for all future time. In other words, for any desired path $\mathcal{P}$ and $\psi_{U}(0)$, we aim to design $\omega_{U}^{y,c}$ such that $\psi_{U}(t) \equiv 0$ for all time $t>T\textsubscript{3}$, where $T\textsubscript{3}$ is some fixed-time. To that end, let us consider $y=\omega_{U}^{y} - \lambda $ as a state variable, where $\lambda$ is the auxiliary input.

\begin{theorem}\label{thm:omega_u_y_c}
Consider the UAV's azimuth rate dynamics as \Cref{eq:omega_u_y_augmented}. If the UAV maintains its commanded angular speed $\omega_{U}^{y,c}$ as
\begin{equation}\label{eq:omega_y_c}
    \omega_{U}^{y,c} = \dfrac{\mathcal{K}\textsubscript{3}\mathcal{K}\textsubscript{4} \omega_{U}^{y} + \dot{\lambda} - \dfrac{\lvert y \rvert \sign(\psi_{U})}{\cos{\theta_{U}}} - \left(\mathcal{M}\textsubscript{3}   y  ^{\alpha\textsubscript{3}} + \mathcal{N}\textsubscript{3}   y  ^{\beta\textsubscript{3}}\right) } {\mathcal{K}\textsubscript{3} \left( 1- f_{y} \right)}
\end{equation}
with the auxiliary control $\lambda$ chosen as
\begin{equation}\label{eq:lambda}
   \lambda = - \cos{\theta_{U}} \left[ \dot{\psi}\tan\theta_U\cos\psi_U\sin\theta-\dot{\psi}\cos\theta-\dot{\theta}\tan\theta_U\sin\psi_U + \left(\mathcal{M}\textsubscript{3} \psi_{U} ^{\alpha\textsubscript{3}} + \mathcal{N}\textsubscript{3}   \psi_{U}  ^{\beta\textsubscript{3}}\right)\right],
   \end{equation}
   where $\mathcal{M}\textsubscript{3}$, $\mathcal{N}\textsubscript{3}$, $\alpha\textsubscript{3}$, $\beta\textsubscript{3}>0$ with $\alpha\textsubscript{3}>1$ and $0<\beta\textsubscript{3}<1$, then, the UAV's azimuth angle $\psi_{U}$ converges to zero within a fixed time $T\textsubscript{3}$ given by
\begin{equation}\label{eq:t3}
    T\textsubscript{3} \leq\dfrac{1}{\bar{\mathcal{M}\textsubscript{3}}(\alpha\textsubscript{3}-1)}+\dfrac{1}{\mathcal{N}\textsubscript{3}(1-\beta\textsubscript{3})},
\end{equation}
with $\bar{\mathcal{M}\textsubscript{3}}=2^{(1-\alpha\textsubscript{3})\mathcal{M}\textsubscript{3}}$   independent of its initial value.
\end{theorem}
\begin{proof}
Please refer to  Appendix IV for proof.
\end{proof}
\begin{remark}
By exercising linear speed command, we can regulate the relative range between the UAV and the pseudo-target to zero. We, however, point out the fact that the orientation of the UAV cannot be controlled through linear speed. Thus, controlling the UAV's lead angle is crucial to control its heading.  A lack of control over the UAV's heading could result in poor tracking performance and oscillations in the regulation of the relative distance.
\end{remark}

One may observe from the expression of auxiliary control $\chi$ in \Cref{thm:vu} that it has the term $\cos{\theta}_{U}\cos{\psi}_{U}$ in the denominator, which is essentially the cosine of the effective heading angle ($\cos{\sigma}_{U}$) of the UAV. Whenever $\cos{\sigma}_{U}=\pm \pi/2$, the auxiliary control will become unbounded, and so will the commanded linear speed. Thus, we now show that the $\sigma = \pm (\pi/2)$ is not an attractor, and hence, it will not cause any singularity in the control input. To that end, we first obtain the dynamics of effective heading angle $\sigma_U$ by differentiating $\cos{\sigma}_{U}$ with respect to time as
\begin{align}
\left(\sin\sigma_U \right)\dot \sigma_U =\left(\cos\theta_U\sin\psi_U\right)\dot\psi_U +\left(\cos\psi_U\sin\theta_U\right)\dot\theta_U. \label{eq:dot_sigma1}
\end{align}
On substituting the values of $\dot\psi_U$ and $\dot\theta_U$ from \Cref{eq:psiUdot} and \Cref{eq:thetaUdot} into \Cref{eq:dot_sigma1} leads us to arrive at
\begin{align*}
 \dot \sigma_U=&~ \dfrac{\cos\theta_U\sin\psi_U\left(\dot{\psi}\tan\theta_U\cos\psi_U\sin\theta-\dot{\psi}\cos\theta-\dot{\theta}\tan\theta_U\sin\psi_U\right)-\cos\psi_U\sin\theta_U\left(\dot{\psi}\sin\theta\sin\psi_U+\dot{\theta}\cos\psi_U\right)}{\sin\sigma_U}\\
&~+\dfrac{\sin\psi_U\omega_U^y+\cos\psi_U\sin\theta_U\omega_U^z}{\sin\sigma_U},
\end{align*}
which, on collecting $\dot\psi$ and $\dot\theta$ terms together, results in
\begin{align}
\nonumber \dot \sigma_U=&~ \dfrac{\dot\psi\left(\cos\theta_U\sin\psi_U\tan\theta_U\cos\psi_U\sin\theta-\cos\theta_U\sin\psi_U\cos\theta-\cos\psi_U\sin\theta_U\sin\theta\sin\psi_U\right)}{\sin\sigma_U} \\
&~-\dfrac{\dot{\theta}\left(\cos\theta_U\sin\psi_U\tan\theta_U\sin\psi_U+\cos\psi_U\sin\theta_U\cos\psi_U\right)}{\sin\sigma_U}+\dfrac{\sin\psi_U}{\sin\sigma_U}\omega_U^y+\dfrac{\cos\psi_U\sin\theta_U}{\sin\sigma_U}\omega_U^z. \label{eq:dot_sigma_2}
\end{align}
After some algebraic simplification, the expression \Cref{eq:dot_sigma_2} may be written in more convenient form as
\begin{equation}\label{eq:dot_sigma_3}
\dot \sigma_U    = \dfrac{\dot\psi}{\sin\sigma_U} \left(-\cos\theta_U\sin\psi_U\cos\theta\right)-\dfrac{\dot{\theta}\sin\theta_U}{\sin\sigma_U} +\dfrac{\sin\psi_U}{\sin\sigma_U}\omega_U^y+\dfrac{\cos\psi_U\sin\theta_U}{\sin\sigma_U}\omega_U^z.
\end{equation}
With $\dot\psi$ and $\dot\theta$ given in \Cref{eq:thetadot,eq:psidot}, the expression in \Cref{eq:dot_sigma_3} becomes
\begin{align*}
\dot \sigma_U =&~ \left(\dfrac{V_T\cos\theta_T\sin\psi_T-V_U\cos\theta_U\sin\psi_U}{r\cos\theta}\right) \left(\dfrac{-\cos\theta_U\sin\psi_U\cos\theta}{\sin\sigma_U}\right)-\left(\dfrac{V_T\sin{\theta_T}-V_U\sin\theta_U}{r}\right)\left(\dfrac{\sin\theta_U}{\sin\sigma_U}\right)\\
&+~\dfrac{\sin\psi_U}{\sin\sigma_U}\omega_U^y+\dfrac{\cos\psi_U\sin\theta_U}{\sin\sigma_U}\omega_U^z,
\end{align*}
which, on collecting similar terms together, yields
\begin{align}\label{eq:sigmaudot}
\nonumber \dot \sigma_U=&~-\dfrac{V_T}{r\sin\sigma_U}\left(\cos\theta_T\sin\psi_T\cos\theta_U\sin\psi_U+\sin\theta_U\sin\theta_T\right)+\dfrac{V_U}{r\sin\sigma_U}\left(\cos^2\theta_U\sin^2\psi_U+\sin^2\theta_U\right)\\
\nonumber&+~\dfrac{\sin\psi_U}{\sin\sigma_U}\omega_U^y+\dfrac{\cos\psi_U\sin\theta_U}{\sin\sigma_U}\omega_U^z.
\end{align}
Equivalently, one has
\begin{align*}
\dot \sigma_U 
=&~-\dfrac{V_T}{r\sin\sigma_U}\left(\cos\theta_T\sin\psi_T\cos\theta_U\sin\psi_U+\sin\theta_U\sin\theta_T\right)+\dfrac{V_U}{r\sin\sigma_U}\left(\cos^2\theta_U\sin^2\psi_U+\left(1-\cos^2\theta_U\right)\right)\\
&+~\dfrac{\sin\psi_U}{\sin\sigma_U}\omega_U^y+\dfrac{\cos\psi_U\sin\theta_U}{\sin\sigma_U}\omega_U^z,\\
=&~-\dfrac{V_T}{r\sin\sigma_U}\left(\cos\theta_T\sin\psi_T\cos\theta_U\sin\psi_U+\sin\theta_U\sin\theta_T\right)+\dfrac{V_U}{r\sin\sigma_U}\left(1-\cos^2\theta_U\left(1-\sin^2\psi_U\right)\right)\\
&+~\dfrac{\sin\psi_U}{\sin\sigma_U}\omega_U^y+\dfrac{\cos\psi_U\sin\theta_U}{\sin\sigma_U}\omega_U^z.
\end{align*}
By using the trigonometric relation $1-\sin^2\psi_U = \cos^2\psi_U$ the above expression of $\dot \sigma_U$ simplifies to
\begin{align*}
\dot \sigma_U =&~-\dfrac{V_T}{r\sin\sigma_U}\left(\cos\theta_T\sin\psi_T\cos\theta_U\sin\psi_U+\sin\theta_U\sin\theta_T\right)+\dfrac{V_U}{r\sin\sigma_U}\left(1-\cos^2\theta_U\cos^2\psi_U\right)\\
\nonumber&+~\dfrac{\sin\psi_U}{\sin\sigma_U}\omega_U^y+\dfrac{\cos\psi_U\sin\theta_U}{\sin\sigma_U}\omega_U^z,\\
=&~-\dfrac{V_T\left(\cos\theta_T\sin\psi_T\cos\theta_U\sin\psi_U+\sin\theta_U\sin\theta_T\right)}{r\sin\sigma_U}+\dfrac{V_U\left(1-\cos^2\sigma_U\right)}{r\sin\sigma_U}+\dfrac{\sin\psi_U}{\sin\sigma_U}\omega_U^y+\dfrac{\cos\psi_U\sin\theta_U}{\sin\sigma_U}\omega_U^z.
\end{align*}
Now, the above expression can be written in a compact form as
\begin{equation}\label{eq:sigmadotufinal}
    \dot\sigma_U= \dfrac{V_U \sin\sigma_U}{r}
    -\dfrac{V_T\left(\cos\theta_T\sin\psi_T\cos\theta_U\sin\psi_U+\sin\theta_U\sin\theta_T\right)}{r\sin\sigma_U}+ \dfrac{\sin\theta_U\cos\psi_U}{\sin\sigma_U} \omega_U^z + \dfrac{\sin\psi_U}{\sin\sigma_U} \omega_U^y.
\end{equation}
It can be readily verified from \Cref{eq:sigmadotufinal} that $\sigma_U = \pm (\pi/2)$ is not an equilibrium point of the dynamics of $\sigma_U$, and thus cannot be an attractor. It is important to note that by exercising the angular velocity control inputs $\omega_U^y$ and $\omega_U^z$ we are nullifying the UAV's lead angles ($\psi_U$, $\theta_U$) to zero. The time taken to make $\sigma_U=0$ can be obtained as $\max\{T\textsubscript{2},T\textsubscript{3}\}$. This can be verified from the UAV's effective heading angle's definition because $\sigma_U=\cos^{-1}(\cos{\theta}_U\cos{\psi}_U)$, which implies $\sigma_U \to 0$ $\iff$ $(\psi_U$, $\theta_U) \to 0$. Therefore, $\sigma_U =0$ in the steady state and the only possibility of $\sigma_U$ to be $\pm (\pi/2)$ is in the transient phase. Additionally, $\sigma_U = \pm (\pi/2)$ is not an attractor, implying that it won't cause any issues during the implementation. With a suitable choice of controller parameters, one can make $T\textsubscript{2},T\textsubscript{3}$ arbitrarily small, and then $\sigma_U$ may not have a chance to become $\pm (\pi/2)$ in the transient phase in most cases. 

If, however, the UAV's effective heading angle is initially at $\pm \pi/2$, that is, $\sigma_U(0) = \pm \pi/2$, we now analyze its behavior. Using the definition of effective heading angle, one can write $\cos\sigma_U(0)=\cos{\psi}_U(0)\cos{\theta}_U(0)=0$, which implies either (1) $\psi_U(0)=\pm \pi/2$ or (2) $\theta_U(0)=\pm \pi/2$, or (3) $\psi_U(0)=\theta_U(0)=\pm \pi/2$. We now separately investigate these cases. For the first case when $\psi_U(0)=\pm \pi/2$ the dynamics of effective heading angle reduces to
\begin{equation}\label{eq:sigmadotu_case1}
    \dot\sigma_U(0)= \dfrac{V_U(0)}{r(0)}
    -\dfrac{V_T(0)\left[\cos\theta_T(0)\sin\psi_T(0)\cos\theta_U(0)+\sin\theta_U(0)\sin\theta_T(0)\right]}{r(0)}+ \omega_U^y \neq 0.
\end{equation}
In the second case when $\theta_U(0)=\pm \pi/2$ the expression in \Cref{eq:sigmadotufinal} can be written as
\begin{equation}\label{eq:sigmadotu_case2}
    \dot\sigma_U(0)= \dfrac{V_U(0)}{r(0)}
    -\dfrac{V_T(0)\sin\theta_T(0)}{r(0)}+ \cos\psi_U(0) \omega_U^z + \sin\psi_U(0)\omega_U^y \neq 0.
\end{equation}
One may write the dynamics of $\sigma_U$ at $t=0$ for $\psi_U(0)=\theta_U(0)=\pm \pi/2$ as
\begin{equation}\label{eq:sigmadotu_case3}
    \dot\sigma_U(0)= \dfrac{V_U(0)}{r(0)}
    -\dfrac{V_T(0)\sin\theta_T(0)}{r(0)}+ \omega_U^y \neq 0.
\end{equation}
One can observe from \Cref{eq:sigmadotu_case1,eq:sigmadotu_case2,eq:sigmadotu_case3} that the dynamics of UAV's effective heading angle, $\dot\sigma_U$, is nonzero for all the possible cases, which will not lead to a singularity in the control input. Therefore if UAV's initial effective heading angle is $\pm (\pi/2)$, that is $\sigma_U(0) \pm (\pi/2)$ , a minuscule change in any other engagement variable will make $\sigma_U(0^{+}) \neq \pm (\pi/2)$, as $\dot\sigma_U \neq 0$ even with the zero control inputs ($V_U=\omega_U^y=\omega_U^z=0$). Consequently, the UAV's commanded linear speed remains bounded. The design procedure can be summarized in \Cref{alg:alg1}.
\begin{algorithm}[!ht]\label{alg:alg1}
  \KwInput{Desired path, UAV's initial position and lead angles, and design parameters.}
  \KwOutput{UAV's current position ($x_U,y_U,x_U$) and lead angles ($\psi_U$, $\theta_U$).}
    Initialize a pseudo-target on the predefined path $\mathcal{P}$.\\ 
    Calculate the relative range between the UAV and pseudo-target $(r)$ using the Euclidean distance formula.\\
    Determine the LOS angle $(\theta)$ between the UAV and the pseudo-target using the trigonometric relation $\theta=\tan^{-1}\left(\dfrac{y_T-y_U}{x_T-x_U}\right)$.\\
    Compute the commanded inputs $\U^c$, $\omega_U^{z,c}$, and $\omega_U^{y,c}$ using \Cref{eq:vu,eq:omega_z_c,eq:omega_y_c}, respectively.\\
    Provide the commanded inputs to the saturation models given in \Cref{eq:vu_saturation_model,eq:omega_saturation_model}, and obtain the UAV's guidance commands $V_U$, $\omega_U^{z}$, and $\omega_U^{y}$.\\
    Update the UAV's position $(x_U,y_U,z_U)$ and  velocity lead angles $(\psi_U,\theta_U).$\\
    Return to step 2 and repeat the process until the mission objective is completed.
\caption{Algorithm for the proposed guidance law.}
\end{algorithm}

One may also observe from \Cref{eq:omega_y_c} in \Cref{thm:omega_u_y_c} that it contains the term $\cos{\theta}_{U}$ in the denominator, which might appear to cause the commanded angular velocity $\omega_{U}^{y,c}$ to become unbounded. Like the previous analysis, here one can also notice from the dynamics of lead angle $\theta_U$ given in \Cref{eq:thetaUdot} that $\theta_{U}=\pm (\pi/2)$ is not an attractor, and thus it won't cause any issue during the implementation. Now for a case when the initial value of lead angle $\theta_{U}$ is $\pm (\pi/2)$, we analyze the behavior of its dynamics at $t=0$ using \Cref{eq:thetaUdot} as 
 \begin{equation}
    \dot{\theta}_U(0)=\omega_U^z-\dot{\psi}(0)\sin\theta(0)\sin\psi_U(0)-\dot{\theta}(0)\cos\psi_U(0) \neq 0.
 \end{equation}
Even $\theta_{U}=\pm (\pi/2)$ at $t=0$, a minuscule change in any other engagement variable will make $\theta_{U}(0^{+}) \neq \pm (\pi/2)$, thus $\omega_{U}^{y,c}$ will not become unbounded.

\begin{figure}[!ht]
    \centering
    \includegraphics[width=0.95\linewidth]{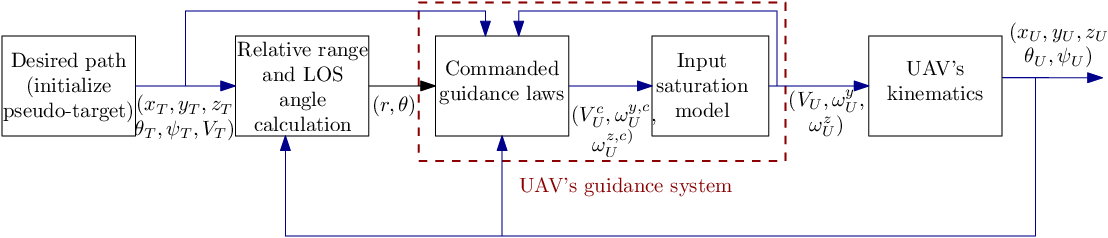}
    \caption{Block-diagram representation of the proposed strategy.}
    \label{fig:block_diagram}
\end{figure}

A block diagram representation of the closed-loop system is shown in \Cref{fig:block_diagram}. First, a pseudo-target is initialized on the desired path. Using the current measurements of the UAV, the relative separation and LOS angle, $r$ and $\theta$, with respect to the pseudo-target are calculated. This data, along with the UAV's lead angles $(\psi_U, \theta_U)$, and the target's linear speed $(V_T)$ and its lead angles $(\psi_T, \theta_T)$, is then provided to the UAV's guidance system. With this information, the UAV's guidance system generates the commanded inputs: the linear speed $(\U^c)$ and angular velocities $(\omega_U^{y,c}, \omega_U^{z,c})$. Finally, these commands are provided to the input saturation model, which, in conjunction with the maximum available control inputs, generates the necessary guidance commands, $V_U$, $\omega_U^{y}$, and $\omega_U^{z}$. Subsequently, the guidance commands are applied to the UAV to adjust its position and heading angles, which will steer the UAV onto the desired three-dimensional path within a fixed time and maintain it there for all future times.

Note that by altering the design parameters, one can change the transient performance even without considering the physical capabilities of the actuators. However, ignoring the physical capabilities of the actuators in the design may be detrimental to the stability and performance of the guidance system and may even drive the UAV into an unsafe mode, potentially causing damage to the actuators. Therefore, for the safe and reliable operation of the UAV, the actuator saturation should be accounted for in the guidance design, and the design parameters must be chosen judiciously for the desired performance.
\begin{remark}
We only require the commanded control inputs to be finite without necessarily assuming their bounds are known. Additionally, we do not require that the commanded inputs always remain within the prescribed bounds. However, we argue that if the commanded inputs are designed to be finite, then the actual control inputs, the linear speed ($V_{U}$) and angular speeds ($\omega_{U}^{y}$ and $\omega_{U}^{z}$), will remain constrained within the sets $\mathcal{S}_{u}$ and $\mathcal{S}_{\omega}$, respectively by using the proposed guidance strategy.
\end{remark}
\begin{remark}
 Since we have mathematically established that the commanded signals ( $\U^{c}$, $\omega_U^{y,c}$, and $\omega_U^{z,c}$) designed in \Cref{eq:vu,eq:omega_y_c,eq:omega_z_c}  remain finite. This, in conjunction with the saturation models, ensures that the actual control inputs remain within their predefined bounds. Since the boundedness of the commanded signals serves as a sufficient condition for \Cref{thm:linear_speed_sat_model,thm:omega_saturation} to guarantee the boundedness of the actual control inputs.
\end{remark}
\section{Performance Evaluations}\label{sec:simulation}
In this section, we demonstrate the efficacy of the proposed three-dimensional path following guidance laws (presented in \Cref{thm:linear_speed_sat_model,thm:vu,thm:omega_saturation,thm:omega_u_z_c,thm:omega_u_y_c}) through various scenarios ranging from rapid curvature changes to UAV's arbitrary initial configuration with respect to path and even assess the performance when its motion changes according to the spline curves. The maximum allowable bounds on the UAV's linear velocity is chosen as $\pm 25$ m/s, and the maximum angular velocity of the UAV in pitch and yaw channels to be $\pm 3$ rad/s, that is, $V_{U}^{\max}= 25$ m/s and $\omega_{U}^{\max} = 3$ rad/s. The input bounds on the control inputs ($V_{U}$, $\omega_{U}^{y}$, and $\omega_{U}^{z}$) are denoted by the black color dotted lines in the corresponding plots. In each trajectory plot, $X$, $Y$, and $Z$, respectively, represent the downrange, cross-range, and altitude in the physical space. The dashed line in black denotes the desired path and variables related to it. A black square marker (on the desired path) represents the initial location of the virtual target, whereas other colored square markers denote the UAV's initial location. The design parameters are chosen as $\mathcal{K}\textsubscript{1}=1$, $\mathcal{K}\textsubscript{2}=0.5$, $\mathcal{K}\textsubscript{3}=1$, $\mathcal{K}\textsubscript{4}=0.5$, $\gamma=2$, $\mathcal{M}\textsubscript{1}=0.1$, $\mathcal{N}\textsubscript{1}=0.3$, $\alpha\textsubscript{1}=1.01$, $\beta\textsubscript{1}=0.99$ $\mathcal{M}\textsubscript{2}=10$, $\mathcal{N}\textsubscript{2}=2$, $\alpha\textsubscript{2}=1.01$, $\beta\textsubscript{2}=0.99$, $\mathcal{M}\textsubscript{3}=10$, $\mathcal{N}\textsubscript{3}=2$, $\alpha\textsubscript{3}=1.01$, and $\beta\textsubscript{3}=0.99$, unless otherwise specified.
Note that the design parameters are chosen metaheuristically by systematically varying their values and selecting the combination that yields the best performance on the validation set.

\begin{figure}[!ht]
    \centering
  \begin{subfigure}{0.5\linewidth}
      \centering
      \includegraphics[width=\linewidth]{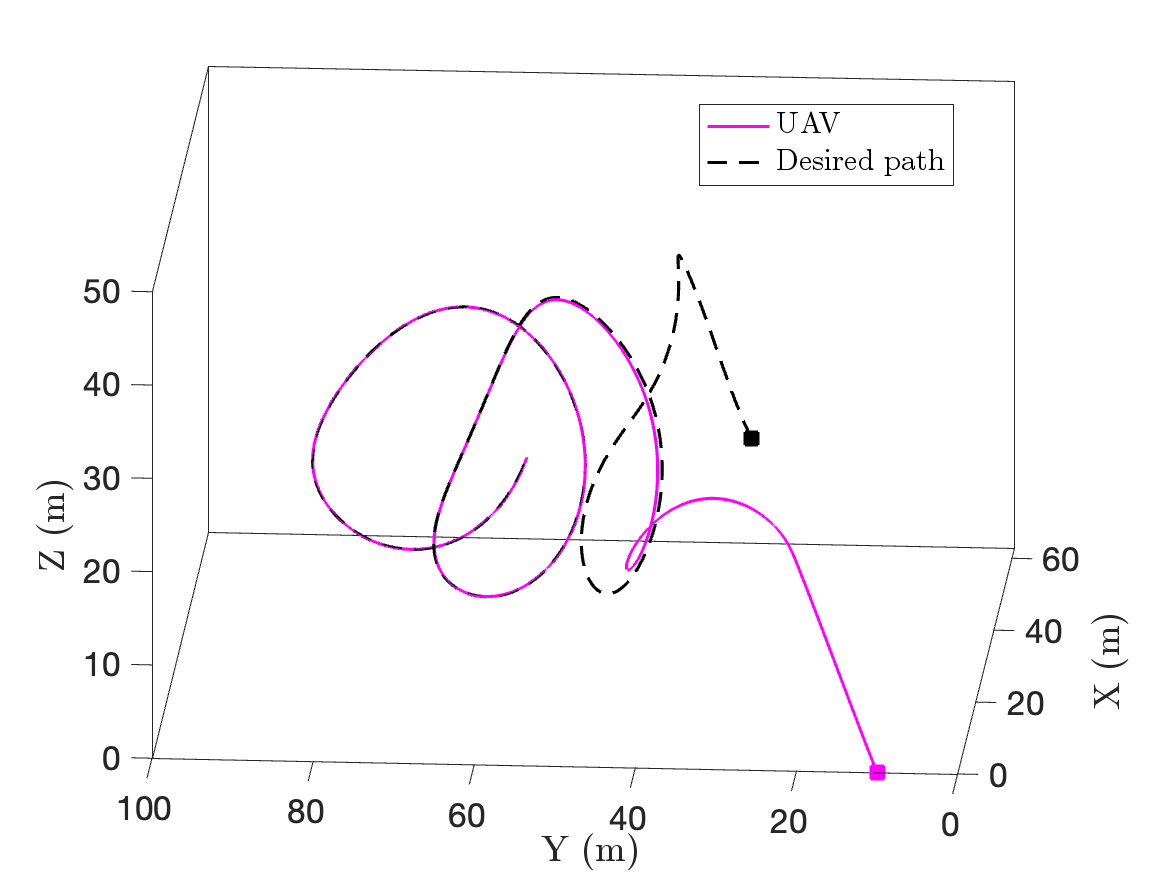}
      \caption{3D path.}
    \label{fig:helix_v0_path}
  \end{subfigure}%
    \begin{subfigure}{0.5\linewidth}
      \centering
      \includegraphics[width=\linewidth]{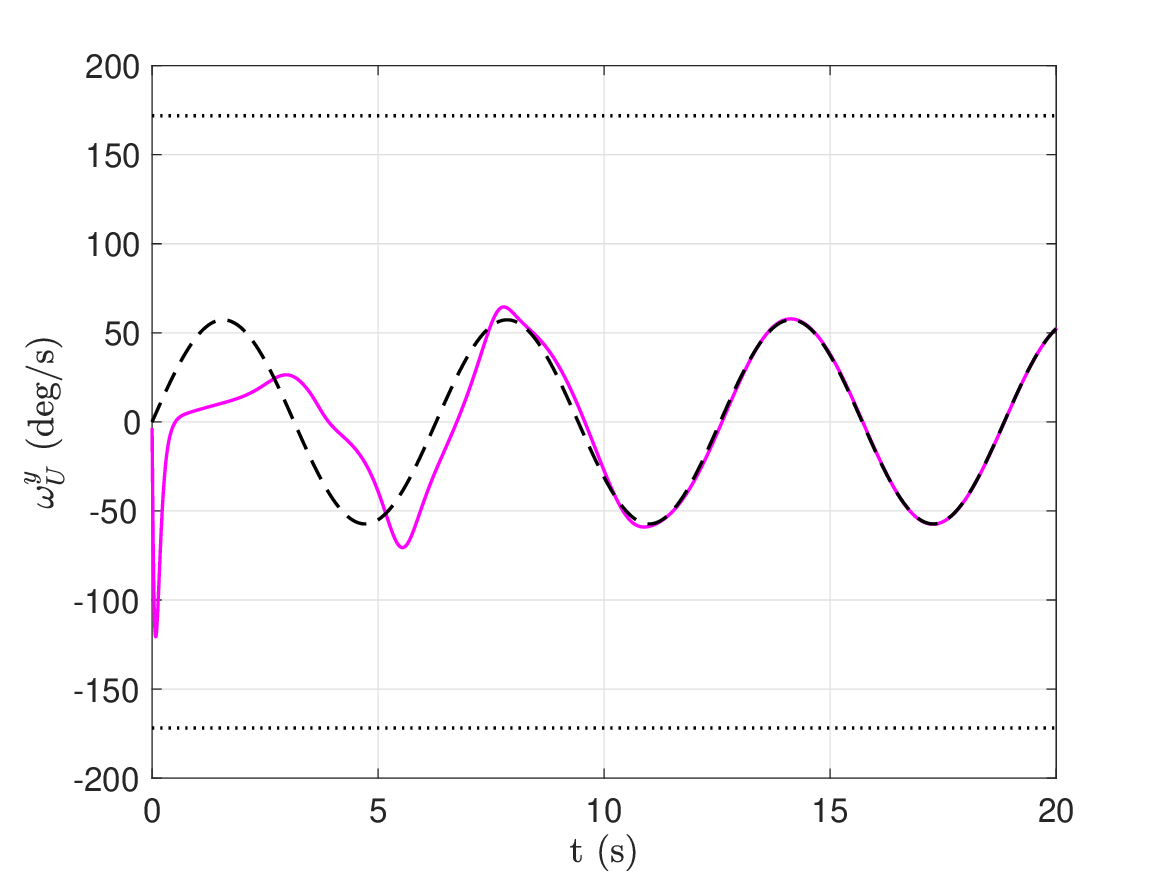}
      \caption{Angular velocities in yaw plane.}
    \label{fig:helix_v0_omega_uy}
  \end{subfigure}
    \begin{subfigure}{0.5\linewidth}
      \centering
      \includegraphics[width=\linewidth]{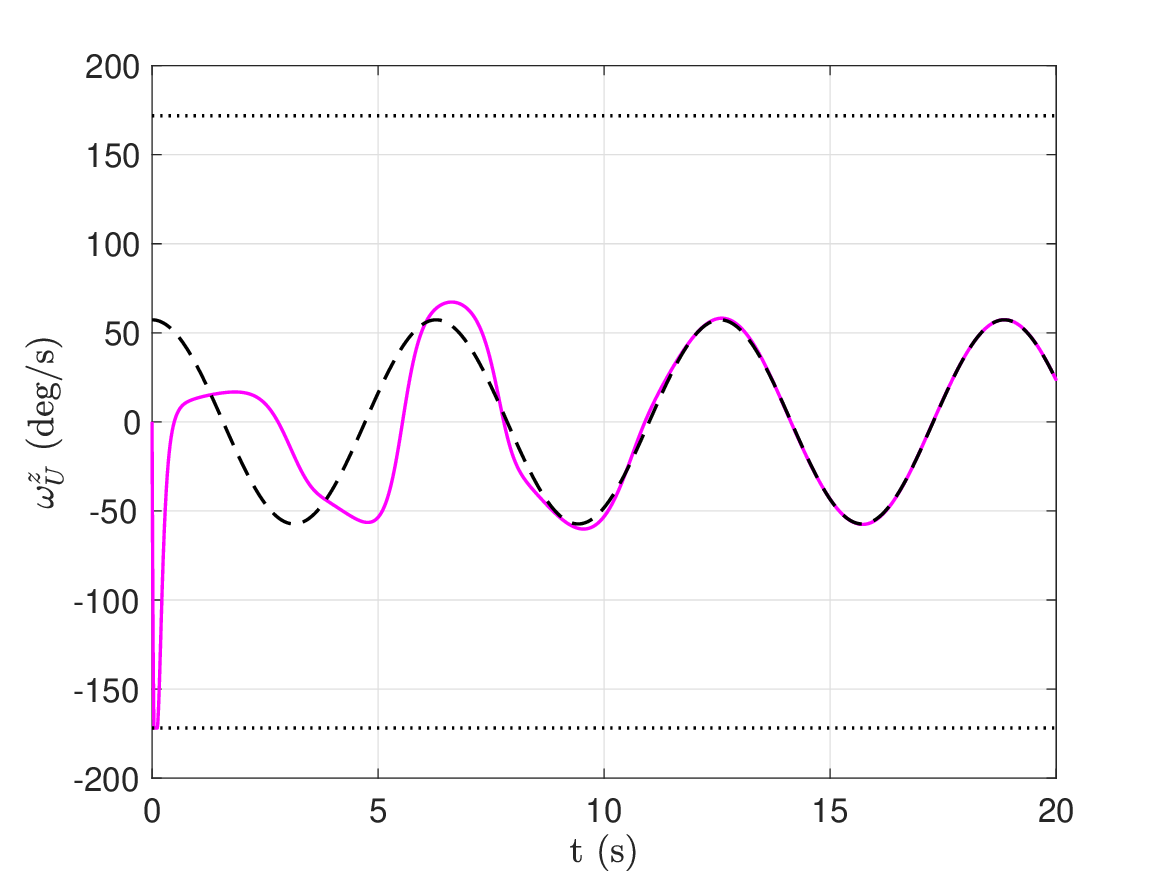}
      \caption{Angular velocities in pitch plane.}
    \label{fig:helix_v0_omega_uz}
  \end{subfigure}%
      \begin{subfigure}{0.5\linewidth}
      \centering
      \includegraphics[width=\linewidth]{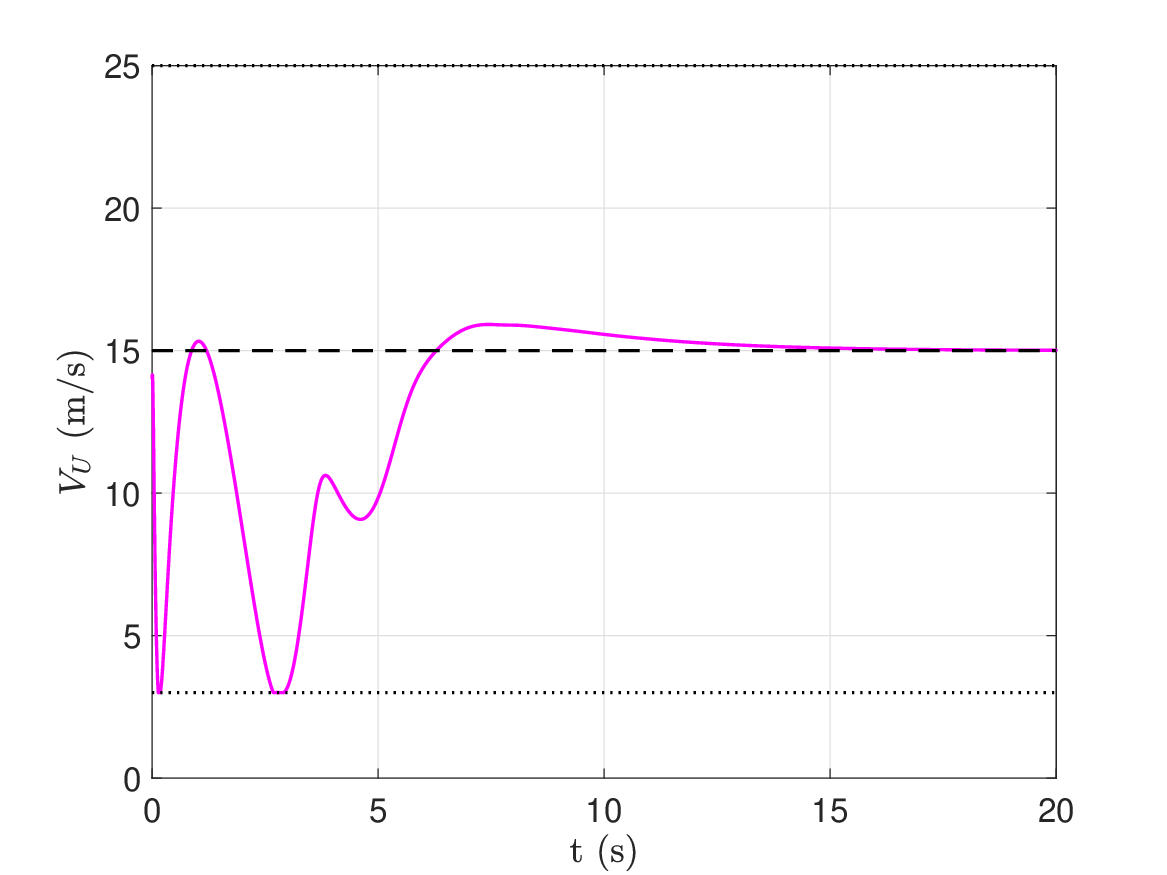}
      \caption{Linear speeds.}
    \label{fig:helix_v0_vu}
  \end{subfigure}
      \begin{subfigure}{0.5\linewidth}
      \centering
      \includegraphics[width=\linewidth]{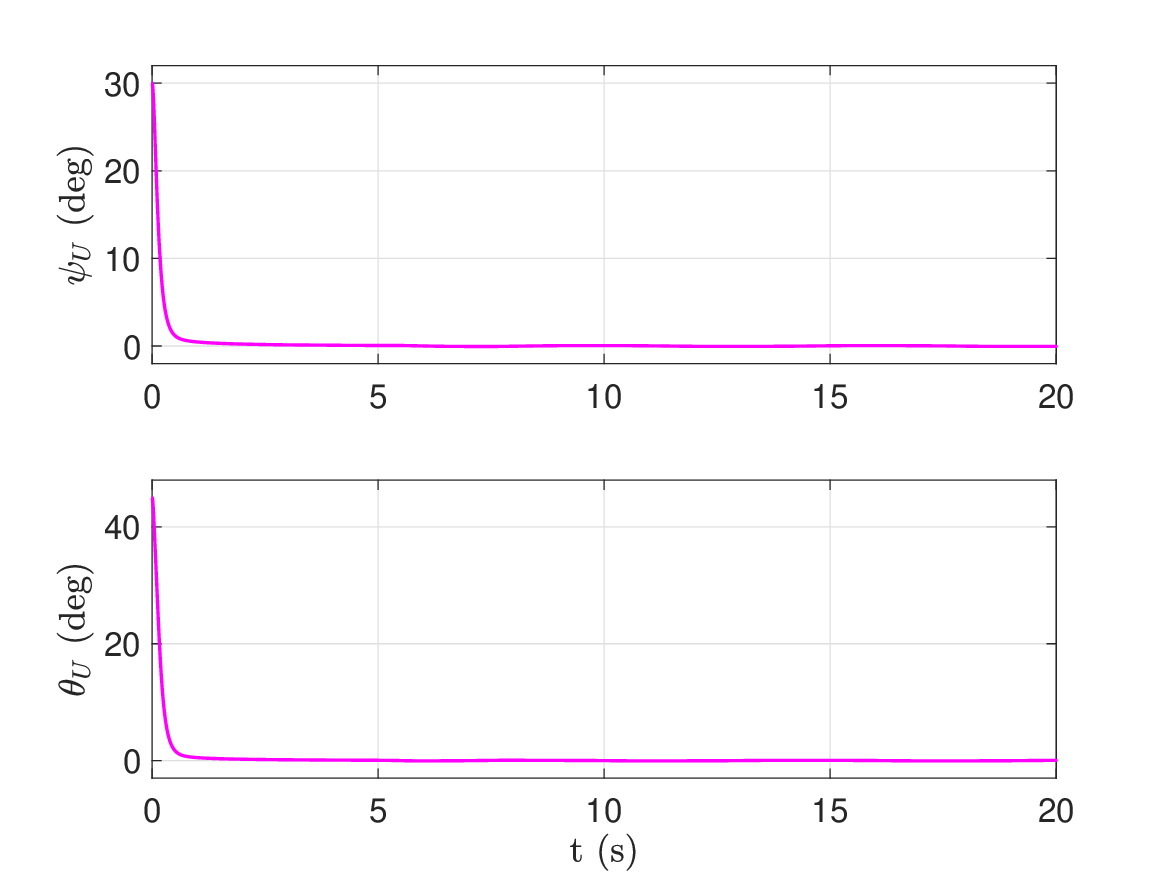}
      \caption{Lead angles.}
    \label{fig:helix_v0_angle_error}
  \end{subfigure}%
  \begin{subfigure}{0.5\linewidth}
      \centering
      \includegraphics[width=\linewidth]{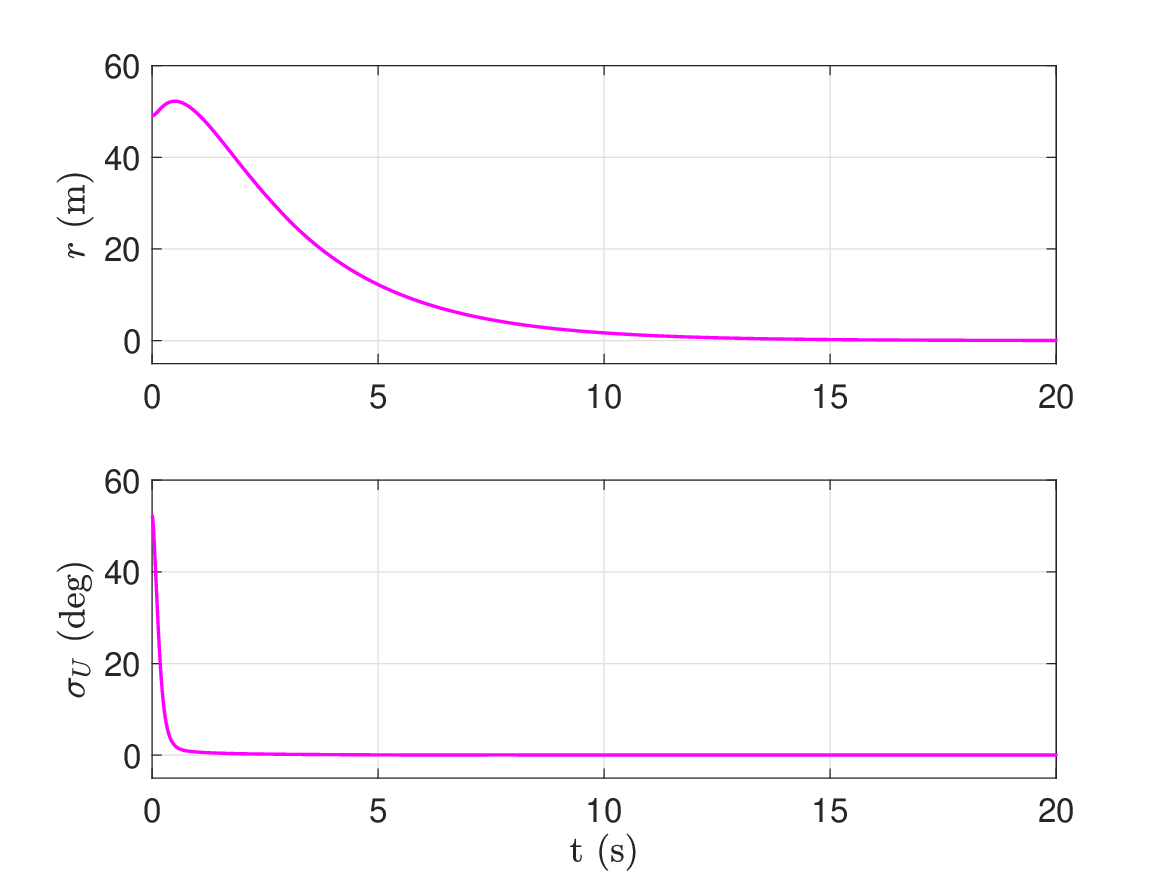}
      \caption{Relative range and effective heading angle.}
    \label{fig:helix_v0_r_sigmau}
  \end{subfigure}
   \caption{UAV following a  helix-like curvilinear path with $V\textsubscript{0}=3$ m/s.}
    \label{fig:helix_v0}
\end{figure}
Under the proposed guidance strategy, we depict the performance of the UAV when it is required to follow a helix-like curvilinear path through \Cref{fig:helix_v0}. The desired path is characterized as the trajectory of a moving virtual target with a constant linear speed of $15$ m/s and time-varying angular speeds $\omega_{T}^{y}= \sin t$ rad/s and $\omega_{T}^{z}= \cos t$ rad/s. In \Cref{fig:helix_v0_path}, the virtual target is initially assumed to be at (40,30,20) m with heading angles of $15^\circ$ in the azimuth and elevation directions. On the other hand, the UAV is initially located at (0,10,0) m with heading angles of $30^\circ$ and $45^\circ$ that is, $\psi_{U}(0)=30^\circ$ and $\theta_{U}(0)=45^\circ$. The minimum allowable linear speed is chosen to be as 3 m/s, that is, $V\textsubscript{0}=3$ m/s. One may observe that the UAV converges to its desired path (around 12 s) and follows the same thereafter for all future times regardless of rapid and periodic curvature changes. The UAV's linear velocity and angular velocities (control inputs) are depicted in \Cref{fig:helix_v0_vu,fig:helix_v0_omega_uy,fig:helix_v0_omega_uz}. We can see that although the commanded values of the control inputs ($V_{U}^c$, $\omega_{U}^{y,c}$, and $\omega_{U}^{z,c}$) may be high, the actual inputs to the UAV ($V_{U}$, $\omega_{U}^{y}$, and $\omega_{U}^{z}$) are much lower. It is also worth observing that the actual control inputs always remain in the predefined bound. This, in turn, validates the claims that the UAV follows its desired trajectory with the bounded inputs. \Cref{fig:helix_v0_angle_error,fig:helix_v0_r_sigmau} depict the lead angles and the relative range between the UAV and the virtual target. One can notice that the UAVs first nullify their lead angles to zero within a fixed time (around 2 s), thereby aligning themselves onto the LOS. Then, they rendezvous with the virtual target by nullifying their relative range value (around 12 s) to zero. In other words, a UAV first orients itself onto the LOS joining it and the virtual target and then follows a tail chase to converge to the virtual target by mimicking the \emph{pure-pursuit's} tail-chase behavior. One may also observe from \Cref{fig:helix_v0_vu,fig:helix_v0_r_sigmau} that slightly higher linear velocity control is required initially. However, as soon as the relative separation between the UAV and the virtual target nullifies, the UAV's linear velocity converges to that of the virtual target. A similar conclusion can be drawn from \Cref{fig:helix_v0_omega_uy,fig:helix_v0_omega_uz,fig:helix_v0_angle_error}.

\begin{figure}[!ht]
    \centering
  \begin{subfigure}{0.5\linewidth}
      \centering
      \includegraphics[width=\linewidth]{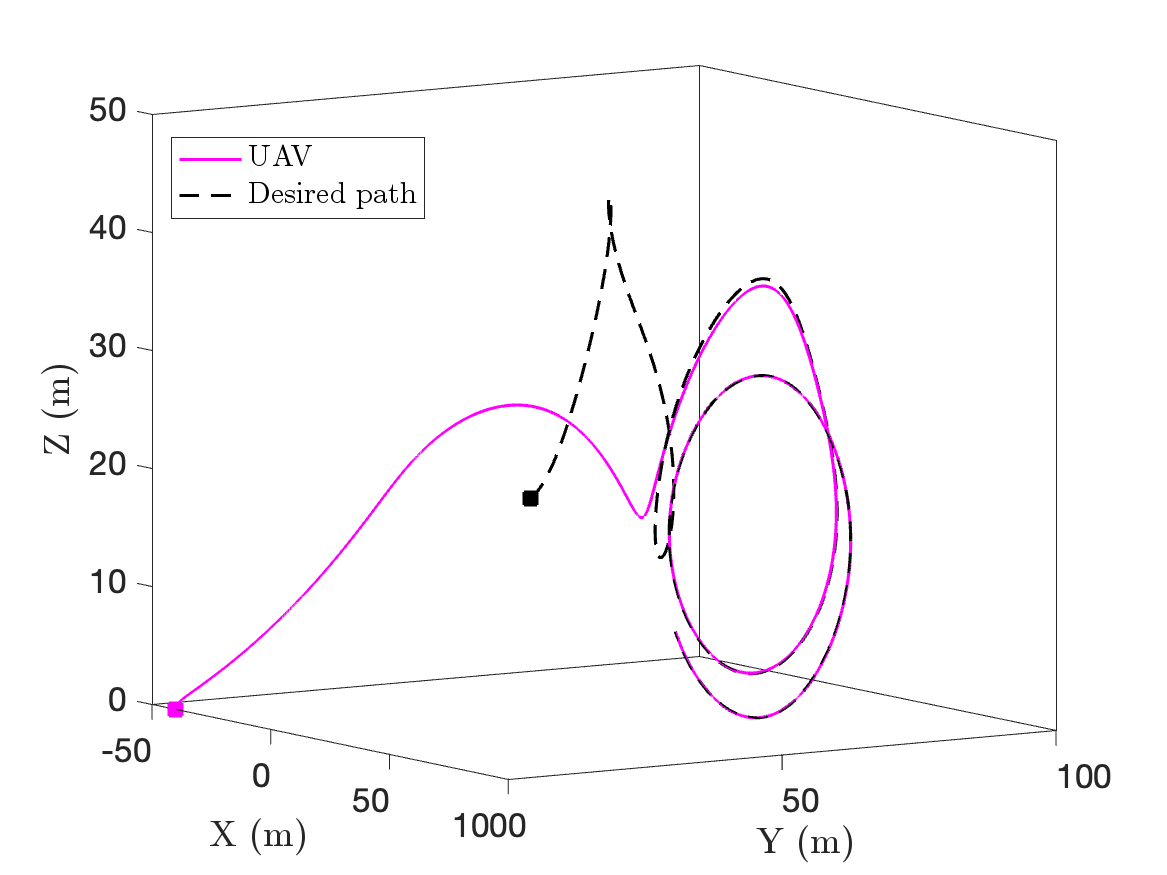}
      \caption{3D path.}
    \label{fig:helix_0_path}
  \end{subfigure}%
    \begin{subfigure}{0.5\linewidth}
      \centering
      \includegraphics[width=\linewidth]{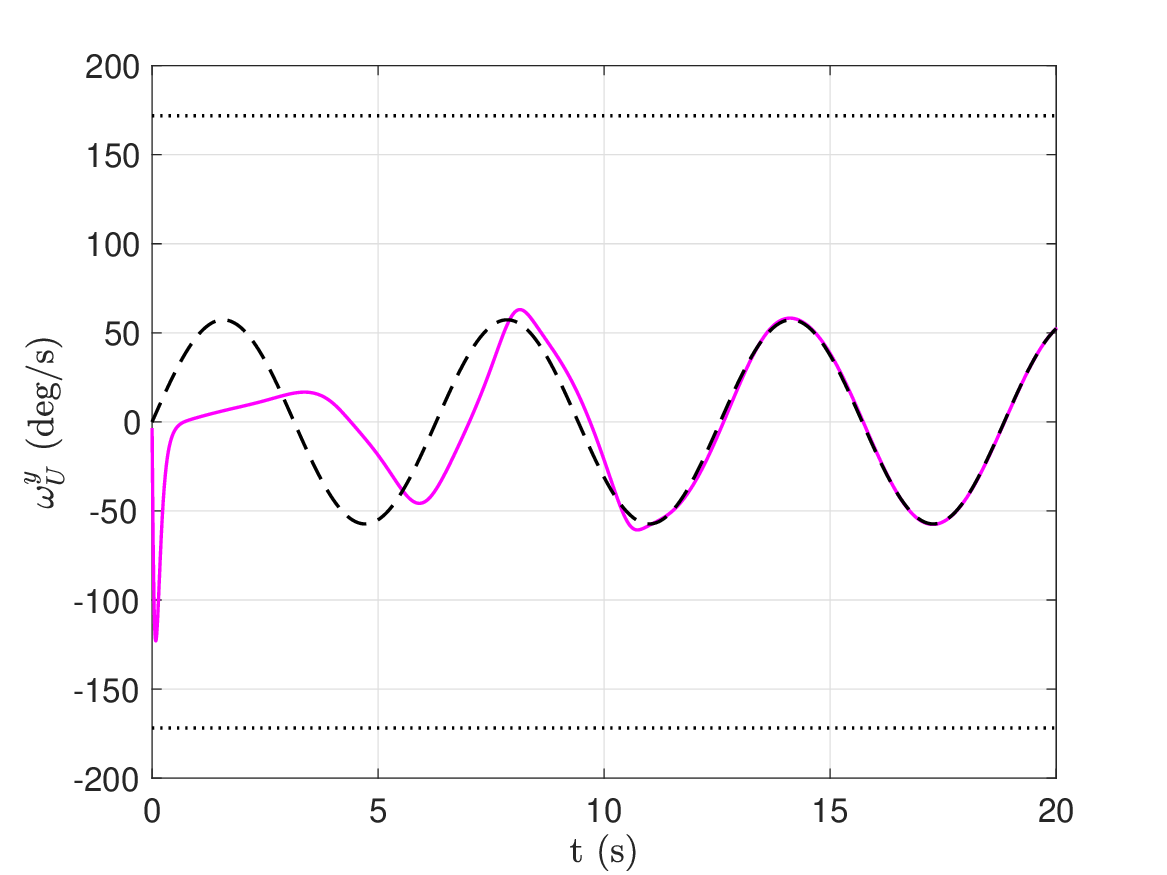}
      \caption{Angular velocities in yaw plane.}
    \label{fig:helix_0_omega_uy}
  \end{subfigure}
    \begin{subfigure}{0.5\linewidth}
      \centering
      \includegraphics[width=\linewidth]{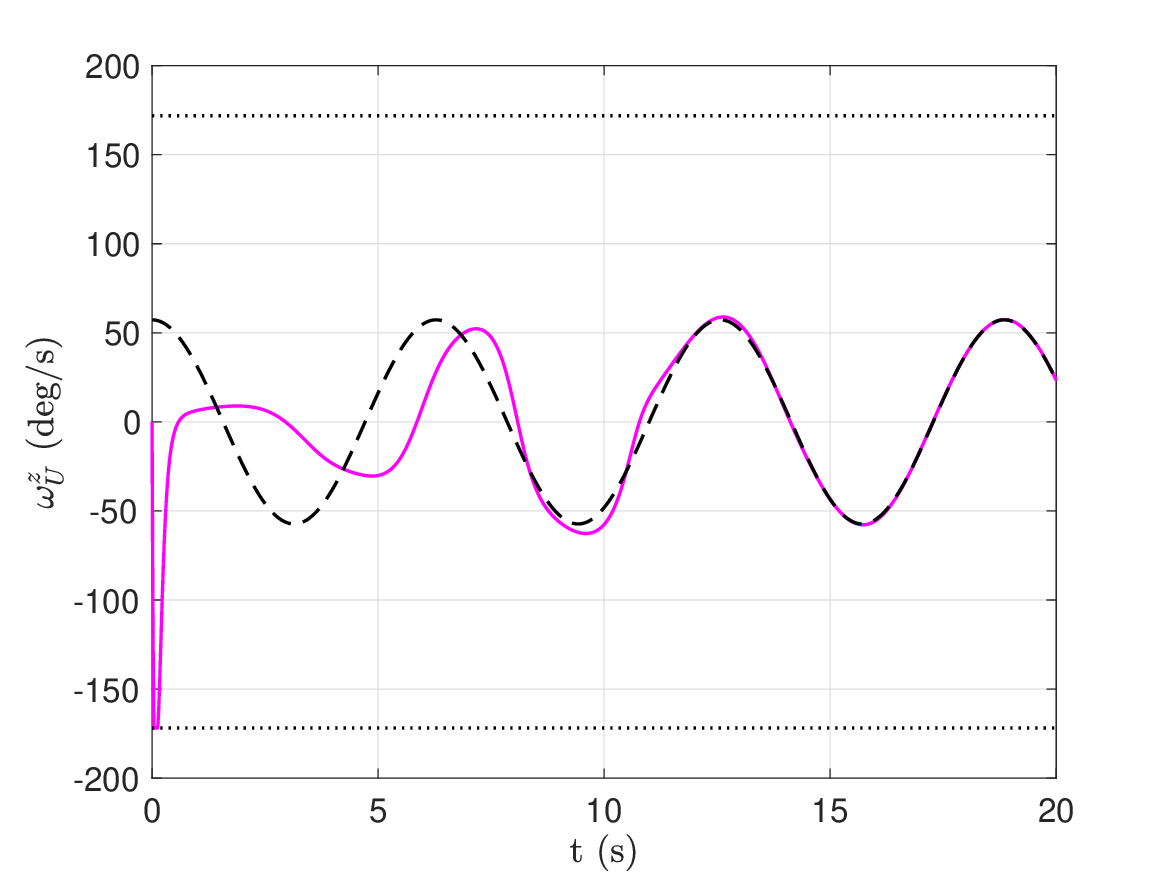}
      \caption{Angular velocities in pitch plane.}
    \label{fig:helix_0_omega_uz}
  \end{subfigure}%
      \begin{subfigure}{0.5\linewidth}
      \centering
      \includegraphics[width=\linewidth]{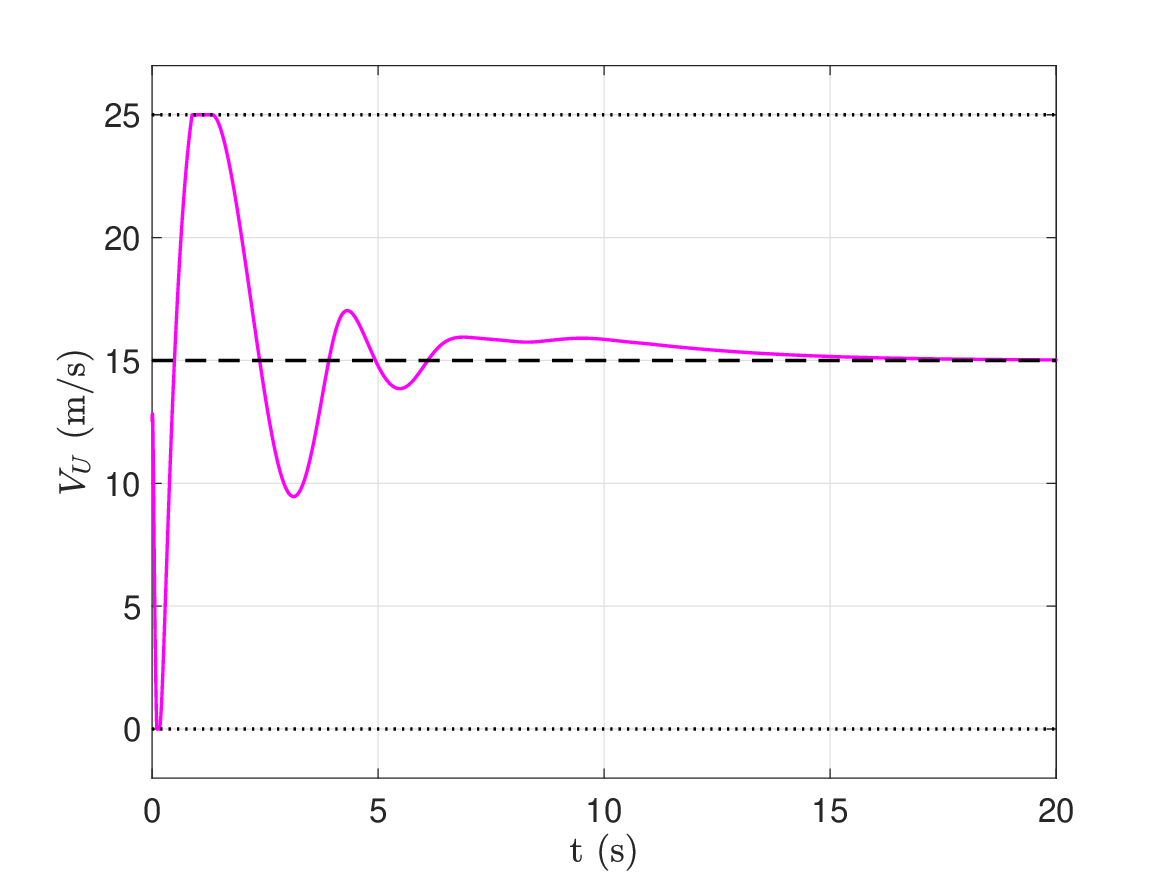}
      \caption{Linear speeds.}
    \label{fig:helix_0_vu}
  \end{subfigure}
      \begin{subfigure}{0.5\linewidth}
      \centering
      \includegraphics[width=\linewidth]{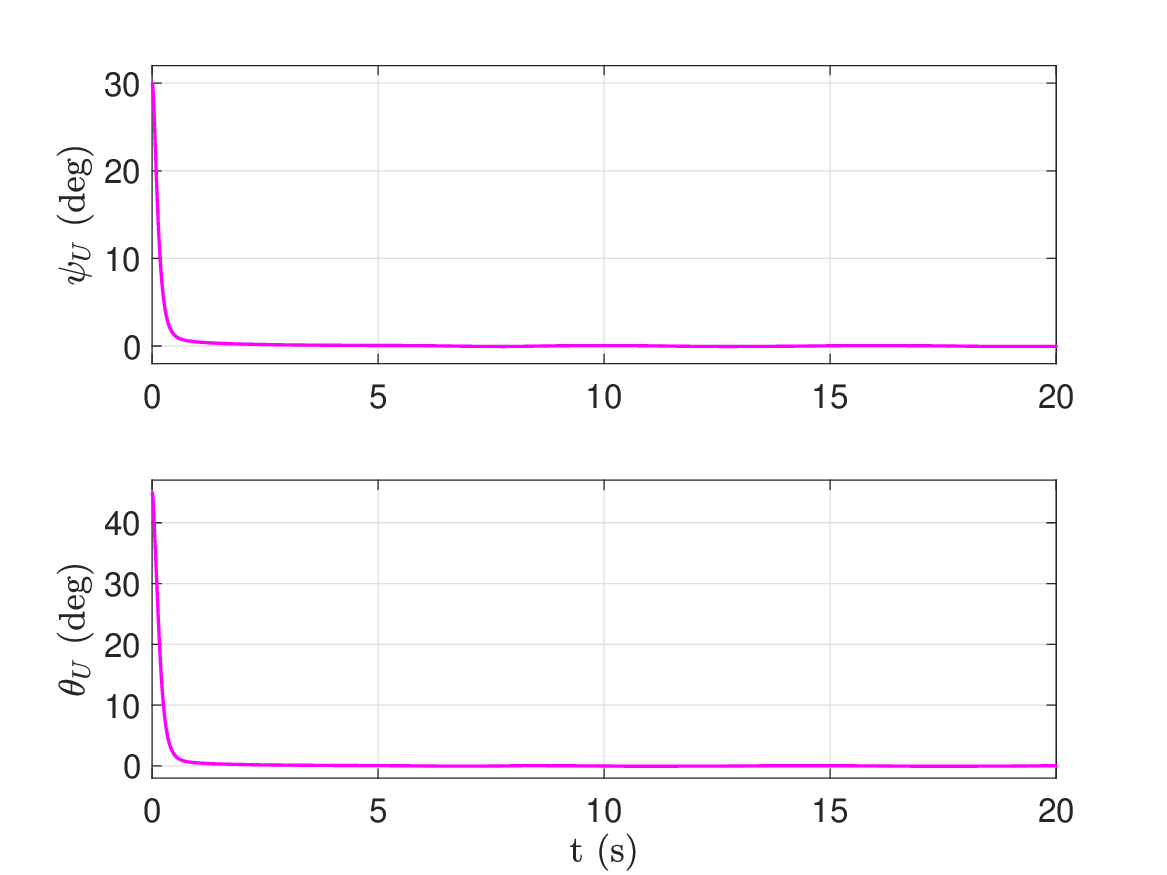}
      \caption{Lead angles.}
    \label{fig:helix_0_angle_error}
  \end{subfigure}%
  \begin{subfigure}{0.5\linewidth}
      \centering
      \includegraphics[width=\linewidth]{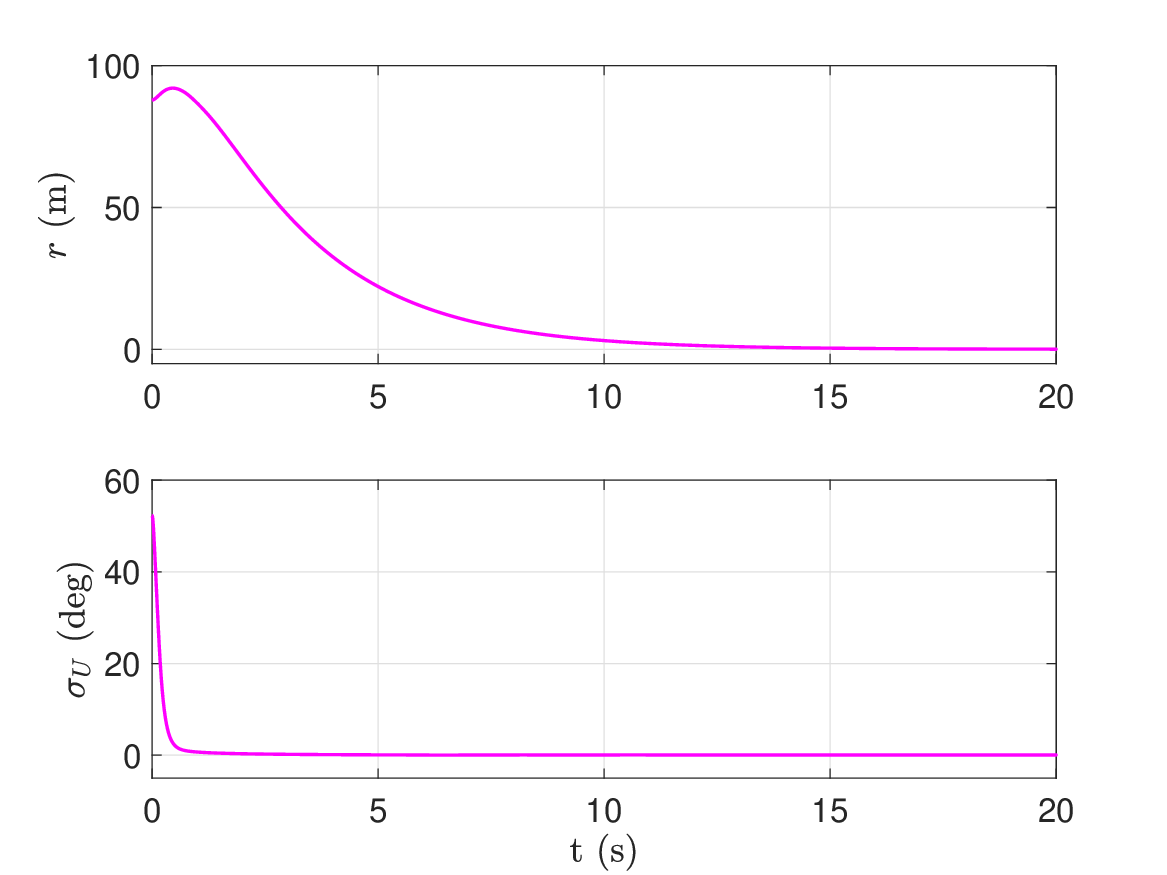}
      \caption{Relative range and effective heading angle.}
    \label{fig:helix_0_rsigmau}
  \end{subfigure}
    \caption{UAV following a  helix-like curvilinear path with $V\textsubscript{0}=0$ m/s.}
    \label{fig:helix_0}
\end{figure}
We now change the minimum speed of the UAV to 0 m/s. Note that this consideration is particularly important for multirotor-type vehicles as they have the ability to hover at a particular position and, thus, they can maintain a zero linear speed. The UAV is initially located at (-40,0,0) m with heading angles of $30^\circ$ and $45^\circ$. While keeping other settings the same as for the helix-like curvilinear path with $V\textsubscript{0}=3$ m/s, we show the performance of the UAV through \Cref{fig:helix_0}. It can be observed that the UAV follows its desired path by exhibiting a similar path-following behavior as for the $V\textsubscript{0}=3$ m/s. Form \Cref{fig:helix_0_vu} one may note that the UAV's linear speed can now reach to zero anytime during the engagement. This is an alluring feature of the proposed guidance strategy that remains applicable to a broader class of vehicles (vehicles that have to maintain a minimum speed as well as those that can maintain zero speed), thereby providing a general treatment of the problem by removing the dependency upon the vehicle class, as well as on the path curvature.

\begin{figure}[!ht]
    \centering
  \begin{subfigure}{0.5\linewidth}
      \centering
      \includegraphics[width=\linewidth]{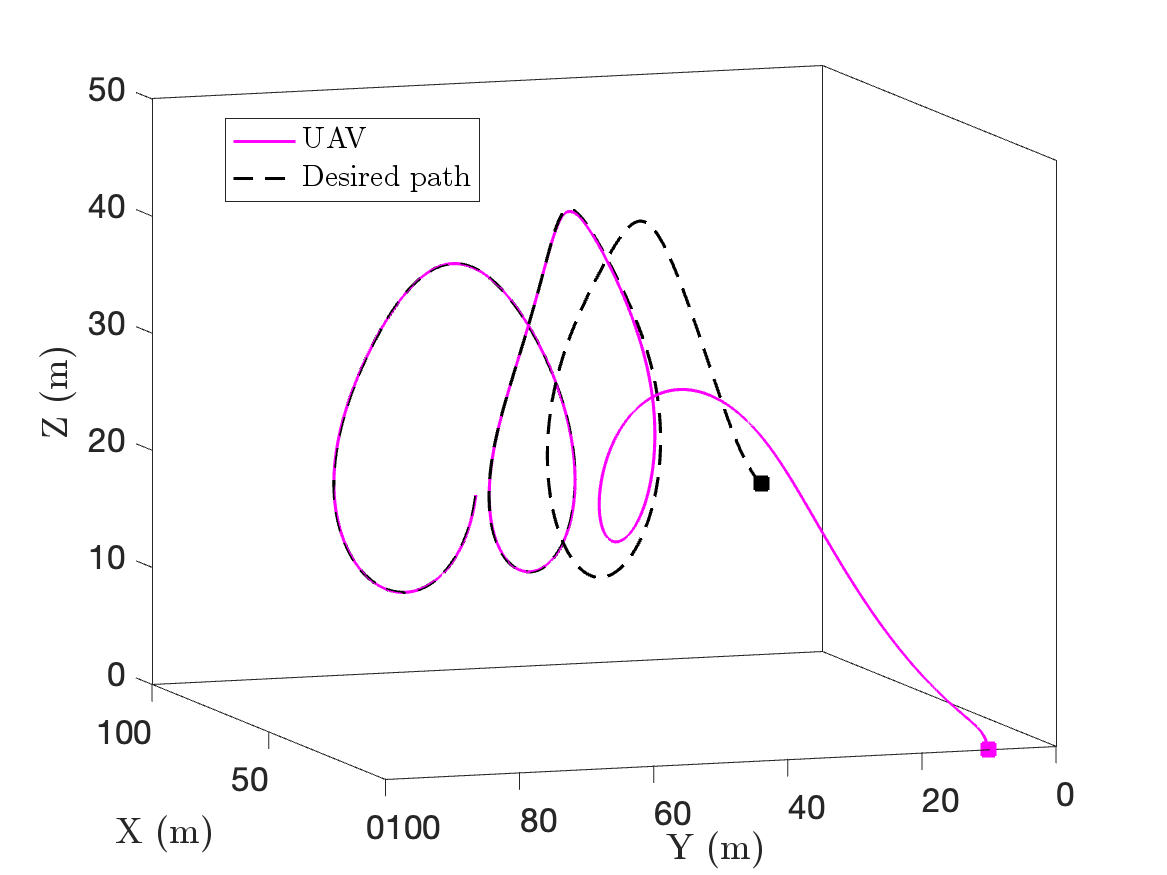}
      \caption{3D path.}
    \label{fig:different_MN_path}
  \end{subfigure}%
    \begin{subfigure}{0.5\linewidth}
      \centering
      \includegraphics[width=\linewidth]{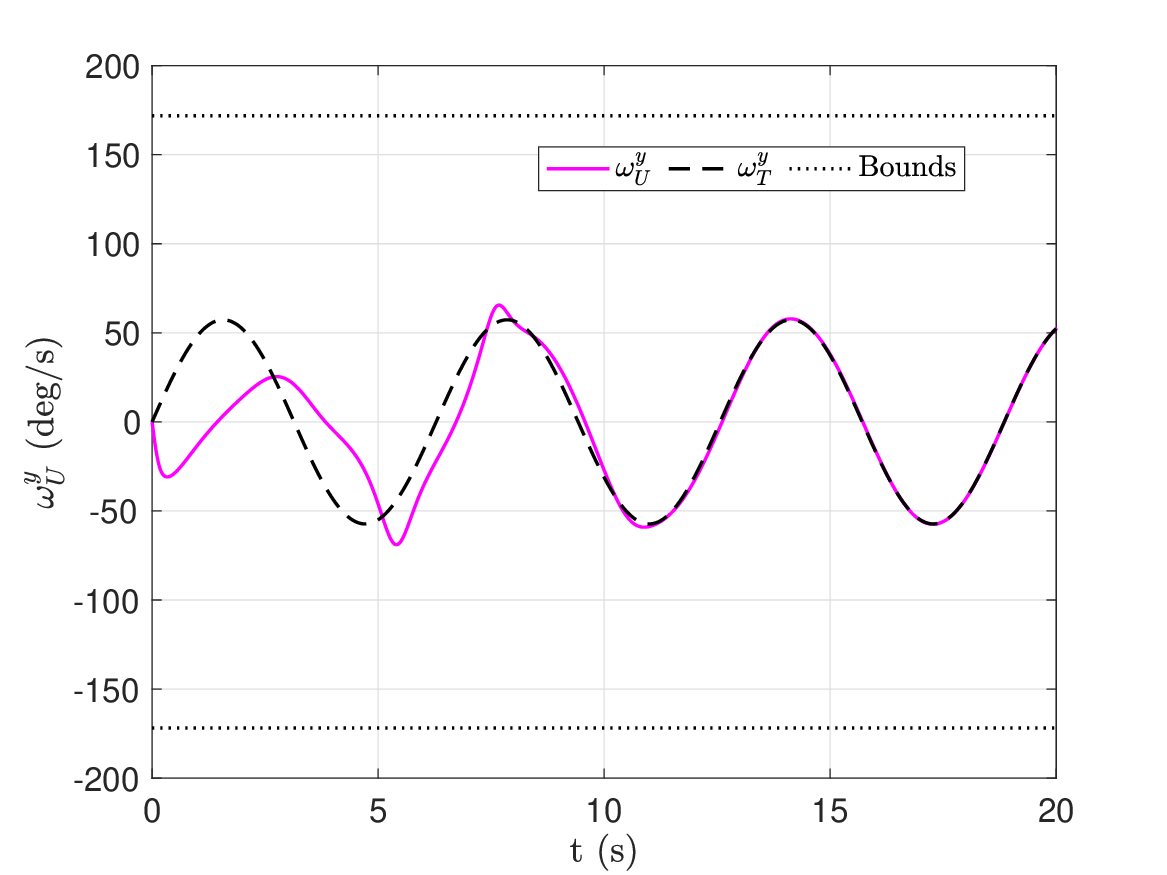}
      \caption{Angular velocities in yaw plane.}
    \label{fig:different_MN_omega_uy}
  \end{subfigure}
    \begin{subfigure}{0.5\linewidth}
      \centering
      \includegraphics[width=\linewidth]{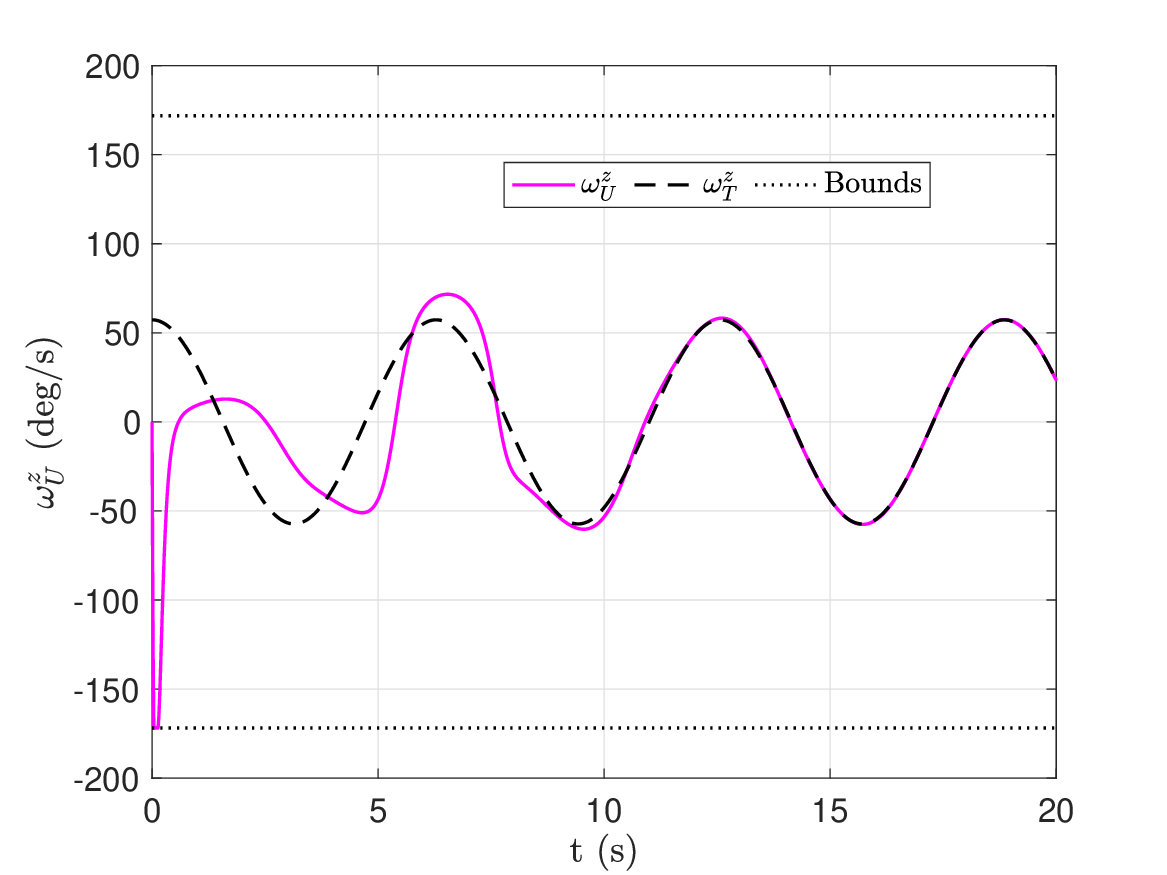}
      \caption{Angular velocities in pitch plane.}
    \label{fig:different_MN_omega_uz}
  \end{subfigure}%
      \begin{subfigure}{0.5\linewidth}
      \centering
      \includegraphics[width=\linewidth]{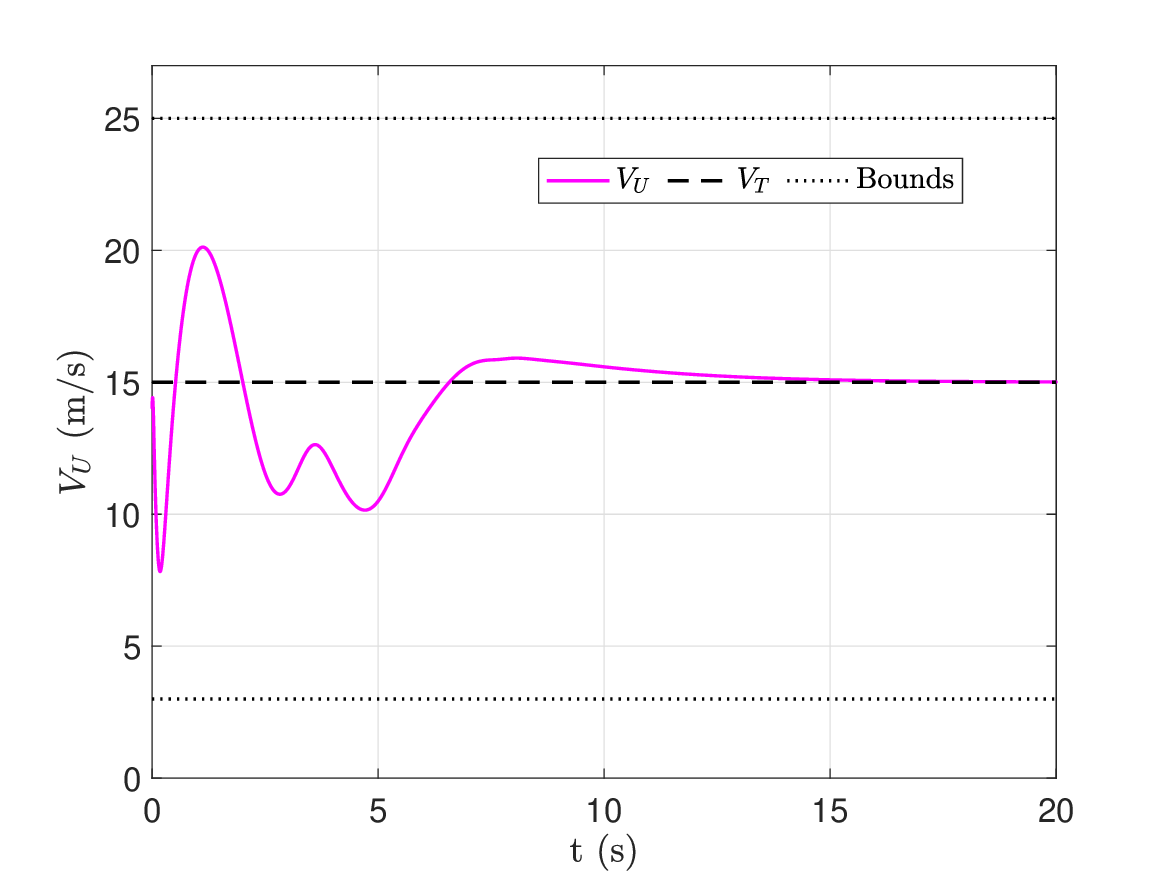}
      \caption{Linear speeds.}
    \label{fig:different_MN_vu}
  \end{subfigure}
      \begin{subfigure}{0.5\linewidth}
      \centering
      \includegraphics[width=\linewidth]{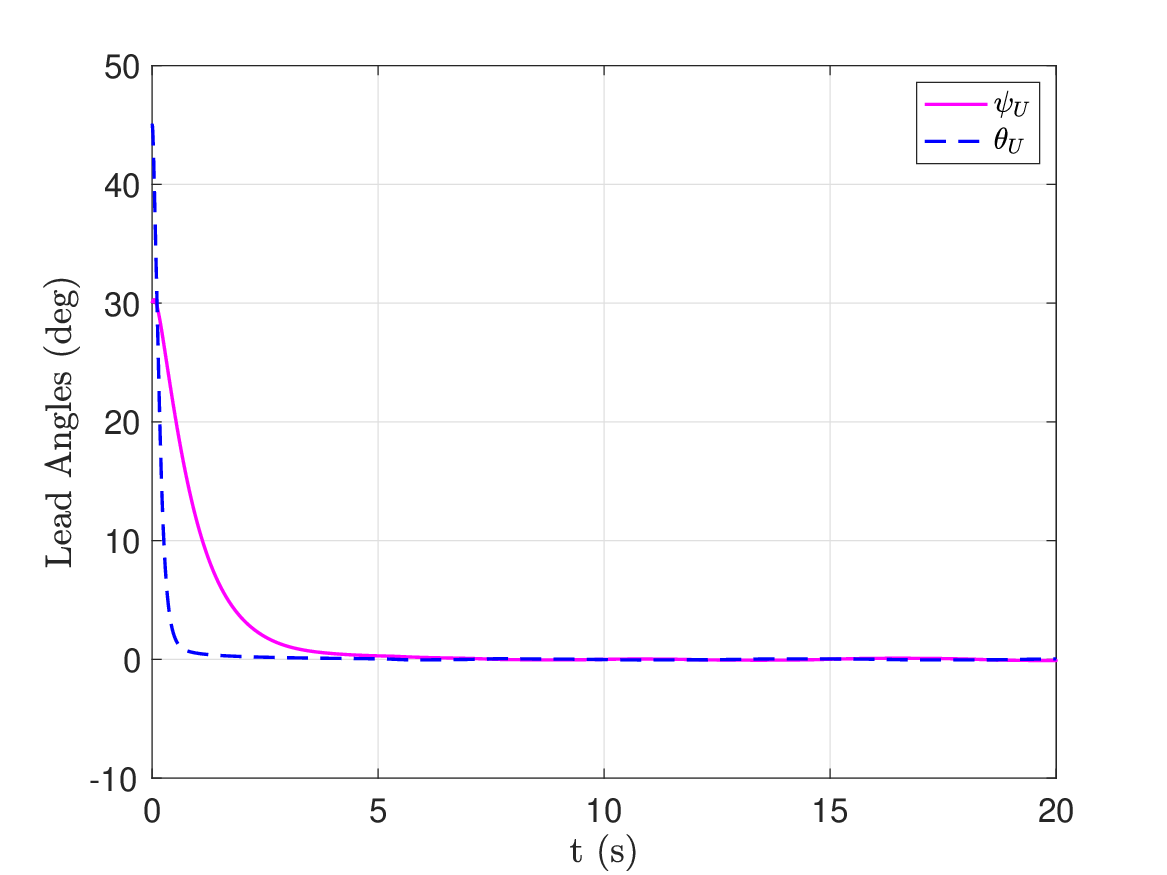}
      \caption{Lead angles.}
    \label{fig:different_MN_leadangles}
  \end{subfigure}%
  \begin{subfigure}{0.5\linewidth}
      \centering
      \includegraphics[width=\linewidth]{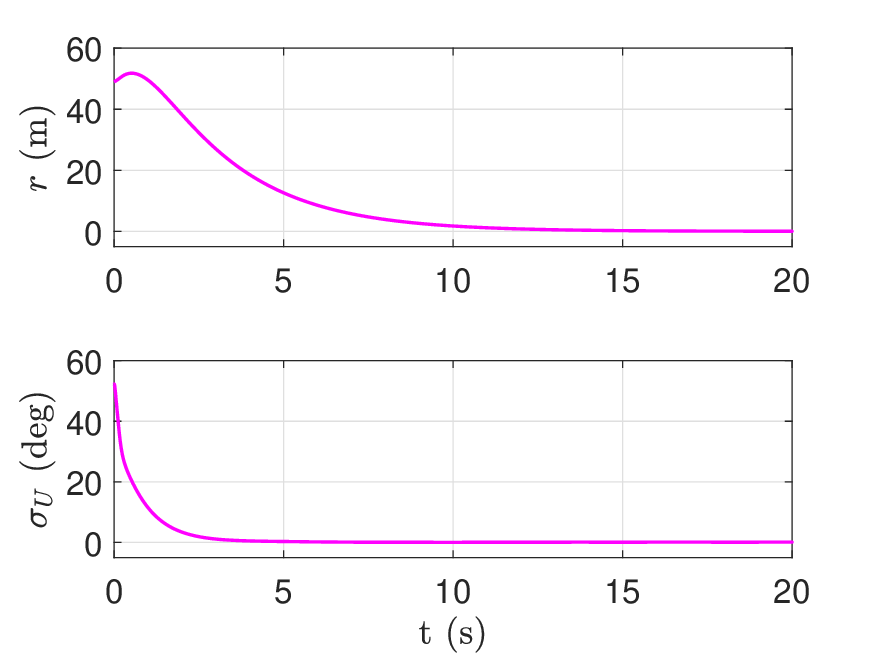}
      \caption{Relative range and effective heading angle.}
    \label{fig:different_MN_r_sigmau}
  \end{subfigure}
   \caption{UAV following a  helix-like curvilinear path with different $T\textsubscript{2}$ and $T\textsubscript{3}$.}
    \label{fig:different_MN}
\end{figure}
Please note that such a choice of convergence time is deliberate rather than co-incidental. The reason behind this is to enable precise tracking capabilities. To elucidate the possibility of different setting times, we have now presented a new simulation where convergence times are different. For $\mathcal{M}\textsubscript{2}=10$, $\mathcal{N}\textsubscript{2}=2$,  $\mathcal{M}\textsubscript{3}=10$, $\mathcal{N}\textsubscript{3}=2$, and keeping the other engagement settings the same as for the helix-like curvilinear path with $V\textsubscript{0}=3$ m/s, we depict the performance of the proposed guidance strategy through \Cref{fig:different_MN}. One can see that the UAV converges to its desired path despite the nonidentical values of the said design parameters. Additionally, it can be observed from \Cref{fig:different_MN_leadangles} that the convergence time of the lead angles in pitch and yaw planes now differs from each other, thereby showing that the convergence time can be adjusted based on the design parameters. Furthermore, from \Cref{fig:helix_v0_angle_error,fig:different_MN_leadangles}, it is evident that reducing $\mathcal{M}_{3}$ and $\mathcal{N}_{3}$ increases the time required for $\psi_U$ to converge to zero. \Cref{fig:helix_v0_omega_uy,fig:different_MN_omega_uy} demonstrate that lower gain values result in smaller control inputs during the transient phase. Importantly, while transient behavior is affected, steady-state control inputs remain unchanged.

\begin{figure}[!ht]
    \centering
  \begin{subfigure}{0.5\linewidth}
      \centering
      \includegraphics[width=\linewidth]{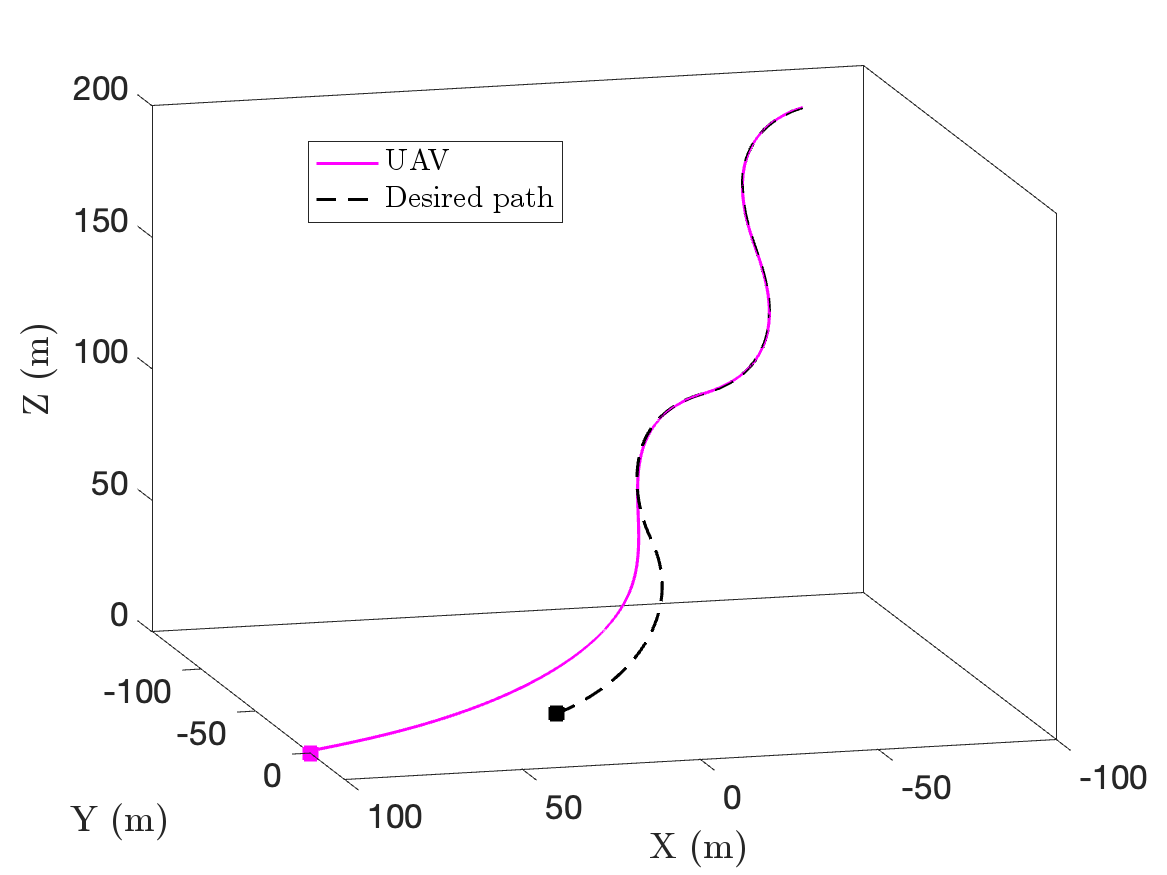}
      \caption{3D path.}
    \label{fig:slowcurve_path}
  \end{subfigure}%
    \begin{subfigure}{0.5\linewidth}
      \centering
      \includegraphics[width=\linewidth]{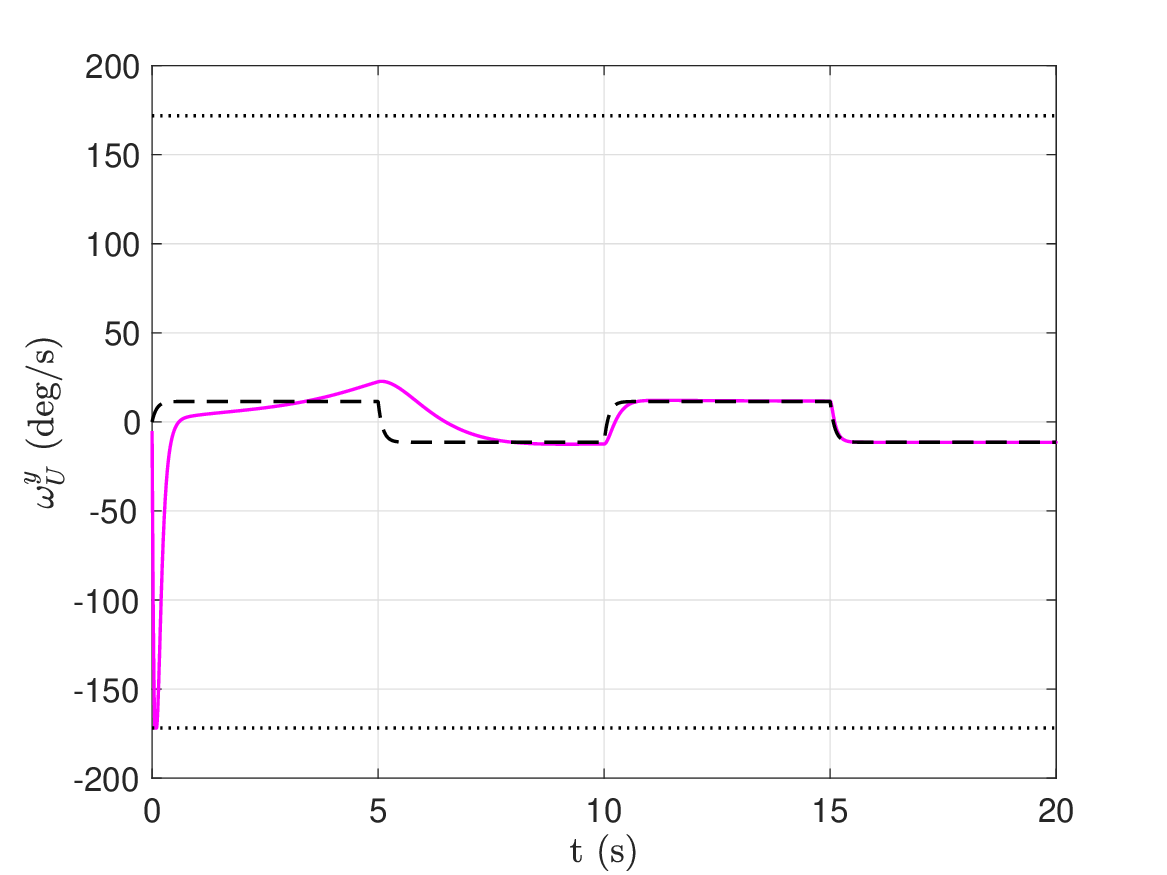}
      \caption{Angular velocities in yaw plane.}
    \label{fig:slowcurve_omega_u_y}
  \end{subfigure}
    \begin{subfigure}{0.5\linewidth}
      \centering
      \includegraphics[width=\linewidth]{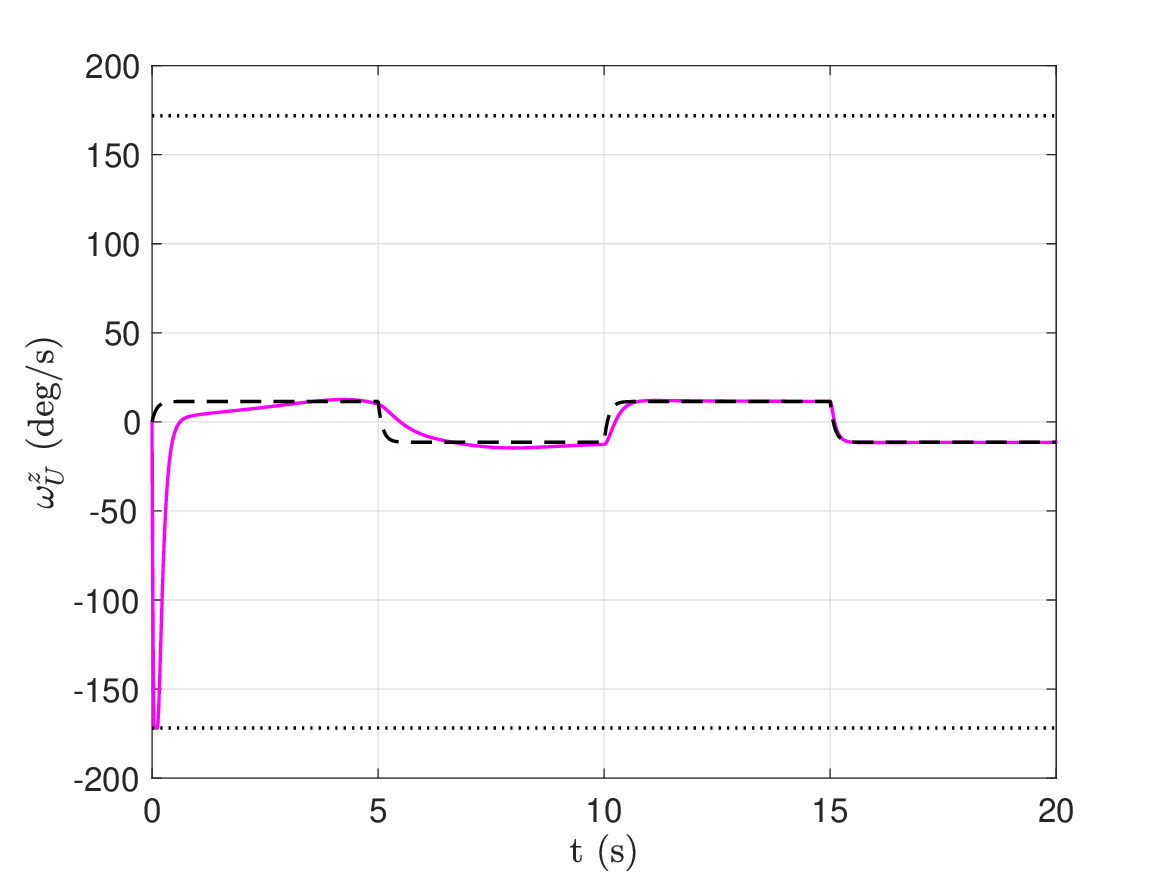}
      \caption{Angular velocities in pitch plane.}
    \label{fig:slowcurve_omega_u_z}
  \end{subfigure}%
      \begin{subfigure}{0.5\linewidth}
      \centering
      \includegraphics[width=\linewidth]{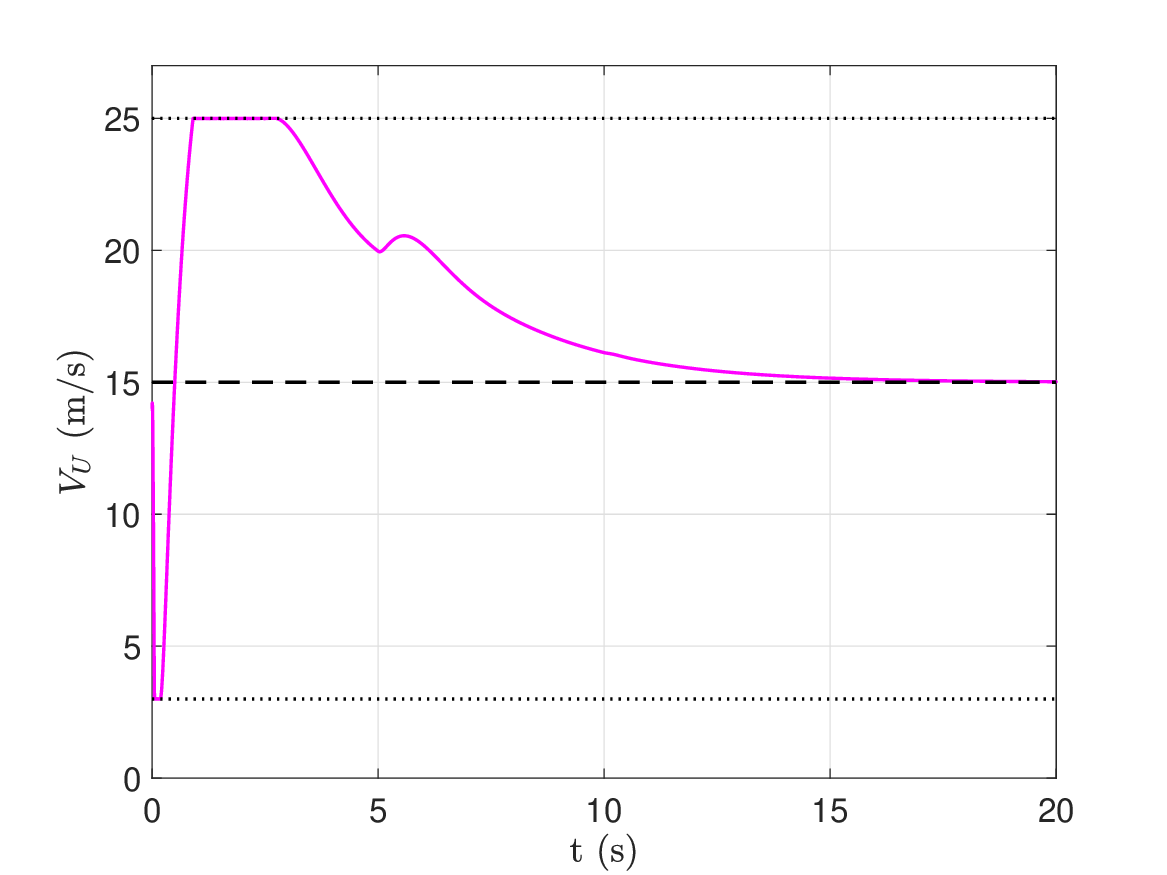}
      \caption{Linear speeds.}
    \label{fig:slowcurve_vu}
  \end{subfigure}
      \begin{subfigure}{0.5\linewidth}
      \centering
      \includegraphics[width=\linewidth]{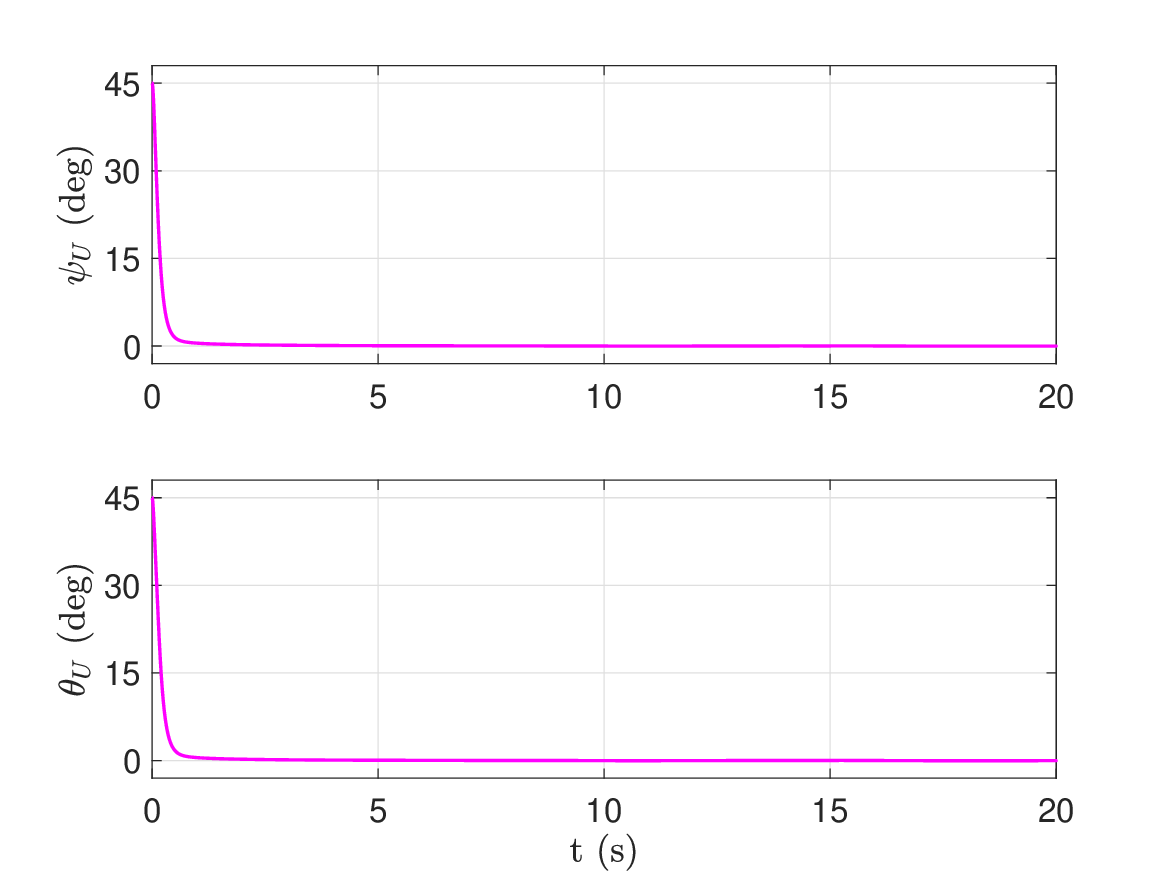}
      \caption{Lead angles.}
    \label{fig:slowcurve_angle_error}
  \end{subfigure}%
  \begin{subfigure}{0.5\linewidth}
      \centering
      \includegraphics[width=\linewidth]{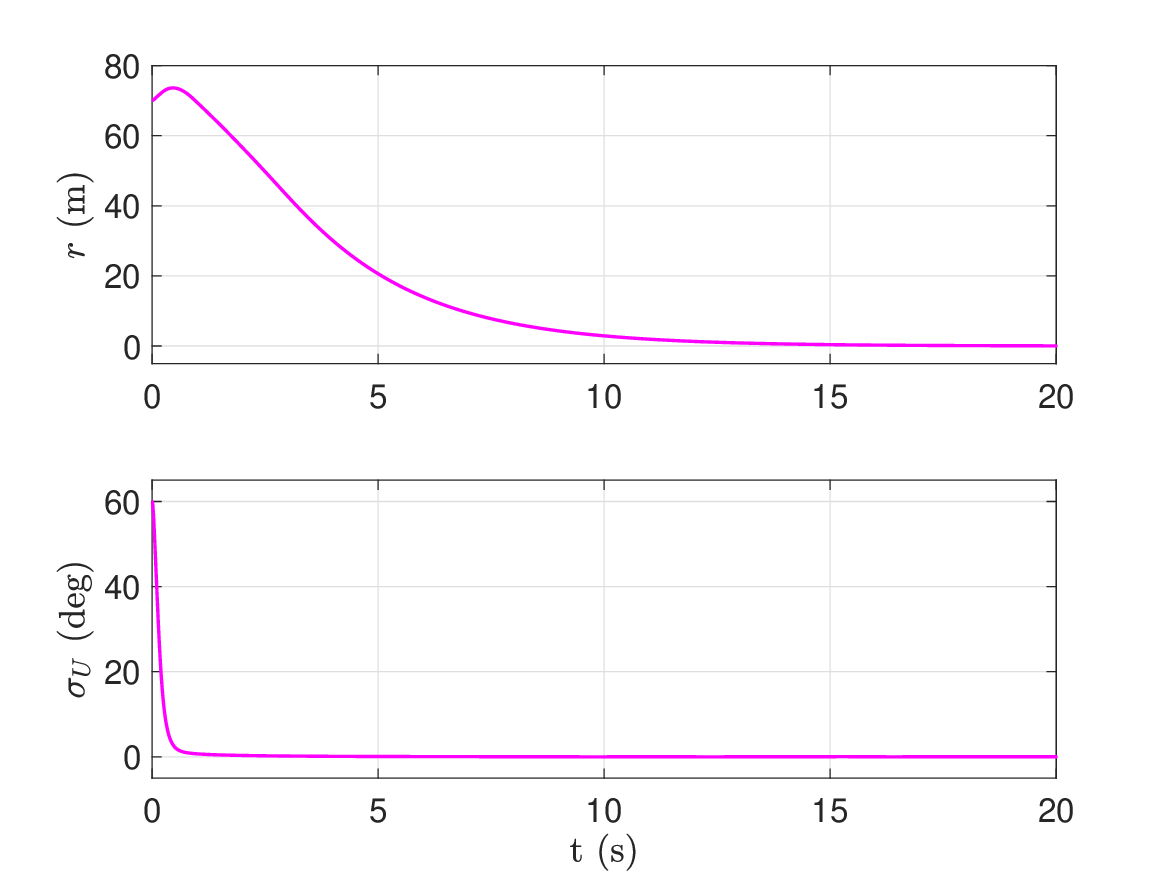}
      \caption{Relative range and effective heading angle.}
    \label{fig:slowcurve_rsigmau}
  \end{subfigure}
    \caption{UAV following a min-radius turn path.}
    \label{fig:slowcurve}
\end{figure}
Most of the smooth curvilinear paths can be thought of as an extension of the S-curve, which may exhibit a minimum turn radius. Such paths may be parametrized as 
\begin{align*}
 V_{T} =& 15~ \text{m/s,}\\
 \omega_{T}^{y} =& \begin{cases}
     0.2; & \text{if}~t \in [0,5)~s,\\
    -0.2; & \text{if}~t \in [5,10)~s,\\
     0.2; & \text{if}~t \in [10,15)~s,\\
    -0.2; & \text{if}~t \in [15,20)~s,
\end{cases},~~ 
\omega_{T}^{z} = \begin{cases}
     0.2; & \text{if}~t \in [0,5)~s,\\
    -0.2; & \text{if}~t \in [5,10)~s,\\
     0.2; & \text{if}~t \in [10,15)~s,\\
    -0.2; & \text{if}~t \in [15,20)~s.
\end{cases}
\end{align*}
In this scenario, the virtual target is initially located at $(40,30,20)$ m with $\psi_{T}(0)=15^\circ$ and $\theta_T(0)=15^\circ$, whereas the UAV is initially located at $(100,0,0)$ m with initial heading angles of $\psi_{U}(0)=45^\circ$ and $\theta_U(0)=45^\circ$. The minimum speed of the UAV is chosen to be $3$ m/s. \Cref{fig:slowcurve} shows that the UAV follows such paths by making necessary turns while simultaneously satisfying the minimum turning radius constraints. It can be observed from \Cref{fig:slowcurve_path} that the UAV converges to its desired path around 12 s by nullifying its lead angles and its relative separation from the virtual target. One can notice from \Cref{fig:slowcurve_omega_u_y,fig:slowcurve_omega_u_z,fig:slowcurve_vu} that once the error variables converge to zero, the UAV modulates its linear and angular velocities accordingly to track the virtual target and remains on the desired path for future times. 

\begin{figure}[!ht]
    \centering
  \begin{subfigure}{0.5\linewidth}
      \centering
      \includegraphics[width=\linewidth]{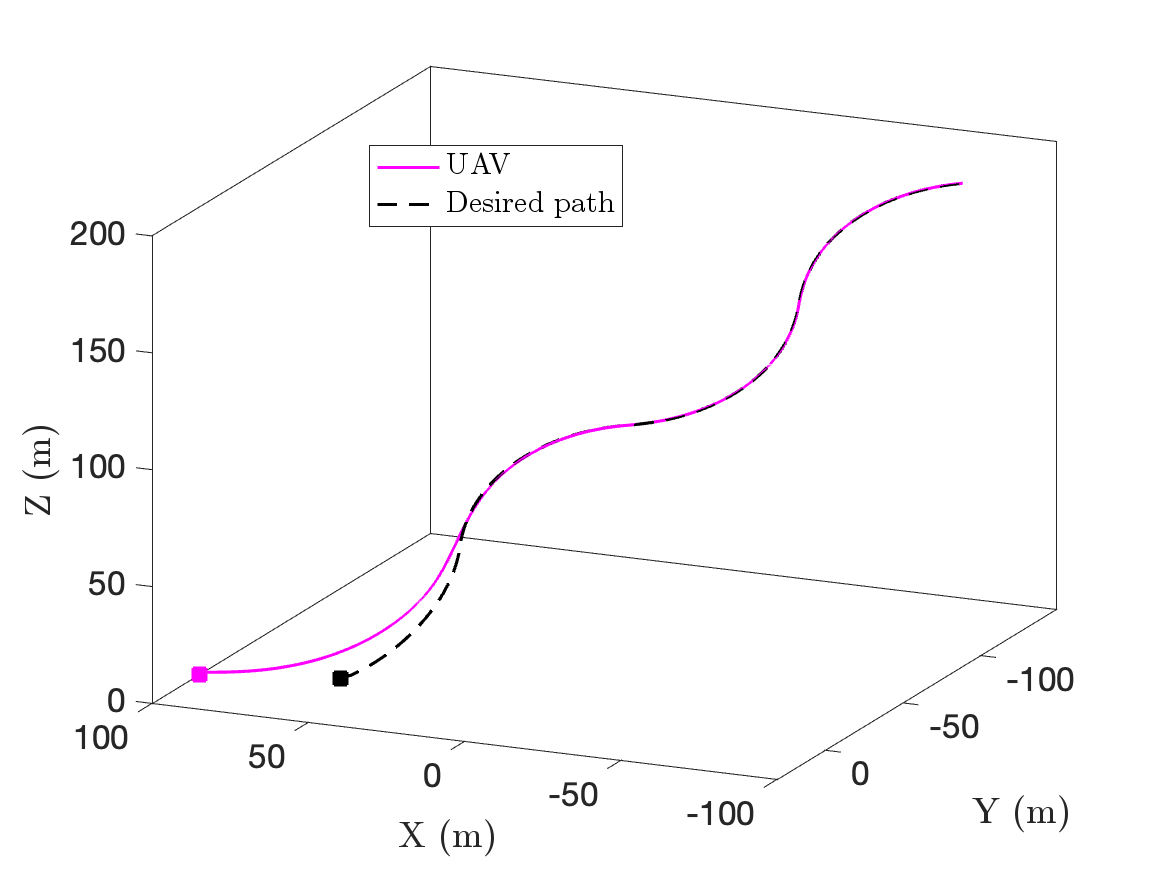}
      \caption{3D path.}
    \label{fig:timevar_vt_path}
  \end{subfigure}%
    \begin{subfigure}{0.5\linewidth}
      \centering
      \includegraphics[width=\linewidth]{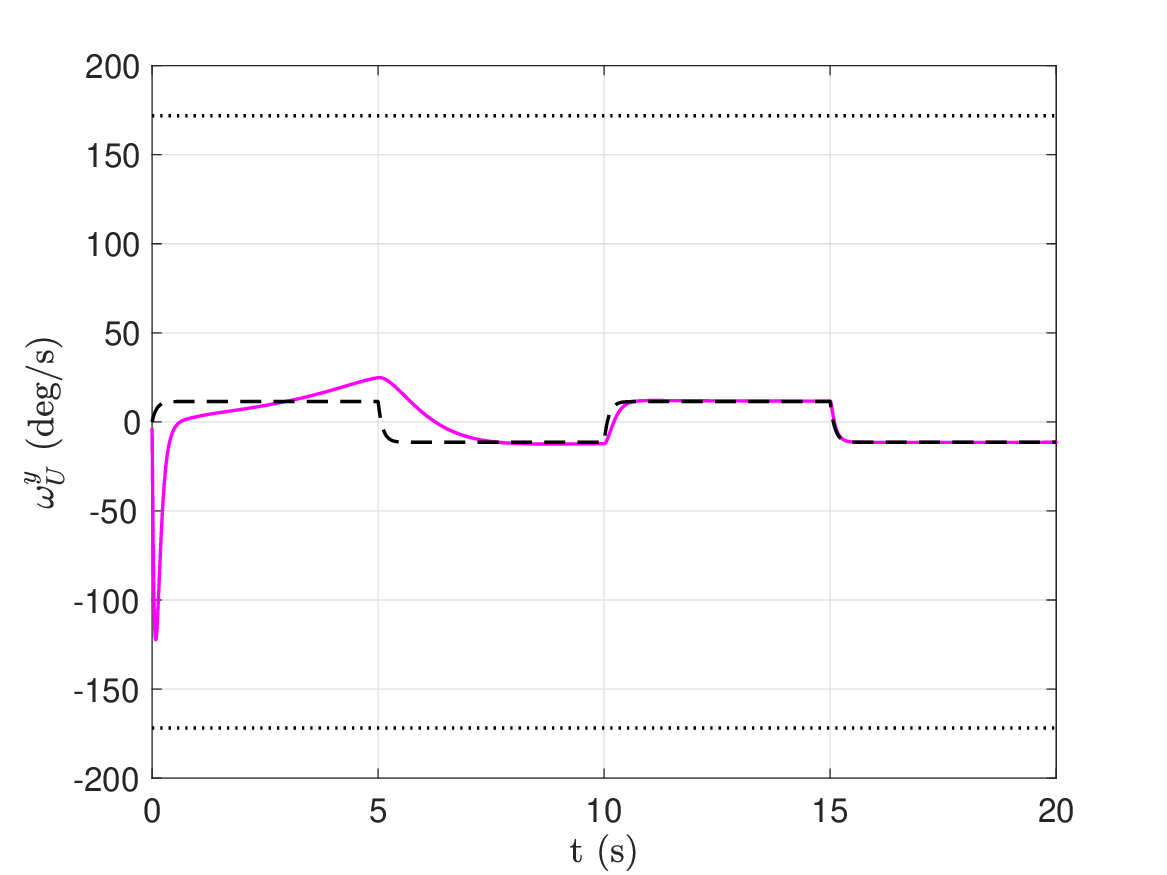}
      \caption{Angular velocities in yaw plane.}
    \label{fig:timevar_vt_omega_uy}
  \end{subfigure}
    \begin{subfigure}{0.5\linewidth}
      \centering
      \includegraphics[width=\linewidth]{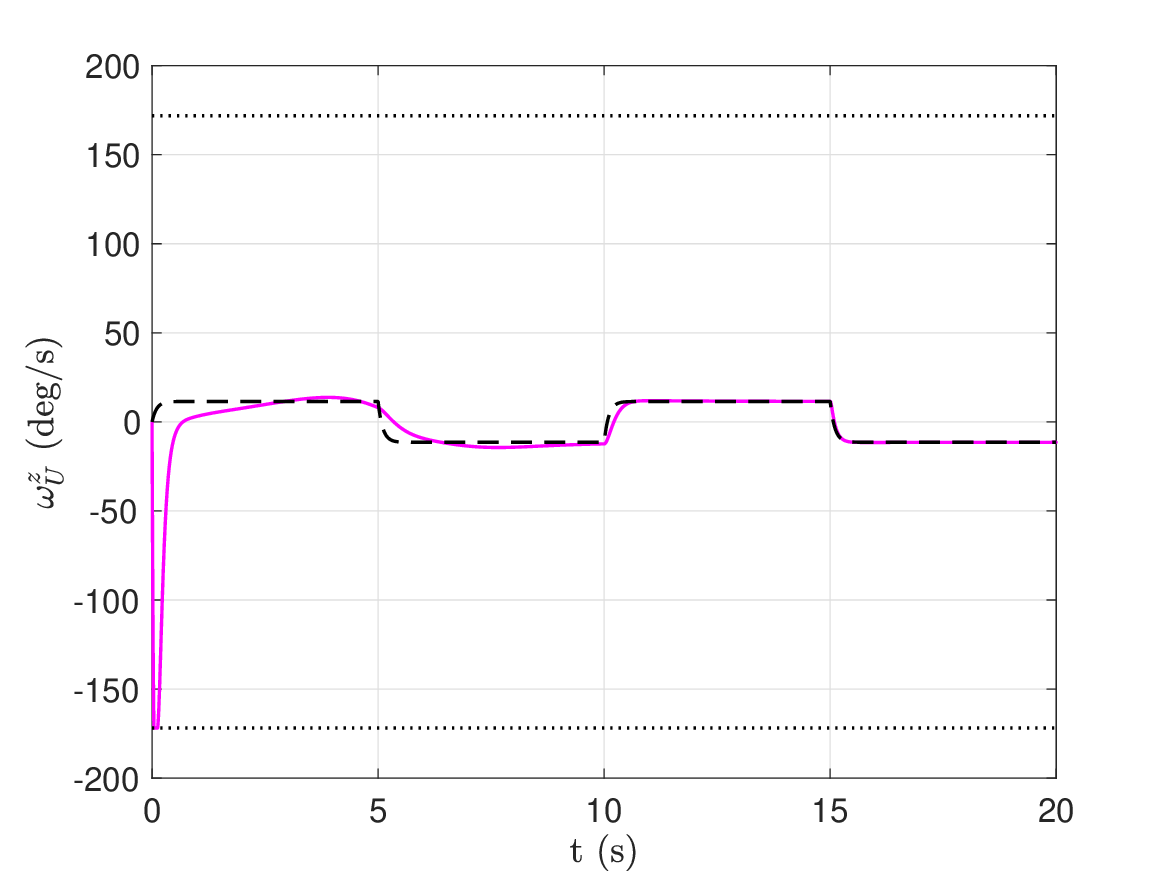}
      \caption{Angular velocities in pitch plane. }
    \label{fig:timevar_vt_omega_uz}
  \end{subfigure}%
      \begin{subfigure}{0.5\linewidth}
      \centering
      \includegraphics[width=\linewidth]{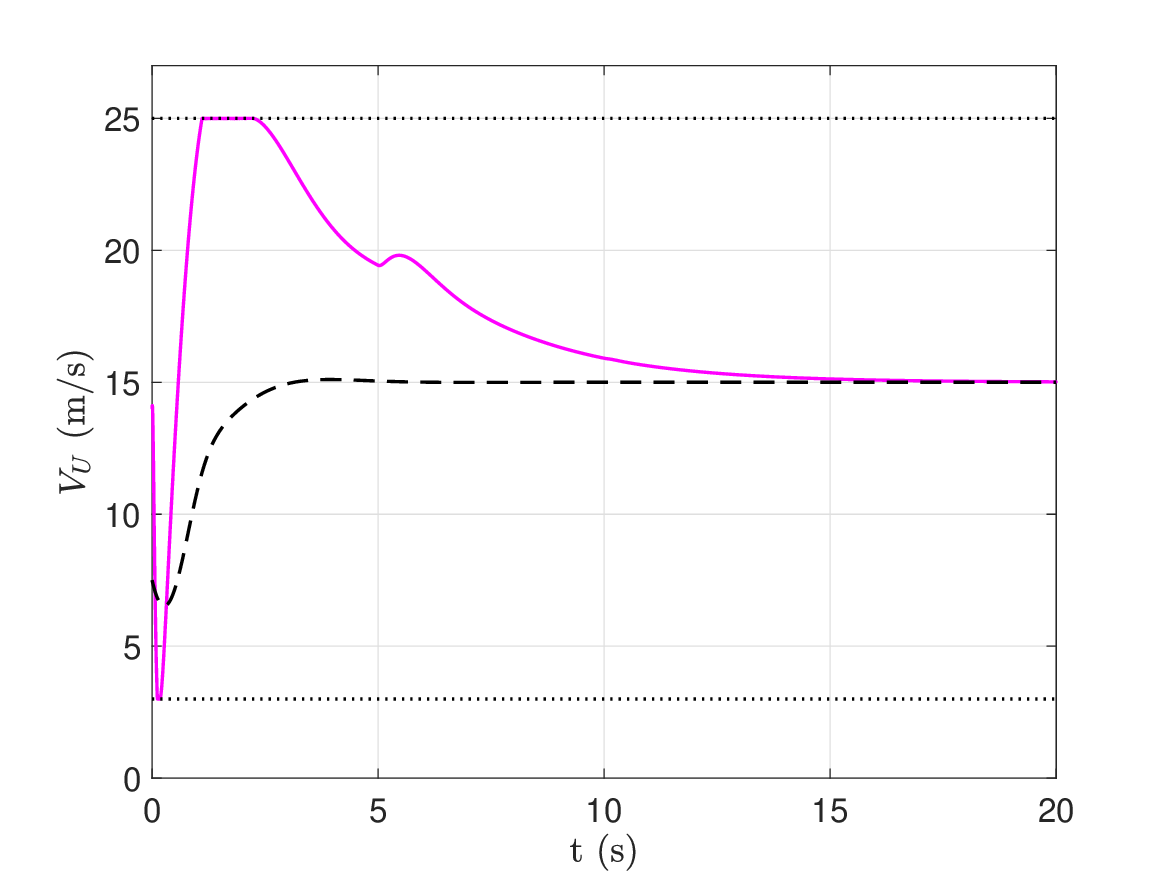}
      \caption{Linear speeds.}
    \label{fig:timevar_vt_vu}
  \end{subfigure}
      \begin{subfigure}{0.5\linewidth}
      \centering
      \includegraphics[width=\linewidth]{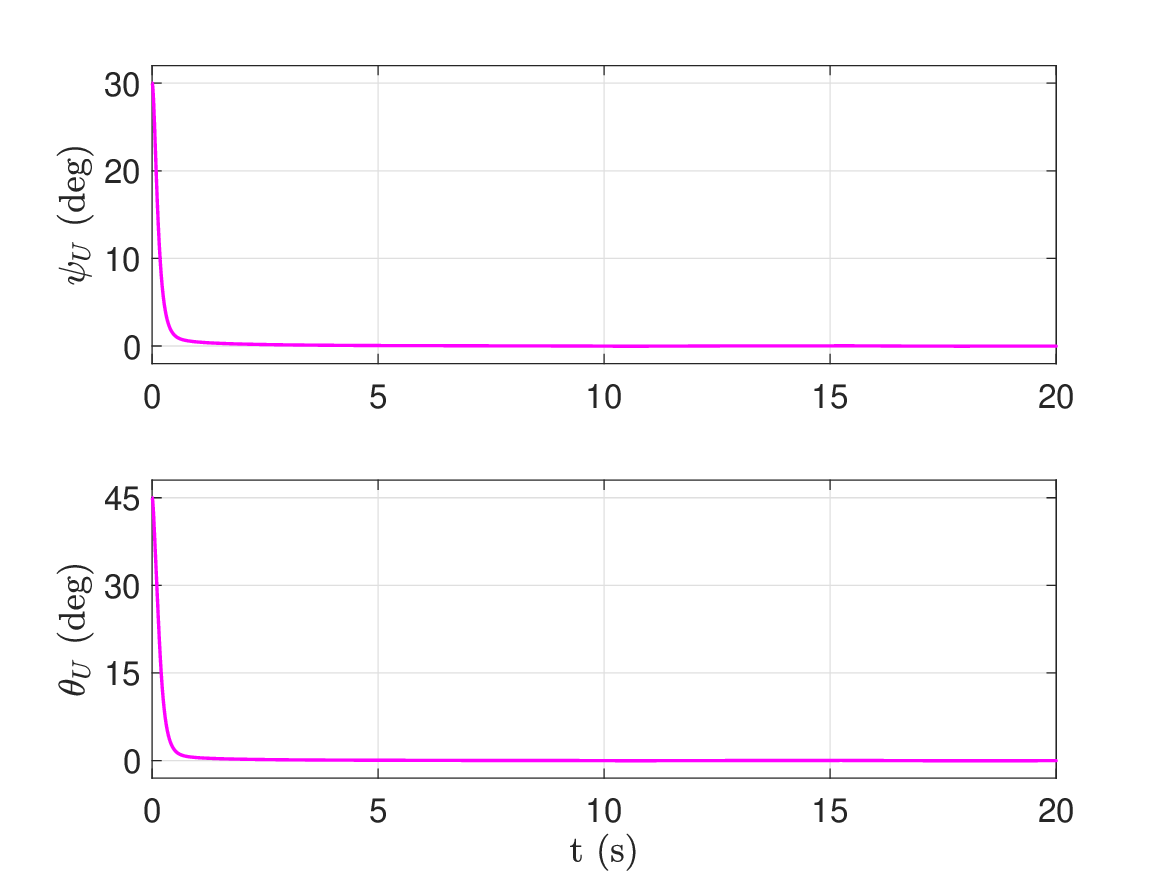}
      \caption{Lead angles.}
    \label{fig:timevar_vt_angle_error}
  \end{subfigure}%
  \begin{subfigure}{0.5\linewidth}
      \centering
      \includegraphics[width=\linewidth]{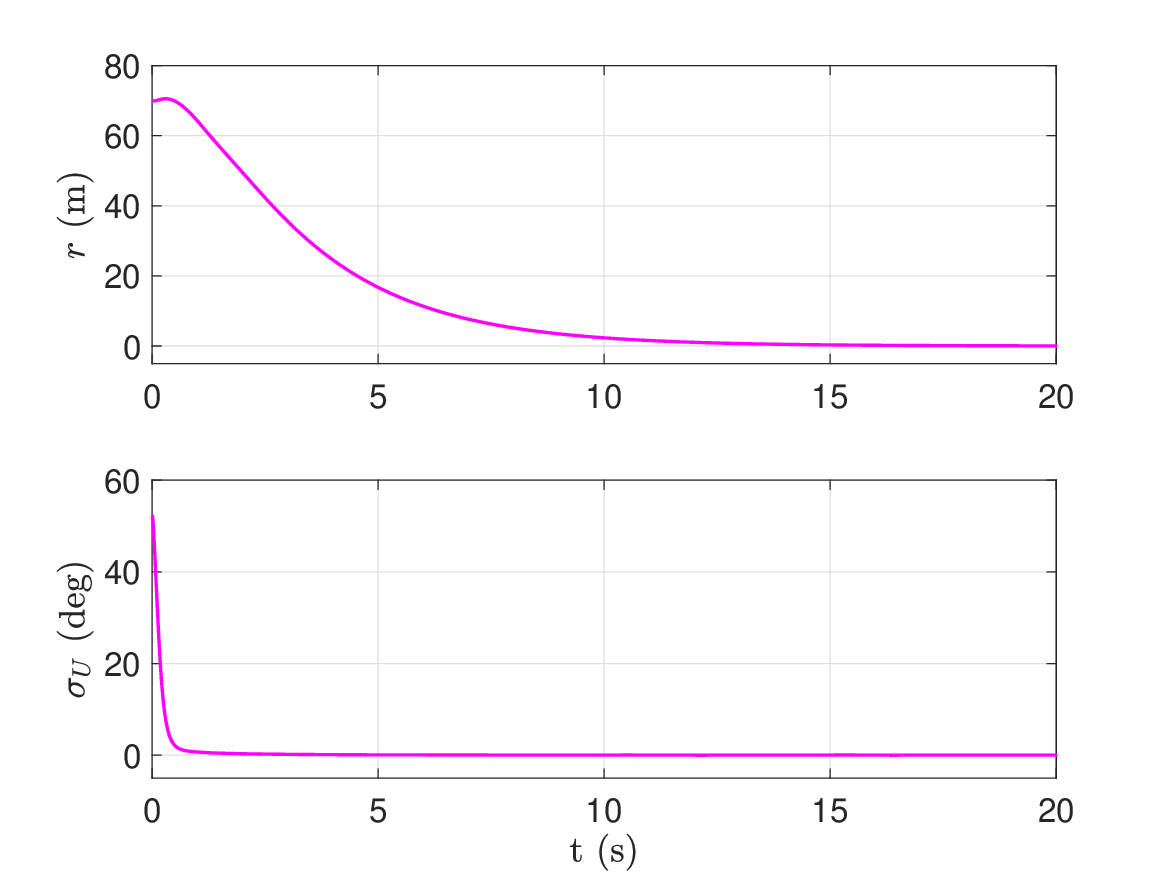}
      \caption{Relative range and effective heading angle.}
    \label{fig:timevar_vt_rsigmau}
  \end{subfigure}
    \caption{Performance under min-radius turns when $V_T$ varies with time.}
    \label{fig:timevar_vt}
\end{figure}
We now consider that the linear velocity of the virtual target becomes time-varying, say $V_T=7.5[ \tanh(t^2) + (1- e^{-t})\sin t ]$ m/s, and the initial position of the UAV is assumed to be $(100,0,0)$ m with $\theta_U(0)=45^\circ$ and $\psi_U(0)=30^\circ$. While keeping the other setting the same in the case of the S-shaped curvilinear path, the performance of the proposed guidance strategy is depicted \Cref{fig:timevar_vt}. We observe a similar performance even when the speed of the pseudo-target changes while also incorporating minimum-turn radius turn constraints. This, in turn, validates the claims that the proposed strategy remains invariant to virtual target speed changes and path curvature changes.

\begin{figure}[!ht]
    \centering
  \begin{subfigure}{0.5\linewidth}
      \centering
      \includegraphics[width=\linewidth]{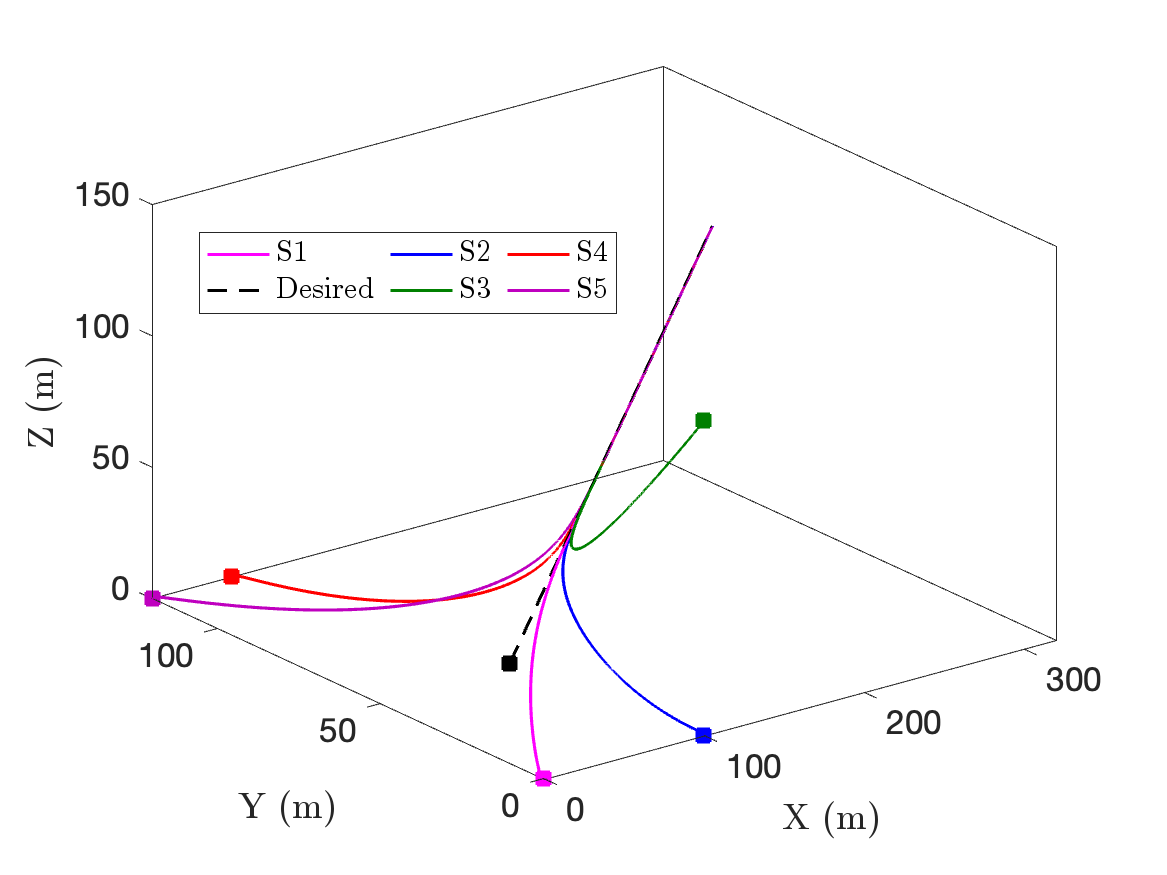}
      \caption{3D path.}
    \label{fig:stline_path}
  \end{subfigure}%
    \begin{subfigure}{0.5\linewidth}
      \centering
      \includegraphics[width=\linewidth]{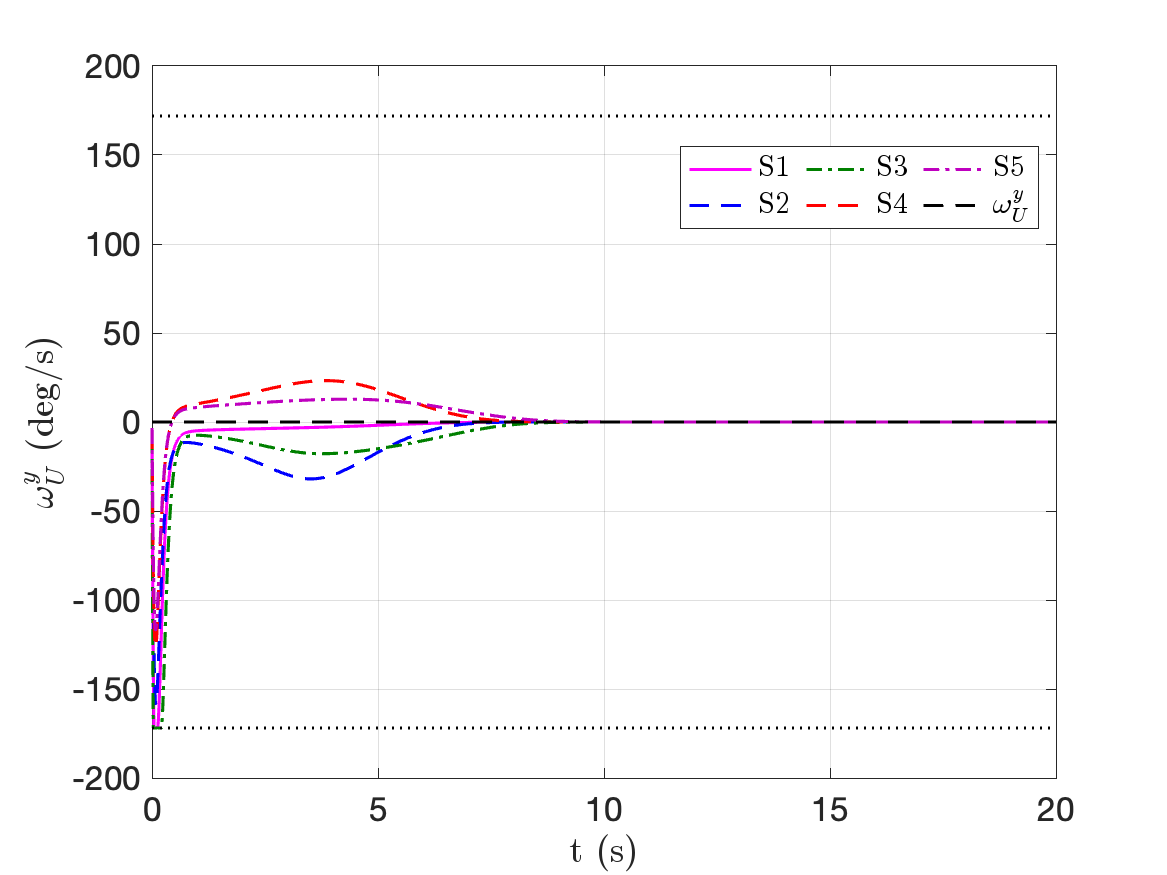}
      \caption{UAV's angular velocities in yaw plane.}
    \label{fig:stline_omega_u_y}
  \end{subfigure}
    \begin{subfigure}{0.5\linewidth}
      \centering
      \includegraphics[width=\linewidth]{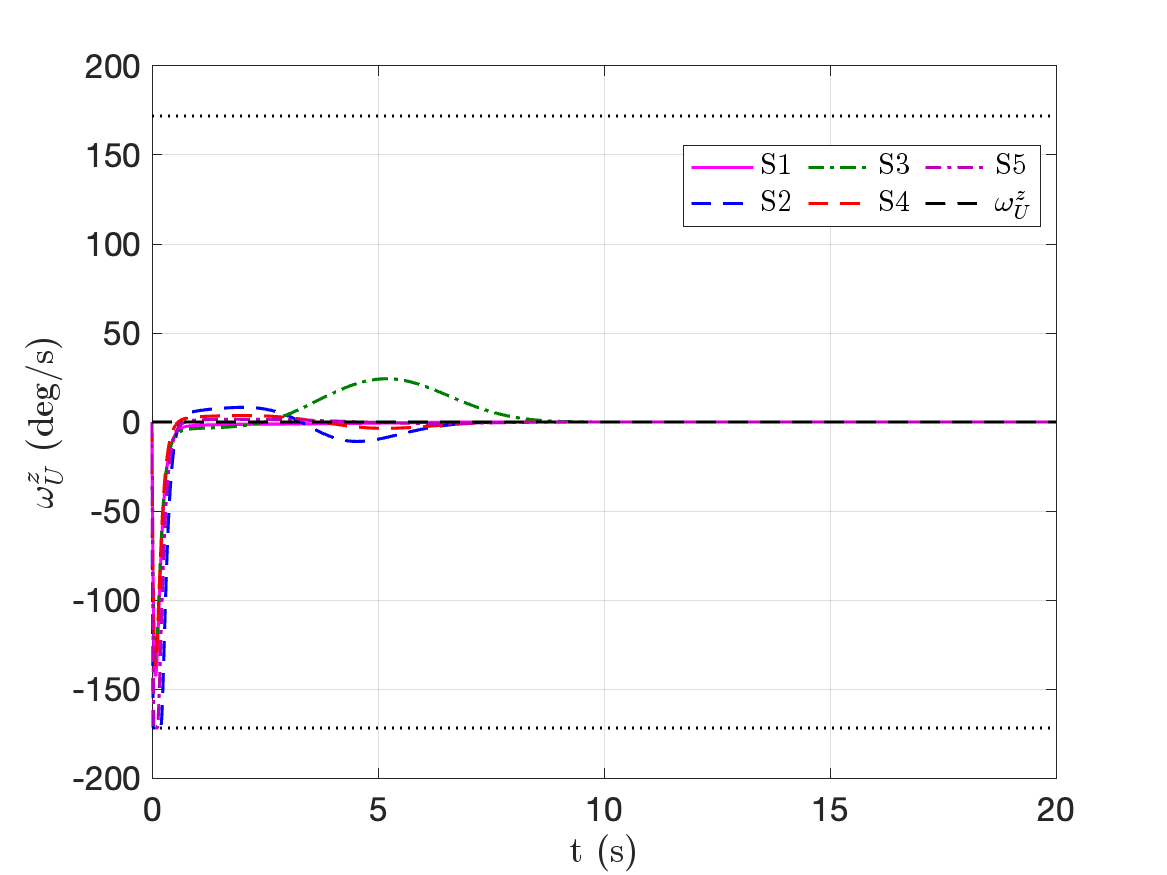}
      \caption{UAV's angular velocities in pitch plane.}
    \label{fig:stline_omega_u_z}
  \end{subfigure}%
      \begin{subfigure}{0.5\linewidth}
      \centering
      \includegraphics[width=\linewidth]{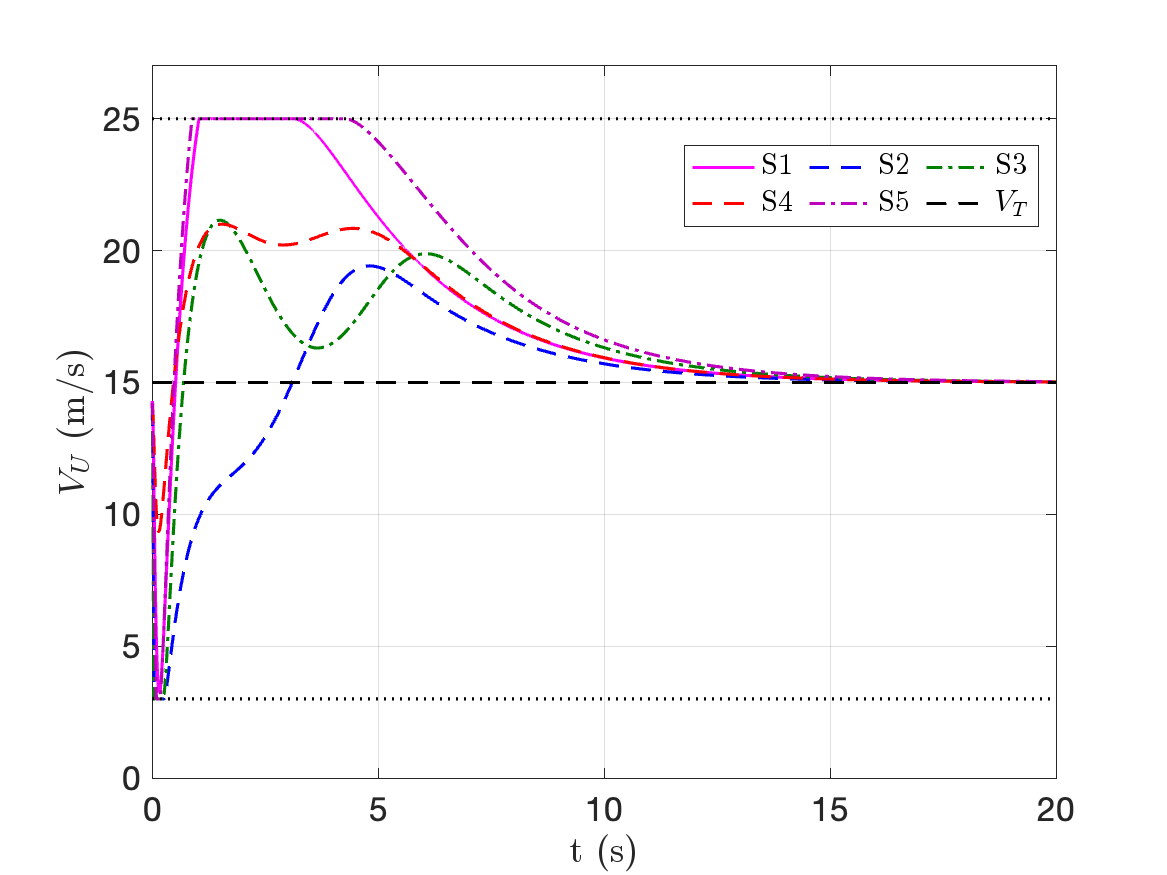}
      \caption{UAV's linear speeds.}
    \label{fig:stline_vu}
  \end{subfigure}
      \begin{subfigure}{0.5\linewidth}
      \centering
      \includegraphics[width=\linewidth]{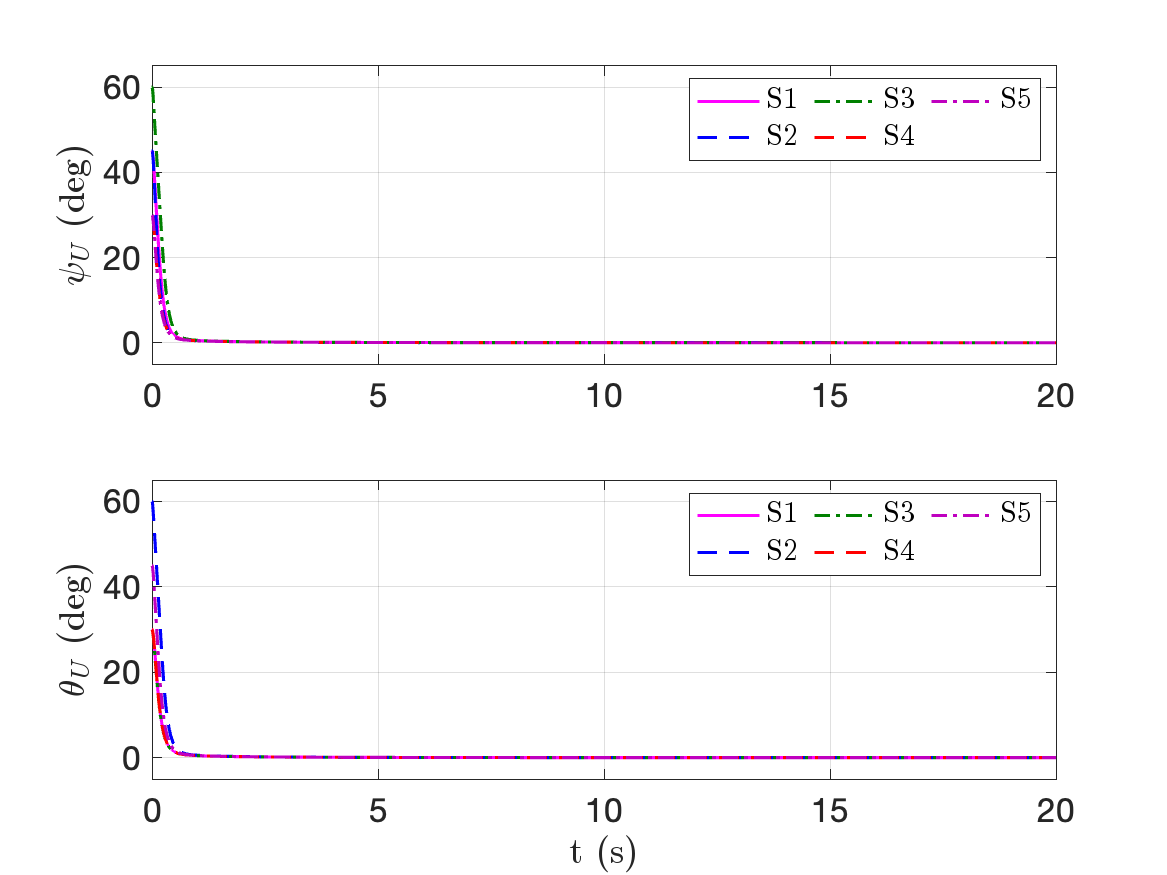}
      \caption{UAV's lead angles.}
    \label{fig:stline_angle_error}
  \end{subfigure}%
  \begin{subfigure}{0.5\linewidth}
      \centering
      \includegraphics[width=\linewidth]{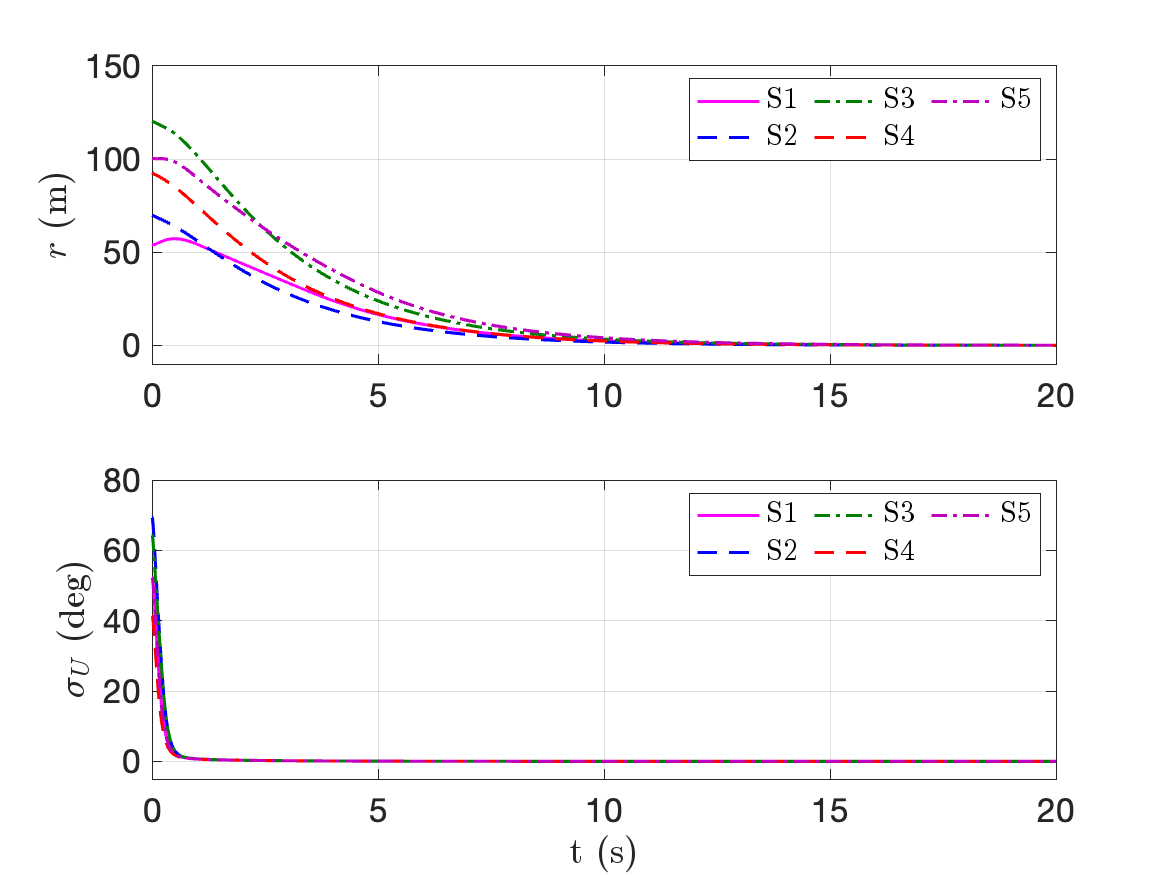}
      \caption{Relative range and effective heading angle.}
    \label{fig:stline_rsigmau}
  \end{subfigure}
    \caption{UAV following a straight-line path from various initial configurations.}
    \label{fig:stline}
\end{figure}
For the initial engagement scenarios given in \Cref{tab:stline_initial_condition}, we demonstrate the merits of the proposed strategy in a straight-line path-following scenario through \Cref{fig:stline}. The straight-line path is generated by moving a pseudo-target, which is initially located at $(40,30,20)$ m with lead angles $15^\circ$ and $15^\circ$ in azimuth and elevation directions, with $V_T=15$ m/s and $\omega_T^y=\omega_T^z=0$ rad/sec. We denote the various engagement scenarios listed in \Cref{tab:stline_initial_condition} though S1--S5 in the corresponding plots. One can observe from \Cref{fig:stline_path} that despite where the UAV starts, it follows the predefined straight-line path while its control inputs remain within the corresponding predefined safe sets $\mathcal{S}_v$ and $\mathcal{S}_{\omega}$ (see \Cref{fig:stline_vu,fig:stline_omega_u_y,fig:stline_omega_u_z}). It is also important to note that the UAV maintains the pursuit guidance philosophy, thereby nullifying lead angles $(\psi_{U},\theta_{U})$ to zero first and then driving the relative range $(r)$ to zero ( refer \Cref{fig:stline_angle_error,fig:stline_rsigmau}), leading to a path-following behavior. In fact, one may also notice from  \Cref{fig:stline_path} that regardless of where the UAV starts, it always follows its desired straight-line path by exhibiting similar path-following behavior. This attests to the global convergent property of the proposed path-following guidance strategy, unlike the asymptotic, exponential, or finite-time guidance strategies in which the performance relies on the initial engagement scenarios.
\begin{table}[!ht]
\centering
\caption{Initial conditions for \Cref{fig:stline}.}
\begin{tabular}{cccccc}
\hline \hline
Scenario &$X(0)$ & $Y(0)$& $Z(0)$& $\psi_U(0)$&$\theta_U(0)$\\
\hline
S1 &   0 m    &  0 m  & 0 m   & $45^\circ$& $30^\circ$\\
S2 &  100 m  &  0 m  &  0 m  & $45^\circ$&  $60^\circ$ \\
S3 &  100 m  &  0 m  & 120 m & $60^\circ$&  $30^\circ$\\
S4 &  50 m   & 120 m & 0 m   & $30^\circ$& $30^\circ$\\
S5 &  0 m    & 120 m & 0 m   & $30^\circ$& $45^\circ$ \\
\hline\hline
\end{tabular}
\label{tab:stline_initial_condition}
\end{table}

It is important to note that the upper bound on the time of error convergence remains approximately the same in these cases so far. This is due to the fact that the performance of the UAV remains unaffected when the geometrical conditions, such as the path’s curvature and the location of the UAV and the path, change. This makes the proposed design robust and appealing and elucidates its wide applicability in scenarios like urban air mobility where the UAV has to traverse under incomplete information by making tight maneuvers.

\section{Conclusions} \label{sec:conclusion}
In this paper, we proposed a simple, effective, and easy-to-implement nonlinear path-following guidance law that steers the UAV onto a desired curvilinear path in the three-dimensional setting within a fixed time, regardless of its initial configurations. The proposed approach accounted for the bounded control authority in the design, which makes it more suitable for real-world deployment. The proposed guidance strategy offered a generic treatment of the path-following problem so that the vehicle (multi-rotors or fixed-wing aircraft) could converge to any predefined path under some mild assumptions. Unlike the existing strategies, the proposed strategy did not require a separate switching algorithm in case the path consists of multiple segments. Furthermore, the proposed strategy only relied on the relative information for implementation, which makes it suitable for scenarios where the complete information is unavailable or hard to obtain, such as urban areas, indoors, GPS-denied environments, etc.

\section*{Appendices}
The detailed proofs of the theorems in this paper are presented here.
\subsection*{Appendix I}
\begin{proof}[Proof of \Cref{thm:linear_speed_sat_model}:]
    The boundedness of the commanded linear speed implies that $\exists\,\xi_{\rm M}>0 : \lvert \U^{c} \rvert \leq \xi_{\rm M}$. For $\lvert \U \rvert = \U^{\max}$, the expression in \Cref{eq:vu_saturation_model} becomes
    \begin{equation*}
         \dot{\U} = - \mathcal{K}\textsubscript{1}\mathcal{K}\textsubscript{2} \U,
    \end{equation*}
and hence $\lvert \U \rvert$ will decrease since $\mathcal{K}\textsubscript{1}$ and $\mathcal{K}\textsubscript{2}$ are strictly positive. Therefore, one may conclude that $\lvert \U \rvert  \leq \U^{\max}$. While the \textit{inequality} is feasible, we now show that the \textit{equality} will never hold. To that end, the saturation model in \Cref{eq:vu_saturation_model} can be written as
\begin{equation}\label{eq:vu_saturation_model_1}
    \dot{\U} =  \mathcal{K}\textsubscript{1}\left[ \left( 1 - \left(\dfrac{\U}{\U^{\max}}\right)^{\gamma} \right)  \U^{c} - \mathcal{K}\textsubscript{2} \U \right].
\end{equation}
Note that the term $\mathcal{K}_1\left[ 1 - \left(\dfrac{\U}{\U^{\max}}\right)^{\gamma} \right]$ is always non-negative. Now, if $\U$ and $\U^c$ have opposite signs, two cases are possible. First, when $\U > 0$ and $\U^c < 0$, $\dot{\U}$ will be negative, implying $\U$ will decrease. Second, when $\U < 0$ and $\U^c > 0$, $\dot{\U}$ will be positive, implying $\U$ to increase. Consequently, $\U$ will remain within the specified bounds. Therefore, the only scenario where $\lvert \U \rvert$ can increase and have a probability to go beyond the specified bounds is when $\U$ and $\U^c$ have the same sign, that is, $\U \U^c \geq 0$. Without loss of generality, let us assume that $\U^{c} > 0$ and $ \U  \geq 0$, then using the fact that $\lvert \U^{c} \rvert \leq \xi_{\rm M}$, one may write \Cref{eq:vu_saturation_model_1} as
\begin{equation}\label{eq:vu_saturation_model_2}
     \dot{\U} \leq  \mathcal{K}\textsubscript{1} \left[ \left( 1 - \left(\dfrac{\U}{\U^{\max}}\right)^{\gamma} \right) \xi_{\rm M} - \mathcal{K}\textsubscript{2} \U \right] = \mathcal{K}\textsubscript{1} \left[ 1 - \left(\dfrac{\U}{\U^{\max}}\right)^{\gamma} -\dfrac{\mathcal{K}\textsubscript{2} \U^{\max} }{\xi_{\rm M}} \left( \dfrac{\U}{\U^{\max}}\right) \right] \xi_{\rm M},
\end{equation}
which, on some algebraic simplification, results into
\begin{equation}\label{eq:vu_saturation_model_3}
    \dot{\U} \leq  \mathcal{K}\textsubscript{1} \xi_{\rm M} \left[ 1 -\left(1 +\dfrac{\mathcal{K}\textsubscript{2} \U^{\max} }{\xi_{\rm M}} \right) \left(\dfrac{\U}{\U^{\max}} \right)^{\gamma} \right].
\end{equation}
Note that, we used the inequality $-\dfrac{\U}{\U^{\max}} \leq -\left( \dfrac{\U}{\U^{\max}} \right)^{\gamma}$ to simplify \Cref{eq:vu_saturation_model_2}. One may observe from \Cref{eq:vu_saturation_model_3} that $ \dot{\U} \leq 0$ if the term $\left[ 1 -\left(1 +\dfrac{\mathcal{K}\textsubscript{2} \U^{\max} }{\xi_{\rm M}} \right) \left(\dfrac{\U}{\U^{\max}} \right)^{\gamma} \right] \leq 0$ . That is,
\begin{equation*}
1 -\left(1 +\dfrac{\mathcal{K}\textsubscript{2} \U^{\max} }{\xi_{\rm M}} \right) \left(\dfrac{\U}{\U^{\max}} \right)^{\gamma} \leq 0,
\end{equation*}
which can be expressed as
\begin{equation}\label{eq:vu_saturation_model_final}
\U \geq \U^{\max} \left[ \dfrac{\xi_{\rm M}}{\xi_{\rm M} + \mathcal{K}\textsubscript{2}\U^{\max}}  \right] ^{1/\gamma}.
\end{equation}
Consider the term $\U^{\max} \left[ \dfrac{\xi_{\rm M}}{\xi_{\rm M} + \mathcal{K}\textsubscript{2}\U^{\max}}  \right] ^{1/\gamma}= \delta_{\rm M}$. It can be observed from \Cref{eq:vu_saturation_model_final,eq:vu_saturation_model_3} that 
$\dot{\U}$ is negative whenever $\U=\delta_{\rm M}$, which is strictly less than $\U^{\max}$. As a result, $\lvert \U\rvert \leq \delta_{\rm M} < \U^{\max}$, that is, the model output never reaches its bounds, and remains confined within the set $\mathcal{S}_{u}$.
\end{proof}

\subsection*{Appendix II}
\begin{proof} [Proof of \Cref{thm:vu}]
Let us choose a continuous, positive definite, and radially unbounded Lyapunov function candidate, $\mathcal{V}\textsubscript{1}(r)=r$. This is a valid Lyapunov function candidate as $\mathcal{V}\textsubscript{1}(r)>0$ if $r\neq 0$, and $\mathcal{V}\textsubscript{1}(0)=0$. Also, $\mathcal{V}\textsubscript{1}(r)\to\infty$ if $r\to\infty$ verifies the radial unboundedness of the chosen Lyapunov function candidate. From hereafter, we drop the argument of $\mathcal{V}\textsubscript{1}(r)$ for brevity. On differentiating $\mathcal{V}\textsubscript{1}$ with respect to time and using \Cref{eq:rdot_agumented}, one may obtain
\begin{equation} \label{eq:dotv1} \dot{\mathcal{V}}\textsubscript{1}=\dot{r}=V_T\cos{\theta_T}\cos{\psi_T}- \left( \U + \U^{\max} + V\textsubscript{0}\right)\cos{\theta_U}\cos{\psi_U}.
\end{equation}
Using the fact that $\U = x + \chi $, the expression in \Cref{eq:dotv1} can be written as 
\begin{equation}
\dot{\mathcal{V}}\textsubscript{1}=V_T\cos{\theta_T}\cos{\psi_T}-\left( x + \chi + \U^{\max} + V\textsubscript{0}\right)\cos{\theta_U}\cos{\psi_U},    
\end{equation}
which on substituting the value of $\chi$ from \Cref{eq:chi} results into
\begin{align} 
\dot{\mathcal{V}}\textsubscript{1}&= -\left( x + \dfrac{V_{T}\cos{\theta}_{T}\cos{\psi}_{T} - \left(\U^{\max}+V\textsubscript{0}\right)\cos{\theta_U}\cos{\psi_U} + \left(\mathcal{M}\textsubscript{1}  r ^{\alpha\textsubscript{1}} + N\textsubscript{1}  r  ^{\beta\textsubscript{1}}\right)}{\cos{\theta}_{U}\cos{\psi}_{U}} + \U^{\max} + V\textsubscript{0}\right)\cos{\theta_U}\cos{\psi_U}\nonumber \\
&+V_T\cos{\theta_T}\cos{\psi_T},\nonumber\\
&= - x \cos{\theta_U}\cos{\psi_U} - \left(\mathcal{M}\textsubscript{1}  r ^{\alpha\textsubscript{1}} + N\textsubscript{1}  r  ^{\beta\textsubscript{1}}\right). \label{eq:vdot_final}
\end{align}
Now consider another Lyapunov function candidate $\mathcal{V}\textsubscript{2}$ as $ \mathcal{V}\textsubscript{2} = \mathcal{V}\textsubscript{1} + \lvert x \rvert$, 
which on differentiating with respect to time results into
\begin{equation} \label{eq:wdot1}
    \dot{\mathcal{V}}\textsubscript{2} = \dot{\mathcal{V}}\textsubscript{1} + \dot{x}\sign(x),
\end{equation}
Using the facts that $\U = x + \chi$ and $\dot{x} = \dot{\U} - \dot{\chi}$ together with  \Cref{eq:vu_agumented,eq:vdot_final}, the expression \Cref{eq:wdot1} becomes
\begin{equation}\label{eq:wdot2}
 \dot{\mathcal{V}}\textsubscript{2} = - x\cos{\theta_U}\cos{\psi_U} - \left(\mathcal{M}\textsubscript{1}  r ^{\alpha\textsubscript{1}} + \mathcal{N}\textsubscript{1}  r  ^{\beta\textsubscript{1}}\right) +  \left(\left\{ \mathcal{K}\textsubscript{1} \left[ 1 - \left(\dfrac{\U}{\U^{\max}}\right)^{\gamma} \right] \right\}  \U^{c} - \mathcal{K}\textsubscript{1}\mathcal{K}\textsubscript{2} \U - \dot{\chi} \right)\sign(x).
\end{equation}
Substituting the value of $\U^{c}$ from \Cref{eq:vu} into \Cref{eq:wdot2} yields
\begin{align} \label{eq:wdot3}
\dot{\mathcal{V}}\textsubscript{2}=&~ \sign(x) \left(\left\{ \mathcal{K}\textsubscript{1} \left[ 1 - \left(\dfrac{\U}{\U^{\max}}\right)^{\gamma} \right] \right\} \times \dfrac{ \mathcal{K}\textsubscript{1}\mathcal{K}\textsubscript{2} \U + \dot{\chi} + \lvert x \rvert\cos{\theta_U}\cos{\psi_U} - \left(\mathcal{M}\textsubscript{1}  x ^{\alpha\textsubscript{1}} + \mathcal{N}\textsubscript{1}  x  ^{\beta\textsubscript{1}}\right) }{\mathcal{K}\textsubscript{1} \left[ 1 - \left(\dfrac{\U}{\U^{\max}}\right)^{\gamma} \right] } - \mathcal{K}\textsubscript{1}\mathcal{K}\textsubscript{2} \U - \dot{\chi} \right)  \nonumber\\
&~-  x \cos{\theta_U}\cos{\psi_U} - \left(\mathcal{M}\textsubscript{1}  r ^{\alpha\textsubscript{1}} + \mathcal{N}\textsubscript{1}  r  ^{\beta\textsubscript{1}}\right)
\end{align}
After some algebraic simplification, the expression in \Cref{eq:wdot3} may be written in a more convenient form as
\begin{equation}\label{eq:wdot_3}
    \dot{\mathcal{V}}\textsubscript{2}= - \left(\mathcal{M}\textsubscript{1}  r ^{\alpha\textsubscript{1}} + \mathcal{N}\textsubscript{1}  r  ^{\beta\textsubscript{1}}\right) - \left(\mathcal{M}\textsubscript{1} \lvert x \rvert^{\alpha\textsubscript{1}} + \mathcal{N}\textsubscript{1} \lvert x \rvert ^{\beta\textsubscript{1}}\right).
\end{equation}
Using the results from \Cref{lem:inequality_zuo} and the relation $r + \lvert x \rvert=\mathcal{V}\textsubscript{2}$, we may write \Cref{eq:wdot_3} as
\begin{equation}  \label{eq:wdot_final}
    \dot{\mathcal{V}}\textsubscript{2}\leq  \left[2^{(1-\alpha\textsubscript{1})}\mathcal{M}\textsubscript{1}\left( r + \lvert x \rvert \right)^{\alpha\textsubscript{1}}  + \mathcal{N}\textsubscript{1}  \left(  r + \lvert x \rvert \right)^{\beta\textsubscript{1}}\right] = - \left[\bar{\mathcal{M}\textsubscript{1}}\left( r + \lvert x \rvert \right)^{\alpha\textsubscript{1}}  + \mathcal{N}\textsubscript{1}  \left(  r + \lvert x \rvert \right)^{\beta\textsubscript{1}}\right] = -\left(\bar{\mathcal{M}\textsubscript{1}} \mathcal{V}\textsubscript{2}^{\alpha_1} + \mathcal{N}\textsubscript{1} \mathcal{V}\textsubscript{2}^{\beta_{1}}\right),
\end{equation}
which is negative definite for $\mathcal{M}\textsubscript{1}$, $\mathcal{N}\textsubscript{1}$, $\alpha\textsubscript{1}$, $\beta\textsubscript{1}>0$ with $\alpha\textsubscript{1}>1$ and $0<\beta\textsubscript{1}<1$. Using the results in \Cref{lem:fixedtime}, it follows from \Cref{eq:wdot_final} that the UAV rendezvouses with the pseudo-target on the desired path within a fixed time, given by \Cref{eq:t1}, (that can be adjusted using the design parameters $\mathcal{M}\textsubscript{1}$, $\mathcal{N}\textsubscript{1}$, $\alpha\textsubscript{1}$, and $\beta\textsubscript{1}$) regardless of the UAV-target initial separation. This concludes the proof.
\end{proof}

\subsection*{Appendix III}
\begin{proof}[Proof of \Cref{thm:omega_u_z_c}]
Consider a Lyapunov function candidate $\mathcal{W}\textsubscript{1} = \lvert \theta_{U} \rvert$, which on differentiating with respect to time along the state trajectories and using \Cref{eq:psiUdot} yields
\begin{equation*}
    \dot{\mathcal{W}}\textsubscript{1} = \dot{\theta}_{U} \sign(\theta_{U}) = \left(\omega_U^z-\dot{\psi}\sin\theta\sin\psi_U-\dot{\theta}\cos\psi_U \right) \sign(\theta_{U}),
\end{equation*}
which can be simplified using the relation $z=\omega_{U}^{z} - \eta$ to
\begin{equation}\label{eq:dot_w1_1}
    \dot{\mathcal{W}}\textsubscript{1} = \left(z + \eta-\dot{\psi}\sin\theta\sin\psi_U-\dot{\theta}\cos\psi_U \right)  \sign(\theta_{U}).
\end{equation}
With the auxiliary control $\eta$ given in \Cref{eq:eta}, the expression in \Cref{eq:dot_w1_1} becomes
\begin{equation}\label{eq:dot_w1_final}
\dot{\mathcal{W}}\textsubscript{1} = z  \sign(\theta_{U}) - \left(\mathcal{M}\textsubscript{2}  \lvert \theta_{U} \rvert ^{\alpha\textsubscript{2}} + \mathcal{N}\textsubscript{2}  \lvert \theta_{U} \rvert  ^{\beta\textsubscript{2}}\right) .
\end{equation}
Consider another Lyapunov function candidate as $\mathcal{W}\textsubscript{2} = \mathcal{W}\textsubscript{1} + \lvert z \rvert$, whose time derivative results in
\begin{equation}\label{eq:dot_w2_1}
\dot{\mathcal{W}\textsubscript{2}} =  \dot{\mathcal{W}}\textsubscript{1} +  \dot{z}\sign(z).
\end{equation}
Using the fact that $\dot{z} = \dot{\omega}_{U}^{z} - \dot{\eta}$ and substituting the value of $\dot{\omega}_{U}^{z}$ from \Cref{eq:dot_omega_u_z} lead us to arrive at
\begin{equation*}
\dot{\mathcal{W}\textsubscript{2}} =  \dot{\mathcal{W}}\textsubscript{1} +   \left( \left( 1- f_{z} \right)  \mathcal{K}\textsubscript{3} \omega_{U}^{c,z} - \mathcal{K}\textsubscript{3}\mathcal{K}\textsubscript{4} \omega_{U}^{z} - \dot{\eta}\right)\sign(z),
\end{equation*}
which on substituting for $\dot{\mathcal{W}}\textsubscript{1}$ and $\omega_{U}^{c,z}$ from \Cref{eq:dot_w1_final,eq:omega_z_c} yields
\begin{align}
\dot{\mathcal{W}\textsubscript{2}} =&~  \left( \left( 1- f_{z} \right)  \mathcal{K}\textsubscript{3}\times\dfrac{\mathcal{K}\textsubscript{3}\mathcal{K}\textsubscript{4} \omega_{U}^{z} + \dot{\eta} -\lvert z \rvert \sign(\theta_{U}) - \left(\mathcal{M}\textsubscript{2}   z  ^{\alpha\textsubscript{2}} + \mathcal{N}\textsubscript{2}   z   ^{\beta\textsubscript{2}}\right) } {\mathcal{K}\textsubscript{3} \left( 1- f_{z} \right)}- \mathcal{K}\textsubscript{3}\mathcal{K}\textsubscript{4} \omega_{U}^{z} - \dot{\eta}\right) \sign(z) \nonumber\\
&~ + z\sign(\theta_{U}) - \left(\mathcal{M}\textsubscript{2}  \lvert \theta_{U} \rvert ^{\alpha\textsubscript{2}} + N\textsubscript{2}  \lvert \theta_{U} \rvert  ^{\beta\textsubscript{2}}\right). \label{eq:dot_w2_2}
\end{align}
After some algebraic simplifications, one may express \Cref{eq:dot_w2_2} as
\begin{equation*}
\dot{\mathcal{W}\textsubscript{2}} = - \left(\mathcal{M}\textsubscript{2}  \lvert \theta_{U} \rvert ^{\alpha\textsubscript{2}} + \mathcal{N}\textsubscript{2}  \lvert \theta_{U} \rvert  ^{\beta\textsubscript{2}}\right)- \left(\mathcal{M}\textsubscript{2}  \lvert z \rvert ^{\alpha\textsubscript{2}} + \mathcal{N}\textsubscript{2}  \lvert z \rvert  ^{\beta\textsubscript{2}}\right),
\end{equation*}
which can be written using the results in \Cref{lem:inequality_zuo} as
\begin{align}
\dot{\mathcal{W}\textsubscript{2}} &\leq - 2^{(1-\alpha\textsubscript{2})}\mathcal{M}\textsubscript{2}  \left(\lvert \theta_{U} \rvert + \lvert z \rvert\right) ^{\alpha\textsubscript{2}} - \mathcal{N}\textsubscript{2}  \left(\lvert \theta_{U} \rvert + \lvert z \rvert\right)^{\beta\textsubscript{2}} \nonumber\\
&=  - \bar{\mathcal{M}\textsubscript{2}}  \left(\lvert \theta_{U} \rvert + \lvert z \rvert\right) ^{\alpha\textsubscript{2}} - \mathcal{N}\textsubscript{2}  \left(\lvert \theta_{U} \rvert + \lvert z \rvert\right)^{\beta\textsubscript{2}} = - \bar{\mathcal{M}\textsubscript{2}} \mathcal{W}\textsubscript{2}^{\alpha\textsubscript{2}} - \mathcal{N}\textsubscript{2}\mathcal{W}\textsubscript{2}^{\beta\textsubscript{2}}. \label{eq:dot_w2_final}
\end{align}
The expression in \Cref{eq:dot_w2_final} is negative definite for $\mathcal{M}\textsubscript{2}$, $\mathcal{N}\textsubscript{2}$, $\alpha\textsubscript{2}$, $\beta\textsubscript{2}>0$ with $\alpha\textsubscript{2}>1$ and $0<\beta\textsubscript{2}<1$. Therefore, using the results in \Cref{lem:fixedtime}, it follows from \Cref{eq:dot_w2_final} that the UAV's elevation angle will be nullified within a fixed-time $T\textsubscript{2}$ irrespective of the UAV's initial configuration relative to the pseudo-target.
\end{proof}

\subsection*{Appendix IV}
\begin{proof}[proof of \Cref{thm:omega_u_y_c}]
We choose a Lyapunov function candidate $\mathcal{W}\textsubscript{3} = \lvert \psi_{U} \rvert$, whose time derivative can be obtained using \Cref{eq:psiUdot} as
\begin{equation*}
    \dot{\mathcal{W}}\textsubscript{3} = \dot{\psi}_{U} \sign(\psi_{U}) = \left(\dfrac{\omega_U^y}{\cos\theta_U}+\dot{\psi}\tan\theta_U\cos\psi_U\sin\theta-\dot{\psi}\cos\theta-\dot{\theta}\tan\theta_U\sin\psi_U \right) \sign(\psi_{U}),
\end{equation*}
and which can be written using the relation $y=\omega_{U}^{z} - \lambda$ as
\begin{equation}\label{eq:dot_w3_1}
    \dot{\mathcal{W}}\textsubscript{1} = \left(\dfrac{y + \lambda}{\cos\theta_U}+\dot{\psi}\tan\theta_U\cos\psi_U\sin\theta-\dot{\psi}\cos\theta-\dot{\theta}\tan\theta_U\sin\psi_U \right) \sign(\psi_{U}).
\end{equation}
If we choose the auxiliary control $\lambda$ as per \Cref{eq:lambda}, then the expression in \Cref{eq:dot_w3_1} becomes
\begin{equation}\label{eq:dot_w3_final}
\dot{\mathcal{W}}\textsubscript{3} = \dfrac{y}{\cos\theta_{U}}\sign(\psi_{U}) - \left(\mathcal{M}\textsubscript{3}  \lvert \psi_{U} \rvert ^{\alpha\textsubscript{3}} + \mathcal{N}\textsubscript{3}  \lvert \psi_{U} \rvert  ^{\beta\textsubscript{3}}\right).
\end{equation}
Now, we consider another Lyapunov function candidate as $\mathcal{W}\textsubscript{4} = \mathcal{W}\textsubscript{3} + \lvert y \rvert $, whose time derivative yields
\begin{equation}\label{eq:dot_w4_1}
\dot{\mathcal{W}\textsubscript{4}} =  \dot{\mathcal{W}}\textsubscript{3} +  \dot{y} \sign(y).
\end{equation}
By using the relation $\dot{y} = \dot{\omega}_{U}^{y} - \dot{\lambda}$ and substituting the value of $\dot{\omega}_{U}^{y}$ from \Cref{eq:dot_omega_u_y}, one may obtain
\begin{equation*}
\dot{\mathcal{W}\textsubscript{4}} =  \dot{\mathcal{W}}\textsubscript{3} +  \left( \left( 1- f_{y} \right)  \mathcal{K}\textsubscript{3} \omega_{U}^{c,y} - \mathcal{K}\textsubscript{3}\mathcal{K}\textsubscript{4} \omega_{U}^{y} - \dot{\lambda}\right)\sign(y),
\end{equation*}
which on substituting for $\dot{\mathcal{W}}\textsubscript{1}$ and $\omega_{U}^{c,z}$ from \Cref{eq:dot_w3_final,eq:omega_y_c} yields
\begin{align}\label{eq:dot_w4_2}
\dot{\mathcal{W}\textsubscript{4}} =&~ 
  \left( \dfrac{\mathcal{K}\textsubscript{3}\mathcal{K}\textsubscript{4} \omega_{U}^{y} + \dot{\lambda} - \dfrac{\lvert y \rvert \sign(\psi_{U})}{\cos{\theta_{U}}} - \left(\mathcal{M}\textsubscript{3}   y  ^{\alpha\textsubscript{3}} + \mathcal{N}\textsubscript{3}   y   ^{\beta\textsubscript{3}}\right) }{\mathcal{K}\textsubscript{3} \left( 1- f_{y} \right)}  \times \left( 1- f_{y} \right)  \mathcal{K}\textsubscript{3}\nonumber - \mathcal{K}\textsubscript{3}\mathcal{K}\textsubscript{4} \omega_{U}^{y} - \dot{\lambda}\right)\sign(y)\\
  &~+\dfrac{y\sign(\psi_{U})}{\cos\theta_{U}} - \left(\mathcal{M}\textsubscript{3}  \lvert \psi_{U} \rvert ^{\alpha\textsubscript{3}} + \mathcal{N}\textsubscript{3}  \lvert \psi_{U} \rvert  ^{\beta\textsubscript{3}}\right).
\end{align}
After some algebraic simplifications, one may write \Cref{eq:dot_w4_2} as
\begin{equation*}
\dot{\mathcal{W}\textsubscript{4}} = - \left(\mathcal{M}\textsubscript{3}  \lvert y \rvert ^{\alpha\textsubscript{3}} + \mathcal{N}\textsubscript{3}  \lvert y \rvert  ^{\beta\textsubscript{3}}\right) - \left(\mathcal{M}\textsubscript{3}  \lvert \psi_{U} \rvert ^{\alpha\textsubscript{3}} + \mathcal{N}\textsubscript{3}  \lvert \psi_{U} \rvert  ^{\beta\textsubscript{3}}\right),
\end{equation*}
which, according to \Cref{lem:inequality_zuo}, becomes
\begin{equation}\label{eq:dot_w4_final}
\dot{\mathcal{W}\textsubscript{4}} \leq - 2^{(1-\alpha\textsubscript{3})}\mathcal{M}\textsubscript{3}\left( \lvert \psi_{U} \rvert +\lvert y \rvert \right)^{\alpha\textsubscript{3}} - \mathcal{N}\textsubscript{3}\left( \lvert \psi_{U} \rvert +\lvert y \rvert \right)^{\beta\textsubscript{3}} =  - \bar{\mathcal{M}\textsubscript{3}} \mathcal{W}\textsubscript{4}^{\alpha\textsubscript{3}} - \mathcal{N}\textsubscript{3}\mathcal{W}\textsubscript{4}^{\beta\textsubscript{3}}.
\end{equation}
The expression in \Cref{eq:dot_w4_final} is negative definite for $\mathcal{M}\textsubscript{3}$, $\mathcal{N}\textsubscript{3}$, $\alpha\textsubscript{3}$, $\beta\textsubscript{3}>0$ with $\alpha\textsubscript{3}>1$ and $0<\beta\textsubscript{3}<1$. Therefore, using the results in \Cref{lem:fixedtime}, it follows from \Cref{eq:dot_w4_final} that the UAV's azimuth angle will be nullified within a fixed-time $T\textsubscript{3}$ irrespective of the UAV-target initial configuration.
\end{proof}
\bibliographystyle{ieeetr}
\bibliography{references.bib}
\end{document}